\documentclass[twocolumn,trackchanges]{aastex7}

\usepackage{hyperref,amsmath, bm, empheq, mathrsfs,  cancel, multirow, xcolor,array,natbib,verbatim} 

\usepackage{placeins}
\allowdisplaybreaks

\makeatletter

\usepackage{etoolbox}
\patchcmd{\thebibliography}{\clearpage}{}{}{}
\patchcmd{\thebibliography}{\newpage}{}{}{}
\AtBeginDocument{
  \setlength{\bibsep}{0pt plus 0.3ex}
}

\makeatother

\begin{document} 

\author[orcid=0000-0001-9001-6118,gname='Loren',sname='Matilsky']{Loren I. Matilsky}
\altaffiliation{U.S. National Science Foundation\\ Astronomy and Astrophysics Postdoctoral Fellow}
\affiliation{Department of Applied Mathematics,
Baskin School of Engineering,
University of California, Santa Cruz, CA 95064-1077, USA}
\email[show]{loren.matilsky@gmail.com}  

\author[0000-0002-5971-8995,gname=Geoffrey, sname=Vallis]{Geoffrey K. Vallis} 
\affiliation{Department of Mathematics and Statistics, 
University of Exeter, 
Exeter, EX4 4QF, UK
}
\email{gkvallis@gmail.com}

\author[0000-0002-8634-1003,gname=Matthew,sname=Browning]{Matthew K. Browning} 
\affiliation{Department of Physics and Astronomy,
University of Exeter,
Exeter, EX4 4QL, UK
}
\email{M.K.M.Browning@exeter.ac.uk}

\author[orcid=0000-0003-4350-5183,gname='Nicholas',sname='Brummell']{Nicholas H. Brummell}
\affiliation{Department of Applied Mathematics,
Baskin School of Engineering,
University of California,
Santa Cruz, CA 95064-1077, USA}
\email{brummell@ucsc.edu}  


\providecommand{\testmacro}{\text{macros.tex v. 2025-01-02 6pm}}
\providecommand{\myemail}{loren.matilsky@gmail.com}

\providecommand{\sn}[2]{#1\times10^{#2}}
\providecommand{\sncomp}[2]{{#1}e{#2}}

\providecommand{\cz}{_{\rm{cz}}}
\providecommand{\rz}{_{\rm{rz}}}
\providecommand{\czt}{_{{\rm cz},t}}
\providecommand{\rzt}{_{{\rm rz},t}}
\providecommand{\fulll}{_{\rm{full}}}
\providecommand{\dimm}{_{\rm{dim}}}
\providecommand{\rad}{_{\rm{r}}}
\providecommand{\nrad}{_{\rm{nr}}}
\providecommand{\loc}{_{\rm{loc}}}
\providecommand{\est}{_{\rm{est}}}

\providecommand{\sgn}[1]{{\text{sgn}}(#1)}

\providecommand{\pderiv}[2]{\dfrac{\partial#1}{\partial#2}}
\providecommand{\matderiv}[1]{\frac{D#1}{Dt}}
\providecommand{\pderivline}[2]{\partial#1/\partial#2}
\providecommand{\parenfrac}[2]{\left(\frac{#1}{#2}\right)}
\providecommand{\brackfrac}[2]{\left[\frac{#1}{#2}\right]}
\providecommand{\bracefrac}[2]{\left\{\frac{#1}{#2}\right\}}
\providecommand{\sign}[1]{{\text{sign}}(#1)}

\providecommand{\av}[1]{\left\langle#1\right\rangle}
\providecommand{\abs}[1]{\left\lvert#1\right\rvert}
\providecommand{\magnitudeof}[1]{\left[#1\right]}
\providecommand{\avsph}[1]{\left\langle#1\right\rangle_{\rm{sph}}}
\providecommand{\avspht}[1]{\left\langle#1\right\rangle_{ {\rm sph}, t}}
\providecommand{\avt}[1]{\left\langle#1\right\rangle_{t}}
\providecommand{\avtpar}[1]{\left(#1\right)_{t}}
\providecommand{\avphi}[1]{\left\langle#1\right\rangle_{\phi}}
\providecommand{\avphit}[1]{\left\langle#1\right\rangle_{\phi,t}}
\providecommand{\avvol}[1]{\left\langle#1\right\rangle_{\rm{v}}}

\providecommand{\avcz}[1]{\left\langle#1\right\rangle\cz}
\providecommand{\avrz}[1]{\left\langle#1\right\rangle\rz}
\providecommand{\avfull}[1]{\left\langle#1\right\rangle\fulll}
\providecommand{\avczt}[1]{\left\langle#1\right\rangle\czt}
\providecommand{\avrzt}[1]{\left\langle#1\right\rangle\rzt}
\providecommand{\avfullt}[1]{\left\langle#1\right\rangle_{{\rm full}, t}}

\providecommand{\avalt}[1]{\langle#1\rangle}
\providecommand{\avaltsph}{\overline}
\providecommand{\avaltspht}[1]{\left(\overline{#1}\right)_{t}}

\providecommand{\dbprime}{^{\prime\prime}}

\providecommand{\define}{\coloneqq}
\providecommand{\definealt}{\equiv}

\providecommand{\five}{\ \ \ \ \ }
\providecommand{\ten}{\five\five}
\providecommand{\twenty}{\ten\ten}
\providecommand{\orr}{\text{or}\five }
\providecommand{\andd}{\text{and}\five }
\providecommand{\where}{\text{where}\five }
\providecommand{\with}{\text{with}\five }


\providecommand{\curl}{\nabla\times}
\providecommand{\Div}{\nabla\cdot}
\providecommand{\lap}{\nabla^2}
\providecommand{\dotgrad}{\cdot\nabla}
\providecommand{\ugrad}{\bm{u}\dotgrad}
\providecommand{\vgrad}{\bm{v}\dotgrad}

\providecommand{\e}{\hat{\bm{e}}}
\providecommand{\erad}{\e_r}
\providecommand{\etheta}{\e_\theta}
\providecommand{\ephi}{\e_\phi}
\providecommand{\elambda}{\e_\lambda}
\providecommand{\ez}{\e_z}
\providecommand{\exi}{\e_\xi}
\providecommand{\eeta}{\e_\eta}
\providecommand{\epol}{\e_{\rm{pol}}}
\providecommand{\emer}{\e_{\rm{mer}}}
\providecommand{\ihat}{\hat{\bm{i}}}
\providecommand{\jhat}{\hat{\bm{j}}}
\providecommand{\khat}{\hat{\bm{k}}}
\providecommand{\eone}{\e_1}
\providecommand{\etwo}{\e_2}
\providecommand{\ethree}{\e_3}

\providecommand{\flux}{{\bm{F}}}

\providecommand{\fluxrad}{\flux_{\rm{r}}}
\providecommand{\fluxnrad}{\flux_{\rm{nr}}}
\providecommand{\fluxcond}{\flux_{\rm{c}}}
\providecommand{\fluxenth}{\flux_{\rm{e}}}

\providecommand{\fluxscalarrad}{F_{\rm{r}}}
\providecommand{\fluxscalarnrad}{F_{\rm{nr}}}
\providecommand{\fluxscalarcond}{F_{\rm{c}}}
\providecommand{\fluxscalarenth}{F_{\rm{e}}}

\providecommand{\Omzero}{\Omega_0}
\providecommand{\twoOmzero}{2\Omega_0}
\providecommand{\Omzerovec}{\bm{\Omega}_0}
\providecommand{\Omsun}{\Omega_\odot}
\providecommand{\Omrz}{\Omega\rz}
\providecommand{\Omcz}{\Omega\cz}
\providecommand{\bruntsun}{N_\odot}
\providecommand{\dOm}{\Delta\Omega}
\providecommand{\dOmsixty}{\dOm_{\rm 60}}
\providecommand{\dOmcz}{\Delta\Omega\cz}
\providecommand{\dOmrz}{\Delta\Omega\rz}
\providecommand{\dOmczt}{\Delta\Omega\czt}
\providecommand{\dOmrzt}{\Delta\Omega\rzt}
\providecommand{\dOmloc}{\Delta\Omega\loc}
\providecommand{\Deltaloc}{\Delta\loc}


\providecommand{\ofr}{(r)}
\providecommand{\rprime}{{r^{\prime}}}
\providecommand{\ofrprime}{(\rprime)}

\providecommand{\cv}{c_{\rm{v}}}
\providecommand{\cp}{c_{\rm{p}}}
\providecommand{\cvcap}{C_{\rm{v}}}
\providecommand{\cpcap}{C_{\rm{p}}}
\providecommand{\cs}{c_{\rm s}}
\providecommand{\gasconst}{\mathcal{R}}
\providecommand{\gammaone}{\Gamma_1}

\providecommand{\tot}{_{\rm{tot}}}
\providecommand{\rhotot}{\rho\tot}
\providecommand{\tmptot}{T\tot}
\providecommand{\prstot}{P\tot}
\providecommand{\stot}{S\tot}
\providecommand{\dsdrtot}{\frac{dS\tot}{dr}}
\providecommand{\dsdrtotline}{dS\tot/dr}

\providecommand{\Dentr}{\Delta s}

\providecommand{\rhoover}{\overline{\rho}}
\providecommand{\tmpover}{\overline{T}}
\providecommand{\prsover}{\overline{P}}
\providecommand{\entrover}{\overline{s}}
\providecommand{\inteover}{\overline{u}}
\providecommand{\enthover}{\overline{h}}
\providecommand{\heatover}{\overline{Q}}
\providecommand{\heatradover}{\overline{Q}_{\rm r}}
\providecommand{\coolover}{\overline{C}}
\providecommand{\bruntsqover}{\overline{N^2}}
\providecommand{\bruntover}{\overline{N}}
\providecommand{\gravover}{\overline{g}}
\providecommand{\nuover}{\overline{\nu}}
\providecommand{\kappaover}{\overline{\kappa}}
\providecommand{\etaover}{\overline{\eta}}
\providecommand{\muover}{\overline{\mu}}
\providecommand{\deltaover}{\overline{\delta}}
\providecommand{\cpover}{\overline{\cp}}
\providecommand{\cvover}{\overline{\cv}}
\providecommand{\csover}{\overline{\cs}}
\providecommand{\cssqover}{\overline{\cs^2}}
\providecommand{\hrhoover}{\overline{H_\rho}}
\providecommand{\Dentrover}{\Delta\entrover}
\providecommand{\Dentroverf}{\Delta\entrover_{\rm f}}

\providecommand{\fluxradover}{\overline{\flux}_{\rm{r}}}
\providecommand{\fluxscalarradover}{\overline{F}_{\rm{r}}}
\providecommand{\fluxnradover}{\overline{\flux}_{\rm{nr}}}
\providecommand{\fluxscalarnradover}{\overline{F}_{\rm{nr}}}
\providecommand{\kradover}{\overline{\kappa}_{\rm rad}}

\providecommand{\rhoa}{\rho_a}
\providecommand{\rhocz}{\rho\cz}
\providecommand{\rhorz}{\rho\rz}
\providecommand{\tmpa}{T_a}
\providecommand{\tmpcz}{T\cz}
\providecommand{\tmprz}{T\rz}
\providecommand{\cpcz}{c_{\rm p,cz}}
\providecommand{\cprz}{c_{\rm p,rz}}
\providecommand{\prsa}{p_a}
\providecommand{\prscz}{p\cz}
\providecommand{\prsrz}{p\rz}
\providecommand{\entra}{s_a}
\providecommand{\intea}{u_a}
\providecommand{\entha}{h_a}
\providecommand{\heata}{Q_a}
\providecommand{\fluxa}{F_a}
\providecommand{\hrhoa}{H_{\rho a}}
\providecommand{\hprsa}{H_{{\rm p} a}}
\providecommand{\heatrada}{\overline{Q}_{{\rm r}a}}
\providecommand{\coola}{C_a}
\providecommand{\bruntsqa}{N^2_a}
\providecommand{\bruntsqrz}{N^2\rz}
\providecommand{\bruntrz}{N\rz}
\providecommand{\grava}{g_a}
\providecommand{\gravcz}{g\cz}
\providecommand{\gravrz}{g\rz}
\providecommand{\nua}{\nu_a}
\providecommand{\nut}{\nu_{\rm t}}
\providecommand{\nurz}{\nu\rz}
\providecommand{\nucz}{\nu\cz}
\providecommand{\kappaa}{\kappa_a}
\providecommand{\kappat}{\kappa_{\rm t}}
\providecommand{\kapparz}{\kappa\rz}
\providecommand{\kappacz}{\kappa\cz}

\providecommand{\etaa}{\eta_a}
\providecommand{\etarz}{\eta\rz}
\providecommand{\etacz}{\eta\cz}
\providecommand{\mua}{\mu_a}
\providecommand{\deltaa}{\delta_a}
\providecommand{\cpa}{c_{{\rm p}a}}
\providecommand{\cva}{c_{{\rm v}a}}
\providecommand{\csa}{c_{{\rm s}a}}
\providecommand{\cssqa}{(\cs^2)_a}
\providecommand{\fluxrada}{F_{{\rm r}a}}
\providecommand{\fluxnrada}{F_{{\rm{nr}}a}}
\providecommand{\fluxnradcz}{F_{{\rm{nr,cz}}}}
\providecommand{\krada}{\kappa_{{\rm rad}a}}
\providecommand{\sigmaa}{\sigma_a}

\providecommand{\dlnrhoover}{\frac{d\ln\rhoover}{dr}}
\providecommand{\dlntmpover}{\frac{d\ln\tmpover}{dr}}
\providecommand{\dlnprsover}{\frac{d\ln\prsover}{dr}}
\providecommand{\dsdrover}{\frac{d\entrover}{dr}}
\providecommand{\dsdroverline}{d\entrover/dr}
\providecommand{\dlnrhooverline}{d\ln\rhoover/dr}
\providecommand{\dlntmpoverline}{d\ln\tmpover/dr}
\providecommand{\dlnprsoverline}{d\ln\prsover/dr}

\providecommand{\rhotilde}{\tilde{\rho}}
\providecommand{\tmptilde}{\tilde{T}}
\providecommand{\rhottilde}{\rhotilde\tmptilde}

\providecommand{\prstilde}{\tilde{p}}
\providecommand{\entrtilde}{\tilde{s}}
\providecommand{\intetilde}{\tilde{u}}
\providecommand{\enthtilde}{\tilde{h}}
\providecommand{\heattilde}{\tilde{Q}}
\providecommand{\heatradtilde}{\tilde{Q}_{\rm r}}
\providecommand{\dentr}{\delta_s}
\providecommand{\dheat}{\delta_Q}
\providecommand{\drad}{\delta_{\rm r}}
\providecommand{\rheat}{r_Q}
\providecommand{\rentr}{r_s}
\providecommand{\lumtilde}{\tilde{L}}
\providecommand{\heatra}{\tilde{Q}_{\texttt{Ra}}}
\providecommand{\cooltilde}{\tilde{C}}
\providecommand{\brunttilde}{\widetilde{N}}
\providecommand{\brunttildesq}{\widetilde{N}^2}
\providecommand{\bruntsqtilde}{\widetilde{N^2}}
\providecommand{\bruntsqtildedim}{\widetilde{N^{*2}}}
\providecommand{\gravtilde}{\tilde{g}}
\providecommand{\nutilde}{\tilde{\nu}}
\providecommand{\kappatilde}{\tilde{\kappa}}
\providecommand{\etatilde}{\tilde{\eta}}
\providecommand{\mutilde}{\tilde{\mu}}
\providecommand{\deltatilde}{\tilde{\delta}}
\providecommand{\cptilde}{\tilde{c}_{\rm p}}
\providecommand{\cvtilde}{\tilde{c}_{\rm v}}
\providecommand{\cptildedim}{\tilde{c}_{\rm p}^*}
\providecommand{\cvtildedim}{\tilde{c}_{\rm v}^*}
\providecommand{\cssqtilde}{\tilde{\cs^2}}
\providecommand{\Dentrtilde}{\Delta\entrtilde}
\providecommand{\Dentrtildef}{(\Delta\entrtilde)_{\rm f}}

\providecommand{\fluxtilde}{\tilde{\flux}}
\providecommand{\fluxscalartilde}{\tilde{F}}
\providecommand{\fluxradtilde}{\tilde{\flux}_{\rm{r}}}
\providecommand{\fluxscalarradtilde}{\tilde{F}_{\rm{r}}}
\providecommand{\fluxnradtilde}{\tilde{\flux}_{\rm{nr}}}
\providecommand{\fluxscalarnradtilde}{\tilde{F}_{\rm{nr}}}
\providecommand{\fluxcondtilde}{\tilde{\flux}_{\rm{c}}}
\providecommand{\fluxscalarcondtilde}{\tilde{F}_{\rm{c}}}

\providecommand{\dlnrhotilde}{\frac{d\ln\rhotilde}{dr}}
\providecommand{\dlntmptilde}{\frac{d\ln\tmptilde}{dr}}
\providecommand{\dlnprstilde}{\frac{d\ln\prstilde}{dr}}
\providecommand{\dsdrtilde}{\frac{d\entrtilde}{dr}}
\providecommand{\dlnrhotildeline}{d\ln\rhotilde/dr}
\providecommand{\dlntmptildeline}{d\ln\tmptilde/dr}
\providecommand{\dlnprstildeline}{d\ln\prstilde/dr}
\providecommand{\dsdrtildeline}{d\entrtilde/dr}

\providecommand{\grav}{g}
\providecommand{\vecg}{\bm{g}}
\providecommand{\geff}{g_{\rm{eff}}}
\providecommand{\vecgeff}{\bm{g}_{\rm{eff}}}
\providecommand{\heat}{Q}
\providecommand{\buoyfreq}{N}
\providecommand{\brunt}{N}
\providecommand{\bruntsq}{N^2}
\providecommand{\hrho}{H_\rho}
\providecommand{\hprs}{H_{\rm{p}}}
\providecommand{\hrhotilde}{\widetilde{H_\rho}}
\providecommand{\hprstilde}{\widetilde{H_{\rm p}}}

\providecommand{\gradrad}{\nabla_{\rm r}}
\providecommand{\gradad}{\nabla_{\rm ad}}

\providecommand{\rhoprime}{{\rho^\prime}}
\providecommand{\tmpprime}{{T^\prime}}
\providecommand{\prsprime}{{p^\prime}}
\providecommand{\entrprime}{{s^\prime}}
\providecommand{\inteprime}{{u^\prime}}
\providecommand{\heatprime}{{Q^\prime}}
\providecommand{\enthprime}{{h^\prime}}
\providecommand{\fradprime}{\bm{F}^\prime_{\rm rad}}
\providecommand{\kradprime}{\kappa^\prime_{\rm rad}}
\providecommand{\fcondprime}{\bm{F}^\prime_{\rm cond}}

\providecommand{\rhohat}{\hat{\rho}}
\providecommand{\tmphat}{\hat{T}}
\providecommand{\prshat}{\hat{p}}
\providecommand{\entrhat}{\hat{s}}
\providecommand{\intehat}{\hat{u}}
\providecommand{\enthhat}{\hat{h}}

\providecommand{\rhoone}{\rho_1}
\providecommand{\tmpone}{T_1}
\providecommand{\prsone}{p_1}
\providecommand{\entrone}{s_1}
\providecommand{\inteone}{u_1}
\providecommand{\enthone}{h_1}

\providecommand{\pomega}{\varpi}

\providecommand{\vecu}{\bm{u}}
\providecommand{\vecv}{\bm{v}}
\providecommand{\veca}{\bm{A}}
\providecommand{\vecb}{\bm{B}}
\providecommand{\vecom}{\bm{\omega}}
\providecommand{\vecj}{\bm{\mathcal{J}}}

\providecommand{\upol}{\vecu_{\rm{pol}}}
\providecommand{\bpol}{\vecb_{\rm{pol}}}
\providecommand{\umer}{\vecu_{\rm{m}}}
\providecommand{\bmer}{\vecb_{\rm{m}}}

\providecommand{\urad}{{u_r}}
\providecommand{\utheta}{{u_\theta}}
\providecommand{\uphi}{{u_\phi}}
\renewcommand{\uphi}{{u_\phi}}

\providecommand{\ulambda}{{u_\lambda}}
\providecommand{\uz}{{u_z}}

\providecommand{\rhoumer}{\av{\rhotilde\umer}}
\providecommand{\rhourad}{\av{\rhotilde\urad}}
\providecommand{\rhoutheta}{\av{\rhotilde\utheta}}
\providecommand{\rhoulambda}{\av{\rhotilde\ulambda}}
\providecommand{\rhouz}{\av{\rhotilde\uz}}
\providecommand{\rhoomphi}{\av{\rhotilde\omphi}}

\providecommand{\omrad}{\omega_r}
\providecommand{\omtheta}{\omega_\theta}
\providecommand{\omphi}{\omega_\phi}
\providecommand{\omlambda}{\omega_\lambda}
\providecommand{\omz}{\omega_z}

\providecommand{\brad}{B_r}
\providecommand{\btheta}{B_\theta}
\providecommand{\bphi}{B_\phi}
\providecommand{\blambda}{B_\lambda}
\providecommand{\bz}{B_z}

\providecommand{\jrad}{\mathcal{J}_r}
\providecommand{\jtheta}{\mathcal{J}_\theta}
\providecommand{\jphi}{\mathcal{J}_\phi}
\providecommand{\jlambda}{\mathcal{J}_\lambda}
\providecommand{\jz}{\mathcal{J}_z}

\providecommand{\vecuprime}{\bm{u}^\prime}
\providecommand{\vecvprime}{\bm{v}^\prime}
\providecommand{\vecuhat}{\hat{\bm{u}}}
\providecommand{\vecvhat}{\hat{\bm{v}}}
\providecommand{\vecbprime}{\bm{B}^\prime}
\providecommand{\vecomprime}{\bm{\omega}^\prime}
\providecommand{\vecjprime}{\bm{\mathcal{J}}^\prime}
\providecommand{\vecuover}{\overline{\bm{u}}}
\providecommand{\vecvover}{\overline{\bm{v}}}
\providecommand{\wprime}{{w^\prime}}
\providecommand{\what}{{\hat{w}}}
\providecommand{\wover}{\overline{w}}
\providecommand{\vecbover}{\overline{\bm{B}}}
\providecommand{\vecomover}{\overline{\bm{\omega}}}
\providecommand{\vecjover}{\overline{\bm{\mathcal{J}}}}

\providecommand{\upolprime}{\vecu_{\rm{pol}}^\prime}
\providecommand{\bpolprime}{\vecb_{\rm{pol}}^\prime}
\providecommand{\umerprime}{\vecu_{\rm{m}}^\prime}
\providecommand{\bmerprime}{\vecb_{\rm{m}}^\prime}

\providecommand{\uradprime}{u_r^\prime}
\providecommand{\uthetaprime}{u_\theta^\prime}
\providecommand{\uphiprime}{u_\phi^\prime}
\providecommand{\ulambdaprime}{u_\lambda^\prime}
\providecommand{\uzprime}{u_z^\prime}

\providecommand{\avurad}{\av{u_r}}
\providecommand{\avutheta}{\av{u_\theta}}
\providecommand{\avuphi}{\av{u_\phi}}
\providecommand{\avulambda}{\av{u_\lambda}}
\providecommand{\avuz}{\av{u_z}}
\providecommand{\aventrhat}{\av{\entrhat}}

\providecommand{\omradprime}{\omega_r^\prime}
\providecommand{\omthetaprime}{\omega_\theta^\prime}
\providecommand{\omphiprime}{\omega_\phi^\prime}
\providecommand{\omlambdaprime}{\omega_\lambda^\prime}
\providecommand{\omzprime}{\omega_z^\prime}

\providecommand{\bradprime}{B_r^\prime}
\providecommand{\bthetaprime}{B_\theta^\prime}
\providecommand{\bphiprime}{B_\phi^\prime}
\providecommand{\blambdaprime}{B_\lambda^\prime}
\providecommand{\bzprime}{B_z^\prime}

\providecommand{\jradprime}{\mathcal{J}_r^\prime}
\providecommand{\jthetaprime}{\mathcal{J}_\theta^\prime}
\providecommand{\jphiprime}{\mathcal{J}_\phi^\prime}
\providecommand{\jlambdaprime}{\mathcal{J}_\lambda^\prime}
\providecommand{\jzprime}{\mathcal{J}_z^\prime}

\providecommand{\cost}{\cos\theta}
\providecommand{\sint}{\sin\theta}
\providecommand{\cott}{\cot\theta}
\providecommand{\rsint}{r\sint}
\providecommand{\orsint}{\frac{1}{\rsint}}
\providecommand{\orsintline}{(1/\rsint)}
\providecommand{\rt}{r\theta}

\providecommand{\amom}{\mathcal{L}}

\providecommand{\minn}{_{\rm{min}}}
\providecommand{\maxx}{_{\rm{max}}}
\providecommand{\inn}{_{\rm{in}}}
\providecommand{\out}{_{\rm{out}}}
\providecommand{\bott}{_{\rm{bot}}}
\providecommand{\midd}{_{\rm{mid}}}
\providecommand{\topp}{_{\rm{top}}}
\providecommand{\bcz}{_{\rm{bcz}}}
\providecommand{\ov}{_{\rm{ov}}}
\providecommand{\rms}{_{\rm{rms}}}
\providecommand{\const}{_{\rm{const}}}

\providecommand{\nr}{N_r}
\providecommand{\nt}{N_\theta}
\providecommand{\np}{N_\phi}
\providecommand{\nmax}{{n_{\rm{max}}}}
\providecommand{\lmax}{{\ell_{\rm{max}}}}

\providecommand{\lsun}{L_\odot}

\providecommand{\msun}{M_\odot}
\providecommand{\rstar}{R_*}
\providecommand{\lstar}{L_*}
\providecommand{\mstar}{M_*}

\providecommand{\rearth}{R_\oplus}
\providecommand{\omearth}{\Omega_\oplus}
\providecommand{\mearth}{M_\oplus}

\providecommand{\taurs}{\tau_{\rm{rs}}}
\providecommand{\taumc}{\tau_{\rm{mc}}}
\providecommand{\tauv}{\tau_{\rm{v}}}
\providecommand{\taurad}{\tau_{\rm{rad}}}
\providecommand{\taums}{\tau_{\rm{ms}}}
\providecommand{\taumm}{\tau_{\rm{mm}}}
\providecommand{\taumag}{\tau_{\rm{mag}}}

\providecommand{\aflux}{\bm{\mathcal{F}}}
\providecommand{\afluxrs}{\aflux^{\rm{rs}}}
\providecommand{\afluxmc}{\aflux^{\rm{mc}}}
\providecommand{\afluxv}{\aflux^{\rm{v}}}
\providecommand{\afluxms}{\aflux^{\rm{ms}}}
\providecommand{\afluxmm}{\aflux^{\rm{mm}}}
\providecommand{\afluxmag}{\aflux^{\rm{mag}}}

\providecommand{\taunu}{\tau_{\nu}}
\providecommand{\taukappa}{\tau_{\kappa}}
\providecommand{\taueta}{\tau_{\eta}}
\providecommand{\tauff}{\tau_{\rm ff}}
\providecommand{\tauomega}{\tau_\Omega}
\providecommand{\taun}{\tau_N}
\providecommand{\taues}{\tau_{\rm ES}}

\providecommand{\pes}{{P_{\rm{ES}}}}
\providecommand{\pburrow}{{P_{\rm{b}}}}
\providecommand{\pessun}{{P_{ {\rm ES}, \odot}}}
\providecommand{\pnu}{{P_{\nu}}}
\providecommand{\pkappa}{{P_{\kappa}}}
\providecommand{\peta}{{P_{\eta}}}
\providecommand{\prot}{{P_{\rm{rot}}}}
\providecommand{\pom}{{P_{\Omega}}}
\providecommand{\pff}{{P_{\rm ff}}}
\providecommand{\pequil}{{P_{\rm{eq}}}}
\providecommand{\pcyc}{{P_{\rm{cyc}}}}
\providecommand{\pcycm}{{P_{{\rm cyc}, m}}}

\providecommand{\pesnd}{{\hat{P}_{\rm{ES}}}}
\providecommand{\pburrownd}{{\hat{P}_{\rm{b}}}}
\providecommand{\pessunnd}{{\hat{P}_{ {\rm ES}, \odot}}}
\providecommand{\pnund}{{\hat{P}_{\nu}}}
\providecommand{\pkappand}{{\hat{P}_{\kappa}}}
\providecommand{\petand}{{\hat{P}_{\eta}}}
\providecommand{\protnd}{{\hat{P}_{\rm{rot}}}}
\providecommand{\pomnd}{{\hat{P}_{\Omega}}}
\providecommand{\pffnd}{{\hat{P}_{\rm ff}}}
\providecommand{\pequilnd}{{\hat{P}_{\rm{eq}}}}
\providecommand{\pcycnd}{{\hat{P}_{\rm{cyc}}}}
\providecommand{\pcycmnd}{{\hat{P}_{{\rm cyc}, m}}}

\providecommand{\tes}{{t_{\rm{es}}}}
\providecommand{\test}{{t_{\rm{es,t}}}}
\providecommand{\tcirc}{{t_{\rm{circ}}}}
\providecommand{\tburrow}{{t_{\rm{b}}}}
\providecommand{\tessun}{{t_{ {\rm es}, \odot}}}
\providecommand{\tnu}{{t_{\nu}}}
\providecommand{\tnut}{{t_{{\nu}\rm,t}}}
\providecommand{\tkappa}{{t_{\kappa}}}
\providecommand{\teta}{{t_{\eta}}}
\providecommand{\trot}{{t_{\rm{rot}}}}
\providecommand{\tomega}{{t_{\Omega}}}
\providecommand{\tbrunt}{t_N}
\providecommand{\tvs}{{t_{\rm vs}}}
\providecommand{\trs}{{t_{\rm rs}}}
\providecommand{\tnucz}{{t_{\nu,{\rm cz}}}}
\providecommand{\tkappacz}{{t_{\kappa,{\rm cz}}}}
\providecommand{\tetacz}{{t_{\eta,{\rm cz}}}}

\providecommand{\tff}{{t_{\rm ff}}}
\providecommand{\tequil}{{t_{\rm{eq}}}}
\providecommand{\tcyc}{{t_{\rm{cyc}}}}
\providecommand{\trun}{{t_{\rm{run}}}}
\providecommand{\tmax}{{t_{\rm{max}}}}
\providecommand{\tcycm}{{t_{{\rm cyc}, m}}}
\providecommand{\tsun}{t_\odot}

\providecommand{\tesdim}{{t_{\rm{es}}^*}}
\providecommand{\testdim}{{t_{\rm{es,t}}^*}}
\providecommand{\tcircdim}{{t_{\rm{circ}}^*}}
\providecommand{\tesshdim}{{t_{\rm{es,sh}}^*}}
\providecommand{\tburrowdim}{{t_{\rm{b}}^*}}
\providecommand{\tnudim}{{t_{\nu}^*}}
\providecommand{\tnutdim}{{t_{{\nu}\rm,t}^*}}
\providecommand{\tkappadim}{{t_{\kappa}^*}}
\providecommand{\tetadim}{{t_{\eta}^*}}
\providecommand{\trotdim}{{t_{\rm{rot}}^*}}
\providecommand{\tomegadim}{{t_{\Omega}^*}}
\providecommand{\tbruntdim}{{t_N^*}}

\providecommand{\tffdim}{{t_{\rm ff}^*}}
\providecommand{\tffdimsq}{{t_{\rm ff}^{*2}}}

\providecommand{\tequildim}{{t_{\rm{eq}}^*}}
\providecommand{\tcycdim}{{t_{\rm{cyc}}^*}}
\providecommand{\trundim}{{t_{\rm{run}}^*}}
\providecommand{\tmaxdim}{{t_{\rm{max}}^*}}
\providecommand{\tsundim}{t_{\odot}^*}
\providecommand{\tvsdim}{t_{\rm vs}^*}
\providecommand{\trsdim}{t_{\rm rs}^*}

\providecommand{\tnuczdim}{{t_{\nu,{\rm cz}}^*}}
\providecommand{\tkappaczdim}{{t_{\kappa,{\rm cz}}^*}}
\providecommand{\tetaczdim}{{t_{\eta,{\rm cz}}^*}}

\providecommand{\omcyc}{{\omega_{\rm{cyc}}}}
\providecommand{\omcycm}{{\omega_{{\rm{cyc}}, m}}}

\providecommand{\ra}{{\rm{Ra}}}
\providecommand{\ratwo}{{\rm{Ra_2}}}
\providecommand{\sigmaeff}{{\sigma_{\rm eff}}}
\providecommand{\sigmaest}{{\sigma_{\rm est}}}
\providecommand{\sigmaeffloc}{{\sigma_{\rm eff,loc}}}
\providecommand{\sigmazero}{{\sigma_0}}
\providecommand{\sigmazerosq}{{\sigma_0^2}}
\providecommand{\sigmashsq}{{\sigma^2_{\rm sh}}}
\providecommand{\sigmaloc}{{\sigma_{\rm loc}}}
\providecommand{\sigmadyn}{{\sigma_{\rm dyn}}}
\providecommand{\sigmadynest}{{\sigma_{\rm dyn,est}}}

\providecommand{\raf}{\ra_{\rm{f}}}
\providecommand{\rafloc}{\ra_{\rm{f,loc}}}
\providecommand{\ramod}{\ra^*}
\providecommand{\rafmod}{\raf^*}
\providecommand{\pr}{{\rm{Pr}}}
\providecommand{\prm}{{\rm{Pr_m}}}
\providecommand{\ek}{{\rm{Ek}}}
\providecommand{\ekt}{{\rm{Ek_T}}}
\providecommand{\ta}{{\rm{Ta}}}
\providecommand{\roc}{{\rm{Ro_c}}}
\providecommand{\rocsq}{{\rm{Ro_c^2}}}
\providecommand{\bu}{{\rm{Bu}}}
\providecommand{\bumod}{{\rm{Bu^*}}}
\providecommand{\buvisc}{{\rm{Bu_{visc}}}}
\providecommand{\burot}{{\rm{Bu_{rot}}}}
\providecommand{\fr}{{\rm Fr}}
\providecommand{\ma}{{\rm Ma}}

\providecommand{\di}{{\rm{Di}}}


\providecommand{\ro}{{\rm{Ro}}}
\providecommand{\lo}{{\rm{Lo}}}

\providecommand{\re}{{\rm{Re}}}
\providecommand{\rem}{{\rm{Re_m}}}
\providecommand{\pe}{{\rm{Pe}}}

\providecommand{\Nrho}{N_\rho}
\providecommand{\nrho}{n_\rho}
\providecommand{\Nrhocz}{N_\rho^{\rm cz}}
\providecommand{\Nrhorz}{N_\rho^{\rm rz}}

\providecommand{\gram}{{\rm{g}}}
\providecommand{\cm}{{\rm{cm}}}
\providecommand{\meter}{{\rm{m}}}
\providecommand{\km}{{\rm{km}}}
\providecommand{\erg}{{\rm{erg}}}
\providecommand{\kelvin}{{\rm{K}}}
\providecommand{\dyn}{{\rm{dyn}}}
\providecommand{\second}{{\rm{s}}}
\providecommand{\radsecond}{{\rm rad\ s^{-1}}}
\providecommand{\minute}{{\rm min}}
\providecommand{\hour}{{\rm hr}}
\providecommand{\ayear}{{\rm yr}}
\providecommand{\aday}{{\rm day}}
\providecommand{\days}{{\rm days}}

\providecommand{\yr}{{\rm{yr}}}
\providecommand{\gauss}{{\rm{G}}}
\providecommand{\kelv}{{\rm{K}}}
\providecommand{\unitentr}{{\rm{erg\ g^{-1}\ K^{-1}}}}
\providecommand{\unitdsdr}{{\rm{erg\ g^{-1}\ K^{-1}\ cm^{-1}}}}
\providecommand{\uniten}{\rm{erg}\ \cm^{-3}}
\providecommand{\unitprs}{\rm{dyn}\ \cm^{-2}}
\providecommand{\unitrho}{\gram\ \cm^{-3}}
\providecommand{\stoke}{\rm{cm^2\ s^{-1}}}

\providecommand{\meanb}{\overline{\bm{B}}}
\providecommand{\flucb}{\bm{B}^\prime}
\providecommand{\totb}{\bm{B}}

\providecommand{\meanv}{\overline{\bm{v}}}
\providecommand{\flucv}{\bm{v}^\prime}
\providecommand{\totv}{\bm{v}}

\providecommand{\emf}{\bm{\mathcal{E}}}
\providecommand{\meanemf}{\overline{\bm{\mathcal{E}}}}
\providecommand{\meanbpol}{\overline{\bm{B}_{\rm{pol}}}}


\providecommand{\rin}{{r_{\rm in}}}
\providecommand{\rout}{{r_{\rm out}}}
\providecommand{\rbot}{{r_{\rm b}}}
\providecommand{\rtop}{{r_{\rm t}}}
\providecommand{\rmin}{{r_{\rm min}}}
\providecommand{\rmax}{{r_{\rm max}}}

\providecommand{\rindim}{r_{\rm in}^*}
\providecommand{\routdim}{r_{\rm out}^*}
\providecommand{\rbotdim}{r_{\rm b}^*}
\providecommand{\rtopdim}{r_{\rm t}^*}

\providecommand{\rbcz}{r_{\rm bcz}}
\providecommand{\rtcz}{r_{\rm tcz}}
\providecommand{\rbrz}{r_{\rm brz}}
\providecommand{\rtrz}{r_{\rm trz}}
\providecommand{\rc}{r_{\rm c}}
\providecommand{\rnrhothree}{r_{3}}
\providecommand{\rtach}{r_{\rm t}}
\providecommand{\rov}{r_{\rm ov}}
\providecommand{\dtach}{\Delta_{\rm t}}
\providecommand{\dtachsun}{\Delta_{\rm t,\odot}}

\providecommand{\rbczdim}{{r_{\rm bcz}^*}}
\providecommand{\rtczdim}{{r_{\rm tcz}^*}}
\providecommand{\rbrzdim}{{r_{\rm brz}^*}}
\providecommand{\rtrzdim}{{r_{\rm trz}^*}}
\providecommand{\rcdim}{{r_{\rm c}^*}}
\providecommand{\rnrhothreedim}{{r_{3}^*}}
\providecommand{\rtachdim}{{r_{\rm t}^*}}
\providecommand{\dtachdim}{{\Delta_{\rm t}^*}}

\providecommand{\rsun}{R_\odot}
\providecommand{\rbczsun}{R_{\rm bcz}}
\providecommand{\rnrhothreesun}{R_{\rm 3}}
\providecommand{\rtachsun}{R_{\rm t}}

\providecommand{\rayleigh}{\texttt{Rayleigh}}
\providecommand{\mesa}{\texttt{MESA}}
\providecommand{\dedalus}{\texttt{Dedalus}}
\providecommand{\rayleigha}{\texttt{Rayleigh 0.9.1}}
\providecommand{\rayleighb}{\texttt{Rayleigh 1.0.1}}

\providecommand{\eulag}{\texttt{EULAG}}
\providecommand{\eulagmhd}{\texttt{EULAG-MHD}}
\providecommand{\ash}{\texttt{ASH}}
\providecommand{\rsst}{\texttt{RSST}}
\providecommand{\rtdt}{\texttt{R2D2}}
\providecommand{\pencil}{\texttt{Pencil}}

\providecommand{\newtext}[1]{{#1}}
\providecommand{\avzt}[1]{\left\langle#1\right\rangle_{z,t}}
\providecommand{\avrt}[1]{\left\langle#1\right\rangle_{r,t}}
\providecommand{\avz}[1]{\left\langle#1\right\rangle_{z}}
\providecommand{\avr}[1]{\left\langle#1\right\rangle_{r}}
\providecommand{\newnewtext}[1]{{#1}}

\defcitealias{Wulff2022}{W22}
\renewcommand{\dimm}{^*}
\newcommand{\dimsq}{^{*2}}
\newcommand{\wl}{_{\rm wl}}

\newcommand{\alphawl}{{\alpha\wl}}
\newcommand{\alphacz}{{\alpha\cz}}

\newcommand{\betatopo}{\beta_{\rm topo}}
\newcommand{\ftopo}{f_{\rm topo}}
\newcommand{\lrtopo}{L_{R,\rm topo}}
\newcommand{\betatwod}{\beta_{\rm 2d}}
\newcommand{\betathin}{\beta_{\rm thin}}

\newcommand{\Nrhowl}{N_\rho^{\rm wl}}

\newcommand{\ztop}{{z_{\rm top}}}
\newcommand{\zbot}{{z_{\rm bot}}}

\newcommand{\lm}[1]{{\color{red}#1}}
\newcommand{\gv}[1]{{\color{blue}#1}}
\newcommand{\mb}[1]{{\color{green}#1}}
\newcommand{\nb}[1]{{\color{orange}#1}}

\newcommand\lmout{\bgroup\markoverwith{\textcolor{red}{\rule[0.5ex]{2pt}{1pt}}}\ULon}
\newcommand\gvout{\bgroup\markoverwith{\textcolor{blue}{\rule[0.5ex]{2pt}{1pt}}}\ULon}
\newcommand\mbout{\bgroup\markoverwith{\textcolor{green}{\rule[0.5ex]{2pt}{1pt}}}\ULon}
\newcommand\nbout{\bgroup\markoverwith{\textcolor{orange}{\rule[0.5ex]{2pt}{1pt}}}\ULon}

\newcommand{\com}[1]{{\color{purple}#1}}
\newcommand{\nbcom}[1]{{\color{orange}[NB: #1]}}
\newcommand{\lmcom}[1]{{\color{red}[LM: #1]}}
\newcommand{\mbcom}[1]{{\color{green}[MB #1]}}
\newcommand{\gvcom}[1]{{\color{blue}[GV: #1]}}

\title{Superrotation and Jet Migration in Simulations of Giant Planet Atmospheres}
%

\begin{abstract}
The mean zonal flow on \newtext{both} Jupiter \newtext{and Saturn} consists of an intricate pattern of jets moving prograde or retrograde compared to the bulk planetary rotation. The strongest \newtext{jet straddles the equator and is prograde, or superrotating. It} is flanked by \newtext{multiple} retrograde/prograde pairs of weaker jets \newtext{at high latitudes}. Two \newtext{candidate} drivers of \newtext{these jets} are ``shallow" quasi-two dimensional turbulence in the outermost stably stratified weather layer (WL) and ``deep" buoyantly-driven Busse columns in the \newtext{underlying} convective zone (CZ). To study \newtext{the respective roles of the WL and CZ}, we implement two rotating, three-dimensional, anelastic convection simulations \newtext{of a giant planet}. In one case, the CZ is isolated, whereas in the other case, the upflows overshoot into a stably stratified near-surface \newtext{layer, driving turbulence there. We regard this layer as a simplified WL, although we include the effects of stable stratification only, neglecting solar irradiation. In each case,} we find that homogenization of \newtext{an appropriately defined} potential vorticity initially \newtext{creates} multiple jets at high latitudes, \newtext{whereas} Busse columns \newtext{in the CZ drive} equatorial superrotation. The presence of a simplified WL significantly alters the thermal wind balance, resulting in large deviations of the contours of the zonal flow from alignment with the rotation axis. Finally, the weaker high-latitude jets slowly migrate poleward and/or merge on a very long time scale ($O(10)$ diffusion time scales or thousands of eddy turnover times).
\end{abstract}



\section{Introduction}\label{sec:intro}
\setcounter{footnote}{0} 
The aim of this paper is to explore how zonal flows arise in simulations of gas-giant planets like Jupiter \newtext{or Saturn}, focusing on the comparison between two-dimensional (2D) and three-dimensional (3D) ideas. We seek to build upon previous work over the last few decades on modeling the \newtext{observed zonal flows on these planets,} which has typically been performed in either a 2D or a 3D context. To move towards unification of the various 2D and 3D concepts, we perform fully nonlinear, 3D simulations of a \newtext{giant} planet and quantitatively assess the results using quasi-2D methods. We first frame the problem by outlining prior observational and theoretical work on \newtext{the zonal flows of Jupiter and Saturn}. We place particular emphasis on the role of potential vorticity (PV) and the distinct types of turbulent eddies that are thought to dominate the dynamics at different depths of \newtext{a giant planet's} interior.  

\subsection{Superrotation and Jets in Planetary Atmospheres}
Mean (longitudinally averaged) zonal flows are a ubiquitous feature of planetary atmospheres. Superrotation---formally defined as a ring of fluid with greater specific angular momentum than that of the planet itself at the equator---is a common but not universal property of such zonal flows for bodies within our solar system (e.g., \citealt{Imamura2020,Vallis2026}). For the purposes of the present study, superrotation (or subrotation) can be thought of simply as mean equatorial zonal flow that moves prograde (or retrograde) relative to the bulk planetary rotation. A key parameter determining the nature of zonal flows is the Rossby number $\ro$ \newtext{of the turbulent eddies}, defined as\footnote{Throughout this work, we reserve the symbol ``$\coloneqq$" for definitions and use ``$\equiv$" to mean ``equal to [some constant] everywhere and for all time.} $\ro\define U/(2\Omega_0L)$, \newtext{where $U$ is a typical eddy speed, $L$ is a typical eddy length scale, and $\Omega_0$ is the bulk rotation rate of the planet. The Rossby number measures the ratio of the magnitudes of the advective force $U^2/L$ to the Coriolis force $2\Omega_0U$, or equivalently, the ratio of the rotational  (Coriolis-force) time scale $(2\Omega_0)^{-1}$ to the eddy turnover time scale $L/U$.} For high Rossby numbers, the eddy turnover time is relatively rapid and the eddies do not ``feel" the rotation very much; the converse is true for low Rossby numbers. The Rossby number thus measures the influence of rotation on the turbulent eddies, or the degree of ``rotational constraint." 

Both rapidly rotating bodies with $\ro\ll1$ (i.e., high rotational constraint, like Jupiter and Saturn) and slowly rotating bodies with $\ro\gg1$ (i.e., low rotational constraint, like Venus and Titan) exhibit permanent strong superrotation (e.g., \citealt{Read2018, Nicolas2026}). The Sun (about mid-way between fast and slow rotation with $\ro\sim1$; e.g., \citealt{Hanasoge2012,Greer2016}) also exhibits strong superrotation, with equatorial regions rotating $\sim$10\% faster than the bulk rotation rate of the deep solar interior (e.g., \citealt{Howe2009}). Other planets, like Mars and Earth ($\ro\sim0.1$--$0.2$), have only a weak and highly variable tendency toward superrotation (e.g., \citealt{Medvedev2011, Vallis2017}), whereas Neptune and Uranus ($\ro\ll1$) exhibit permanent subrotation (e.g., \citealt{Soyuer2022}). Observations of superrotation or subrotation on exoplanets is naturally very difficult, however there is \newtext{evidence that some tidally locked hot Jupiters tend to superrotate (e.g., \citealt{Knutson2007,Heng2015,Brogi2016,Song2021}).}


%
In addition to superrotation/subrotation, \newtext{high-latitude} jets---meridionally localized bands of strong zonal flow \newtext{moving prograde or retrograde relative to the bulk planetary rotation}---are observed on Jupiter and Saturn (e.g., \citealt{Fletcher2020,Kaspi2020}). \newtext{Spatially resolved maps of the jets can be obtained from cloud tracking in the outer weather layer (WL; e.g., \citealt{GarciaMelendo2011,Tollefson2017}).} \newtext{Apart from the large superrotating equatorial jet,} Jupiter has 6--7 prograde/retrograde \newtext{high-latitude} jet pairs in each hemisphere (about \newtext{24--28 high-latitude} jets total), whereas Saturn has 4--5 pairs of jets per hemisphere (about \newtext{16}--20 total). 

\newtext{In the last decade or so, measurements of Jupiter and Saturn's gravitational fields from the Juno and Cassini spacecraft have yielded new information on how the zonal flows extend into the convective zone (CZ) deep below the WL (e.g., \citealt{Kaspi2020}). Forward models of the the interior jet structure can be fit to the gravity measurements via a type of thermal wind balance (e.g., \citealt{Kaspi2009,Matilsky2023}), yielding a nonunique inference of the interior zonal flow from the directly measured gravity field. Although these inferences have some degeneracy, they strongly suggest several attributes of the interior zonal flow, including: (1) The jets extend well below the weather layer (about 3,000 km below the WL for Jupiter and about 9,000 km below the WL for Saturn) and (2) The jets extend inward primarily cylindrically, i.e., along the direction parallel to the rotation axis. Because of the key role convection is thought to play in driving the jets (see Section \ref{sec:driving_mechanisms}), the values $3,000$ and $9,000$ km are often simply taken to be the depths of the CZs of Jupiter and Saturn in global simulations.} 



\subsection{Energy Sources in Giant Planets}\label{sec:energy_sources}
In this work, we focus on zonal flow generation in a \newtext{gas-giant} planet \newtext{like Jupiter or Saturn with} strong rotational constraint ($\ro\ll1$). Jupiter \newtext{and Saturn are} particularly interesting systems because \newtext{they have} two roughly equal energy sources of free thermal energy that can be tapped to produce the zonal flow. The first is from radiation by the Sun, corresponding to an absorbed energy flux of about $8.1\ \rm W\ m^{-2}$ for Jupiter and $2.6\ \rm W\ m^{-2}$ for Saturn, which occurs in the relatively shallow WL (e.g., \citealt{Guillot2005}). The WL corresponds to the directly observable topmost layers of Jupiter's interior/atmosphere, which contain several types of clouds and long-lived storms. The WL is strongly stable to convection (or ``stably stratified"), as inferred from atmospheric temperature profiles obtained by occultation measurements from the Voyager Spacecraft \citep{Lindal1981} and\newtext{, on Jupiter,} in-situ measurements at one latitude/longitude from the Galileo Probe \citep{Seiff1996}. Because the solar heating is strongest near \newtext{each giant planet's} equator, this ``shallow" energy source induces a pole-to-equator temperature gradient \newtext{(baroclinicity), which can potentially drive quasi-2D eddies through baroclinic instabilities.} A second energy source is the combined effect of Jupiter \newtext{and Saturn's} gravitational contraction, precipitation and settling of helium, and internal heat release (i.e., bulk planetary cooling). These effects lead to an outward energy flux of about $5.4\ \rm W\ m^{-2}$ from \newtext{Jupiter's} interior \newtext{and about 2.0 $\rm W\ m^{-2}$ from Saturn's interior} (e.g., \citealt{Guillot2005}; see Table \ref{tab:redim}). This ``deep" energy source heats the outer layers of \newtext{the giant planets} from below and drives overturning 3D eddies in the CZ, often in the form of columnar rolls parallel to the rotation axis owing to the strong rotational constraint.

Figure \ref{fig:schematic} shows a schematic of \newtext{a giant planet's} interior layers (not to scale) from an inner radius $\rin$ to an outer radius $\rout\define a$ \newtext{(where $a$ is the giant planet's dimensional radius, which we take to correspond to the top of the WL; see Table \ref{tab:redim})}. There is a CZ (with its deep rotationally aligned columnar eddies or ``Busse columns;" e.g., \citealt{Busse2002}) beneath a WL (with its shallow vertically constricted ``pancake eddies" of predominantly horizontal motion), with the boundary between CZ and WL denoted by $\rc$. We also show the coordinate systems we employ in this work: spherical coordinates $\phi$ (azimuth angle), $\theta$ (latitude), and $r$ (spherical radius) and cylindrical coordinates $\lambda\define r\cos\theta$ (cylindrical radius, or moment arm), $\phi$, and $z\define r\sin\theta$ (axial coordinate). The corresponding unit vectors are denoted with $\e$ (for example, the axial unit vector is $\ez$). Note that $(\ephi,\etheta,\erad)$ and $(\elambda,\ephi,\ez)$ are both right-handed sets of unit vectors. The system rotates at the frame rate $\Omzerovec$ parallel to $\ez$ and feels inward gravity $\tilde{\bm{g}}$ parallel to $-\erad$. The ``tangent cylinder" (i.e., the cylinder $\lambda\equiv\rin$ which circumscribes the base of the CZ) separates high latitudes (inside the tangent cylinder, where the Busse columns can exist in only one hemisphere or the other) from low latitudes (outside the tangent cylinder, where the columns can extend across the equator).

\begin{figure}
	\centering
	\includegraphics[width=3.4375in]{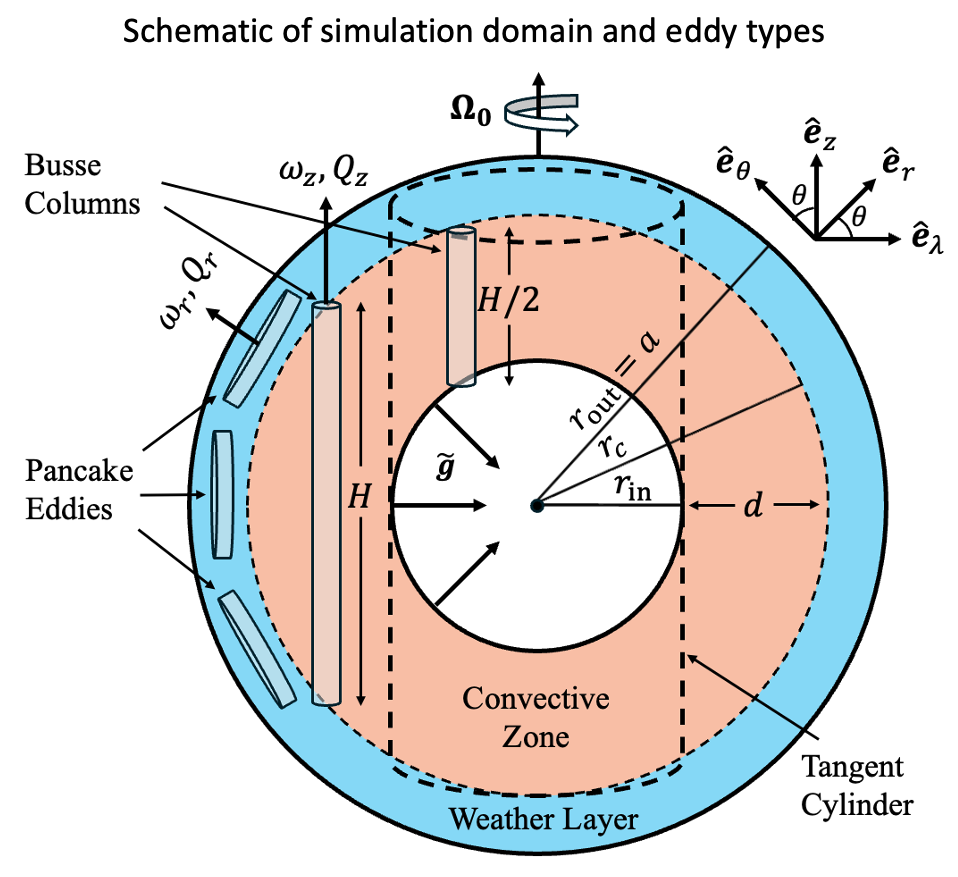}
	\caption{Schematic of our simulation domain (not to scale), showing a spherical shell with a CZ (brown layer) underneath a stably stratified \newtext{simplified} WL (blue layer) with the different types of eddies sketched as rotating cylinders. Each type of eddy mixes a corresponding type of potential vorticity (PV): in the CZ, rotationally aligned Busse columns of height $H(\lambda)$ (see Equation \ref{eq:hlambda_dim}) mix $Q_z$ (Equation \ref{eq:pvz_dim}), while in the WL, the flattened pancake eddies of primarily horizontal flow mix $Q_r$ (Equation \ref{eq:pvr_dim}). The tangent cylinder, spherical and cylindrical unit vectors, and parameters defined in the text ($\rin,\rc,\rout\define a$, $d\define\rc-\rin$, $\tilde{\bm{g}}$, and $\Omzerovec$) are also shown.    }
	\label{fig:schematic}
\end{figure}

In ideal circumstances (very strong stable stratification in the shallow WL and very strong rotational constraint in the deep CZ), we might expect each type of eddy in Figure \ref{fig:schematic} to be effectively 2D, with the pancake eddies composed of primarily vertical (radial) vorticity $\omega_r$ (which is vertically invariant for a pancake eddy, or independent of $r$) and the Busse columns composed of primarily axial vorticity $\omega_z$ (which is axially invariant within a column, or independent of $z$). Here, $\vecom\define\curl\vecu$ is the relative vorticity, with $\vecu$ the fluid velocity measured in the rotating frame. In either case, the equations of motion can be averaged, in either $r$ or $z$, to obtain a set of equations similar to those for 2D turbulence. Ideas from 2D turbulence theory can then be used to explain the formation of multiple jets and superrotation. \newtext{We use exactly this approach in quantifying the appropriate dynamics in Sections \ref{sec:jets_cz} and \ref{sec:jets_wl} below.} 

\subsection{Deep and Shallow Driving Mechanisms for Mean Zonal Flow} \label{sec:driving_mechanisms}
Because both types of eddies are effectively 2D in idealized circumstances (because there is symmetry in either $r$ or $z$), either could drive mean zonal flows according to 2D turbulence theory (e.g., \citealt{Rhines1975, Vallis1993}). If the origin of the jets is shallow (i.e., in the WL), the pancake eddies primarily do the driving, whereas if the origin of the jets is ``deep" (i.e., in the CZ), the Busse columns primarily do the driving. \newtext{Because the energy sources driving the different types of eddy are roughly equal,} this has led to an ongoing ``deep versus shallow" driving debate in the community \newtext{regarding the origins of Jupiter and Saturn's jets} (e.g., \citealt{Vallis2026}). 

Whatever its form (in our case, either pancake eddies or Busse columns), stochastically driven quasi-2D turbulence in a rotating system is thought to drive zonal flows by a turbulent cascade process. Injection of energy at small, rotationally unconstrained length scales leads to an inverse cascade of energy to larger and larger scales (e.g., \citealt{Kraichnan1967}). When the eddies become large enough, they begin to feel the effects of rotation and, in part, start to take the form of Rossby waves. Nonlinear interactions (i.e., Reynolds stresses) then drive mean zonal flows (e.g., \citealt{Dickinson1969,Thompson1971, Rhines1975, Vallis1993}). 

\newtext{Depending on the parameter regime, simulations of 2D turbulence show the formation of zonal jets and, if the domain is large enough, multiple jets, typically with a width commensurate with an appropriate Rhines scale, $\sqrt{U/\beta}$, where $U$ is some eddy velocity and $\beta$ is a background vorticity gradient or equivalent quantity. (We define the Rhines scale more precisely and make quantitative estimates of it for our simulations in Sections \ref{sec:jets_cz} and \ref{sec:jets_wl}.) Although jet formation from both types of eddy can be described in the context of quasi-2D turbulence, the different types of symmetry (in either $r$ or $z$) leads to several differences. In particular, we expect jets arising from deep driving (Busse columns) to be cylindrically aligned (i.e., have $z$-independent zonal flow), whereas jets arising from shallow driving (pancake eddies) are expected to be vertically aligned (have $r$-independent zonal flow). Measurements of Jupiter \newtext{and Saturn}'s gravity field by the Juno \newtext{and Cassini} spacecraft have been argued to favor cylindrical alignment of the jets in the CZ (e.g., \citealt{Galanti2021,Kaspi2023}), but the gravimetric signature of the WL is too weak  to give useful constraints on form of the jets in that region.} 

In rotating fluid mechanics, a quantity of key interest is the potential vorticity (PV) of the flow (e.g., \citealt{Rossby1936,Ertel1942a}). In many contexts, PV is completely or partially conserved during fluid motion, which results in strong constraints on how global-scale circulations (including circulation in the meridional $r$-$\theta$ plane, as well as mean zonal flows like jets and superrotation) can evolve in a rotating system. In quasi-2D turbulence, PV can become partially homogenized; this can lead to a ``staircase" in mean PV with respect to latitude $\theta$ (for shallow driving) or moment arm $\lambda$ (for deep driving), for which mean PV is roughly constant on ``steps" (each of width the order of an appropriate Rhines scale) and has sharp gradients between the steps. Because the mean PV is related to the $\theta$- or $\lambda$-derivative of mean zonal flow, a staircase in mean PV corresponds to multiple zonal jets. 

For vortex tubes, such as the idealized pancake vortices and Busse columns sketched in Figure \ref{fig:schematic}, the PV is typically defined as the absolute vorticity (the planetary vorticity $2\Omega_0\ez$ added to the relative vorticity $\vecom$) divided by the length of the tube (e.g., \citealt{Vallis2017}). For the pancake eddies, however, the vertical motion is minimal and stretching and compression effects can be neglected, so the PV reduces simply to the absolute vorticity. Furthermore, because of the low aspect ratio of the pancake eddies, only the vertical component of the angular velocity, $\Omega_0\sin\theta$, affects the motion of the fluid via the Coriolis force (e.g., \citealt{Pedlosky1987}). The relevant PV thus only has a component in the vertical direction ($\erad$) and its scalar form is given by 
\begin{align}\label{eq:pvr_dim}
    Q_r \define \omrad + f,
\end{align}
where $f\define \twoOmzero\sin\theta$ is the Coriolis parameter or background planetary vorticity (also the background PV in this case). The variation of $f$ with latitude ($\beta\define (1/a)df/d\theta$) gives rise to the ``geometric $\beta$-effect," which determines how Rossby waves propagate in a shallow atmosphere (e.g., \citealt{Rossby1939,Haurwitz1940}) and thus the form of the Rhines scale. 

For the Busse columns, stretching and compression effects are significant, because the column height quantity $H$ varies significantly as the columns move about in $\lambda$ (see Figure \ref{fig:schematic}). Furthermore, \newtext{extension of the columns} in $z$ does not allow for any of the components of the Coriolis force to be neglected. From Figure \ref{fig:schematic}, the suggested form of PV for a Busse column only has a component in the axial direction ($\ez$) given by 
 \begin{align}\label{eq:pvz_dim}
 Q_z \define \frac{\omz+\twoOmzero}{H} = \frac{\omz}{H} + \ftopo>0,
 \end{align}
 where, in anology to $f$, $\ftopo\define \twoOmzero/H$ is the background planetary PV due to the topography, which here comes from the slopes of the bounding spherical shells. Similarly to the pancake vortices, the variation of the planetary PV $\ftopo$ with moment arm ($\betatopo\define -Hd\ftopo/d\lambda=(2\Omega_0/H)dH/d\lambda$) gives rise to a ``topographic $\beta$-effect," which is thought to determine a different Rhines scale and staircase/jet structure in the deep-driving case compared to the shallow-driving case.
 
 Assuming the form of PV in Equation \eqref{eq:pvz_dim} presupposes that the columns extend all the way through the CZ, which is not in general true if there is significant density stratification across the layer. Strong density stratification leads to yet another form of PV due to the ``compressional $\beta$-effect," \newtext{which can be characterized by the compressional $\beta$-parameter $\beta_{\rm comp}\define\twoOmzero/H_\rho$, where $H_\rho\define-(d\ln\rhotilde/dr)^{-1}$ is the local density scale-height and $\rhotilde(r)$ is the background density} (e.g., \citealt{Gastine2012}). \newtext{In fact, in the models discussed here, the density scale-height is roughly equal to the shell-depth, resulting in similar orders of magnitude for $\beta_{\rm comp}$ and $\betatopo$.} \newnewtext{Nevertheless, we find that vorticity generation due to the background density stratification is substantially weaker than that due to the topography in our models (see Appendix \ref{ap:beta}). Thus, we neglect the compressional $\beta$-effect throughout this work. However, we acknowledge that density stratification is quite weak in our models compared to the real giant planets, and thus compressional effects in Jupiter or Saturn may well be important.}

\subsection{Prior Work on Modeling Zonal Flows}

\newtext{The formation of mean zonal flows and jets has been studied extensively in 2D, quasi-geostrophic, and shallow-water contexts, using both internal instabilities and stochastic forcing as driving mechanisms. Various reviews can be found in the volume by \citet{Galperin2019} and, more recently, the review article by \citet{Vallis2026}. In shallow models on the $\beta$-plane, multiple jets typically remain steady as long as some form of mechanical friction (e.g., Rayleigh drag) is included to arrest the inverse cascade (e.g., \citealt{Scott2007,Schubert2009}). When such models are used in a planetary context, both superrotation and subrotation of the equatorial jet can arise (e.g., \citealt{Cho1996,Showman2007}), depending somewhat sensitively on parameters. For example, strong radiative damping and weak bottom drag tend to lead preferentially to superrotation (e.g., \citealt{Saito2015}) but can lead to difficulties in reproducing the high-latitude jets \citep{Warneford2017}.}





\newtext{In some contrast, simulations of convection in a rotating spherical shell, using the 3D equations of motion with a full treatment of the Coriolis effect, robustly lead to superrotation in a rapidly rotating regime (e.g., \citealt{Gilman1981,Yamazaki2005,Gastine2013,Heimpel2022,Camisassa2022}). For high levels of both turbulence and rotational constraint, multiple high-latitude jets are also achieved. However, because of their significant computational expense, thus far 3D simulations not been run to full equilibration.}\footnote{By ``equilibration," we mean the absence of either secular growth of the zonal flows' kinetic energy (e.g., \citealt{Takehiro2024}) or any change in the latitudinal positions of the jets (e.g., \citealt{Wulff2022}).}

\newtext{That said, the 3D simulations have yielded important new insights into the dynamics of the deep driving mechanism. For instance, \citet{Duer2023} noted a 3D eddy-momentum transport of angular momentum and associated cells of meridional circulation and (at low latitudes) Busse columns from the tilted boundaries driving equatorial superrotation. \citet{Wulff2022} explored the influence of a stably stratified layer \textit{below} the CZ in a 3D simulation. They found that the stable stratification could substantially attenuate the jet amplitude with depth (without relying on a bottom drag from Lorentz forces) and that the jet contours were significantly tilted away from cylindrical alignment, suggesting a key change in thermal wind balance. Finally, \citet{Takehiro2024} found that after long time integrations, there was high-latitude jet migration in the global simulations, which eventually yielded a quasi-asymptotic state containing very few high-latitude jets. Even \citet{Takehiro2024} stopped short of full equilibration of the simulations, suggesting that the robust jet structures of prior fully 3D models (which were much less evolved) may have corresponded to a transient state.}

\newtext{In order to explore 3D effects in a more economical way, quasi-2D simulations have been extended to include a parameterized deep driving mechanism. \citet{Smith2004} considered the convective forcing to be stochastic in nature and incorporated a geometric $\beta$-effect, and \citet{Zeng2026} allowed for a topographic $\beta$-effect, mimicking the effect of the variable length of the Busse columns (as  discussed by \citealt{Yano2005} and \citealt{Vallis2019}). This work builds upon the 2D simulations of \citet{Brummell1993}, extending it to include curved boundaries. Superrotation is robustly found in these simulations if the topographic $\beta$-parameter is negative at low latitudes, as is suggested by the varying length of the Busse columns.}

\newtext{To resolve the deep versus shallow driving debate, a comparison of 2D and 3D effects would ideally be performed using a coupled CZ--WL system. Recently, fully 3D models have begun to couple the CZ to a neighboring stable layer into which the flows can overshoot (e.g., \citealt{Wulff2022}, who put the stable layer \textit{below} the CZ). \citet{Heimpel2015,Heimpel2022} implemented thin region of stable stratification \textit{above} a CZ, which allowed the first study of effects like convective overshoot and the inhibition of vertical motion in a region beginning to resemble a WL. \citet{Heimpel2022} also performed some of the first direct comparisons to 2D theory, quantifying the presence of staircases and discussing the two forms of PV.  }

\newtext{In this work, we aim to pick up where recent global CZ--WL simulations (e.g., \citealt{Wulff2022,Heimpel2022,Duer2023}) left off, by precisely quantifying the effects of a stable layer above the CZ and making a comprehensive comparison to 2D theory. We consider two 3D spherical-shell anelastic simulations of rapidly rotating thermally driven convection. In the ``CZ-only case," the CZ is isolated with impenetrable boundaries, whereas in the ``CZ--WL case," the convection can overshoot into a thin, stably stratified layer atop the CZ. We consider this top layer to be a simplified ``weather layer." In this first study, the motion in the WL is not driven by external radiation, but by overshooting convection (i.e., there is no external large-scale baroclinic driving). Our WL is further simplified by considering only a single fluid (i.e., no moist convection or conversions between different molecular species, as might be important in the real Jovian cloud deck; e.g., \citealt{Ingersoll2004}).}


\newtext{Our numerical setup is similar to  \citet{Heimpel2015,Heimpel2022} and \citet{Wulff2022}. The fundamental novelty of our study lies in the analysis of the simulations, which essentially has three parts. First, we identify the distinct forms of PV in the CZ and WL and perform quantitative comparisons of the jet length scale and topographic or geometric Rhines scale separately for each layer. We also perform a torque analysis to explicitly compare 2D ideas like PV homogenization with the 3D effects like eddy-momentum-flux convergence (similar to the analysis of \citealt{Duer2023}). Second, we reproduce \citet{Wulff2022}'s result that a stably stratified layer tilts the contours of zonal flow away from the rotation axis, and perform an explicit thermal wind analysis to identify the mechanism behind the contour tilt and evaluate whether we might expect such a tilt in a real giant planet's WL. Third, and most importantly, we perform one of the first quantitative analyses of simulated jet migration. We run our models for an exceedingly long time interval, verify equilibration, and report on the surprising physics leading to the asymptotic state of these simulations.}

In Section \ref{sec:num}, we describe the numerical details of the two simulations. In Section \ref{sec:jets_structure}, we describe the basic flow properties of the simulations, including the alternating zonal jets achieved early in the simulation. We then describe the formation of these jets at early times via PV homogenization separately for the CZ (Section \ref{sec:jets_cz}) and WL (Section \ref{sec:jets_wl}). In Section \ref{sec:superrot}, we show that the superrotation outside the tangent cylinder is a direct result of angular momentum transport by Busse columns, but only weakly consistent with homogenization of PV. In Section \ref{sec:migration}, we describe the slow poleward migration and eventual disappearance of the high-latitude jets. In Section \ref{sec:disc}, we discuss our results and offer some concluding remarks.

\section{Numerical Setup}\label{sec:num}
We evolve the anelastic fluid equations in 3D, rotating spherical shells using the open-source {\rayleigh} code \citep{Featherstone2016a,Matsui2016,Featherstone2024}. {\rayleigh} is pseudospectral, evolving nonlinear terms on the right-hand sides of the equations of motion in physical space (a grid in $\phi$, $\theta$, and $r$) and linear terms in spectral space (a truncated set of Chebyshev basis functions in the radial direction and spherical harmonics in the horizontal directions). 

\subsection{Simulation Geometry}
We consider two spherical-shell simulations. Each shell contains a CZ in the region $\rindim\leq r\dimm\leq \rcdim$, where $\rindim$ is the inner boundary of the domain and $\rcdim$ is the top of the CZ ($\rindim$ and $\rcdim$ are shared constants for both simulations; see Figure \ref{fig:schematic}).\footnote{From this section onward, asterisks denote explicitly dimensional variables while the lack of asterisks denote their nondimensional counterparts. Our chosen nondimensionalization is given in Section \ref{sec:nondim}.} In the CZ-only case, there is an isolated CZ and the domain's outer boundary $\routdim$ is equal to $\rcdim$. The other case contains a stably stratified, \newtext{simplified} WL above the CZ. In this CZ--WL case, $\routdim>\rcdim$ and the convection is allowed to overshoot across $r\dimm=\rcdim$ into the WL. 

This geometry is characterized by the aspect ratio of the CZ,
\begin{equation}\label{eq:alphacz}
\alphacz\define\frac{\rindim}{\rcdim}=0.8
\end{equation}
and the aspect ratio of the WL,
\begin{equation}\label{eq:alphawl}
\alphawl \define \frac{\rcdim}{\routdim}=
\begin{cases} 
1 &\five\text{CZ-only case}\\
0.952 &\five\text{CZ--WL case.}
\end{cases}
\end{equation}

\newtext{We choose the aspect ratio of the CZ rather arbitrarily to match the gas-giant simulations of \citet{Wulff2022}. We note that this aspect ratio is quite a bit smaller than the inferred aspect ratio of Jupiter's CZ (about 0.96) and is closer to the aspect ratio of Saturn's CZ (about 0.84; e.g., \citealt{Kaspi2020}). Because the aspect ratio is believed to play an important role in planetary dynamics (e.g., \citealt{Duer2024}), our simulations might be considered closer to Saturn than Jupiter. However, we make such comparisons with caution because many of the other parameters are very far from the true regime of any real gas giant. For example, the WL of a giant planet is very thin (e.g., $\alphawl=O(0.999)$ for both Jupiter and Saturn; see Table \ref{tab:input_nond}). We choose a much thicker WL ($\alphawl=0.952$) to more easily spatially resolve the differing dynamics in the CZ and WL.  }

We choose the CZ depth $d\define\rcdim-\rindim$ as our unit of length, leading to the following nondimensional radii: 
\begin{subequations}\label{eq:radii_nd}
\begin{align}
\rin &\define \frac{\alphacz}{1-\alphacz}=4,\\
\rc &\define \frac{1}{1-\alphacz}=5,\five\andd\\
\rout &\define \frac{1}{\alphawl(1-\alphacz)}=
    \begin{cases}
    \begin{aligned}
    5&\five\text{CZ-only case}\\
    5.25 &\five \text{CZ--WL case.}
    \end{aligned}
    \end{cases}
\end{align}
\end{subequations}
For $a=70$,000 km \newtext{(58,000 km)}, this results in a unit of length of $d=a/5.25=13$,000 km \newtext{(11,000 km)} \newtext{for Jupiter (Saturn).}

\subsection{Background State}
The anelastic approximation consists of assuming a solenoidal mass flux and thermal perturbations that are small relative to a hydrostatically and thermally balanced background state (e.g., \citealt{Ogura1962,Gough1969}). The background state consists of the background density $\rhotilde\dimm$, temperature $\tmptilde\dimm$, pressure $\prstilde\dimm$, gravitational acceleration $\gravtilde\dimm$, buoyancy frequency $\brunttilde\dimm$, prescribed internal heating/cooling $\heattilde\dimm$, viscosity $\nutilde\dimm$, and thermal diffusivity $\kappatilde\dimm$. Ideally, we would choose these profiles based on an accurate structure model \newtext{of either Jupiter or Saturn}. However, given the inherent uncertainty in these structure models (e.g., \citealt{Guillot2005,French2012}) and the necessarily great separation in parameter space between our simulations and the appropriate astrophysical environment to begin with \newtext{(due to computational constraints, as discussed below)}, we opt for a simpler approach.

This approach is described in detail in Appendix \ref{ap:ref}, but the essential assumptions are as follows. We assume a perfect gas, hydrostatic balance, $\gravtilde\dimm\propto1/r\dimsq$, adiabatic background stratification ($\brunttilde\dimm\equiv0$) in the CZ and stable background stratification ($\brunttilde\dimsq>0$) in the WL. 

The parameters controlling the background state are then the specific heat ratio
\begin{equation}\label{eq:gamma}
\gamma\define \dfrac{\cp}{\cv} \equiv \frac{3}{2}
\end{equation}
(where the specific heat at constant pressure $\cp$ and the specific heat at constant volume $\cv$ are constants), the number of  density scale heights across the CZ,
\begin{equation}\label{eq:nrhocz}
\Nrhocz\define\ln\left[\frac{\rhotilde\dimm(\rindim)}{\rhotilde\dimm(\rcdim)}\right]=1.2
\end{equation}
and the number of scale heights across the WL,
\begin{equation}\label{eq:nrhowl}
\Nrhowl\define\ln\left[\frac{\rhotilde\dimm(\rcdim)}{\rhotilde\dimm(\routdim)}\right]=0.530.
\end{equation}

These choices, along with the geometric parameters $\alphacz$ and $\alphawl$, fully determine the thermal profiles $\rhotilde\dimm$, $\prstilde\dimm$, and $\tmptilde\dimm$ (see Equations \ref{eq:tmptilde} and \ref{eq:rhotilde}.) In the CZ, these choices result in the polytropic stratification $\rhotilde\dimm\propto\tmptilde^{*n}$, where $n\define1/(\gamma-1)\equiv2$, which \newtext{corresponds to the value assumed by the standard Jovian convection benchmark} \citep{Jones2011}.\footnote{\newtext{The actual equation of state of a real giant planet is more complex; however, $n=2$ is a good intermediate value between Jupiter and Saturn's convecting layers (where $n=2.5$ is a good approximation for molecular hydrogen) and their deep layers (where $n=1.5$ is a good approximation for pressure-ionized metallic hydrogen; see \citealt{Guillot1999,Guillot2005}).}} Note that the dimensional background entropy $\entrtilde\dimm$ is given by 
\begin{equation}\label{eq:dsdrdim}
\left( \dsdrtilde\right) \dimm =\frac{\cp}{\gravtilde\dimm}\brunttilde\dimsq,
\end{equation}
where we set the arbitrary integration factor such that $\entrtilde\dimm\equiv0$ in the CZ (see Equations \ref{eq:entrtilde} and \ref{eq:dsdrquart}.)

\newtext{The value $\Nrhocz=1.2$ is chosen to match the CZ-only simulation of \citet{Wulff2022}, whereas the value $\Nrhowl=0.530$ is a consequence of the chosen WL aspect ratio $\alphawl=0.952$ and the arbitrary choice to have $\dsdrtildeline$ be a step-like function. Both these values differ significantly from actual values of either Jupiter or Saturn (where $\Nrhocz\sim10$ and $\Nrhowl\sim5$; see Table \ref{tab:input_nond}), once again because of computational limitations.}

We drive the convection by choosing $\heattilde\dimm$ to heat the bottom regions of the CZ and cool the top regions in two narrow boundary layers (see Equation \ref{eq:heating}.) Driving the convection by heating and cooling layers has the advantage of yielding ``diffusion-free" scalings for the heat transport in certain regimes (e.g., \citealt{Hadjerci2024,JoshiHartley2025,Lewis2026}). Physically, $\heattilde\dimm$ represents the incoming flux from \newtext{the simulated giant planet's} interior due to effects like gravitational contraction, as well as radiative cooling at the surface. In this first study, we neglect the irradiance from the Sun, which would induce an additional latitudinally varying energy source and associated baroclinicity in the WL. Note that the flux $\fluxscalartilde\dimm$ driven through the system via $\heattilde\dimm$, which is mostly carried by convection in the steady state, is 
\begin{equation}\label{eq:fnrad}
\fluxscalartilde\dimm\define\frac{1}{r\dimsq}\int_{\rindim}^{r\dimm}\heattilde\dimm\ofrprime \rprime^2d\rprime.
\end{equation}
We choose $\nutilde\dimm\equiv\nu$ and $\kappatilde\dimm\equiv\kappa$ to be constants.


Each simulation's control parameters are summarized in Table \ref{tab:input_nond}. The parameters of the CZ-only case are chosen such that the simulation is as close as possible to the single CZ-only case considered by \citealt{Wulff2022} (see Figure 3d from that paper). The CZ--WL case is qualitatively similar to the simulations of \citet{Heimpel2015,Heimpel2022}, who implemented a thin region of stable stratification atop a CZ in a spherical shell by injecting heat at both the top and bottom of the shell. The CZ--WL case considered here differs primarily in that we drive convection in the inner layer via the fixed heating/cooling function and enforce the stable layer with an imposed background stable stratification. This allows the precise control of the location of the boundary between the CZ and stable layer and the magnitude of the stable stratification in the WL. Another key difference is that we include the CZ-only case as a control, which allows us to explicitly isolate the effect of a stably stratified WL on the convection and mean zonal flows. 


\begin{table*}
    \caption{\newtext{Nondimensional input parameters for both the CZ-only and CZ--WL cases, compared to the appropriate values for Jupiter and Saturn. We give the corresponding dimensional input parameters for the simulations in Table \ref{tab:redim}. For Jupiter and Saturn: $\Nrhowl$ is defined to be 4.6 (see Appendix \ref{ap:redim}), $\Nrhocz$ is computed from $\rhotilde\dimm$ at the base of the CZ (at $r=0.95a$ and $r=0.85a$ for Jupiter and Saturn, from \citealt{French2012} and \citealt{Preising2023}, respectively) and $\rhotilde\dimm$ at $\prstilde\dimm=10\ {\rm bar}$ \citep{Guillot2005}, and $\roc$ and $\raf$ are estimated using the values from Table \ref{tab:redim} and then setting $\roc=\tomegadim/\tffdim$ and $\raf=\pr\rocsq/\ek^2$.}}
\label{tab:input_nond}
\centering
\begin{tabular}{*{6}{l}}
\hline
Parameter & Definition & CZ-only case & CZ--WL case & Jupiter & Saturn \\
\hline
$\alphacz$ & $\rindim/\rcdim$ & 0.8 & 0.8 & 0.96 & 0.84\\
$\alphawl$ & $\rcdim/\rindim$ & 1 & 0.952 & 0.998 & 0.997\\
$\rin$ & $\alphacz/(1-\alphacz)$ & 4 & 4 & - & -\\
$\rc$ & $1/(1-\alphacz)$ & 5 & 5 & - & - \\
$\rout$ & $1/[\alphawl(1-\alphacz)]$ & 5 & 5.25 & - & - \\
$\gamma$ & $\cp/\cv$ & 1.5 & 1.5 & 1.4 & 1.4\\
$\Nrhocz$ & $\ln[\rhotilde\dimm(\rindim)/\rhotilde\dimm(\rcdim)]$ & 1.2 & 1.2 & 8.1 & 7.8 \\
$\Nrhowl$ &$\ln[\rhotilde\dimm(\rcdim)/\rhotilde\dimm(\routdim)]$ & 0 & 0.530 & 4.6 & 4.6 \\
$\di$ & $\gravcz d/\cp\tmpcz$ & 0.599 & 0.599 & -  & - \\
\hline
$\pr$ & $\nu/\kappa$ & 0.5 & 0.5 & 0.43 & 0.15 \\
$\raf$ & $\dfrac{\gravcz d^3}{\nu\kappa}\dfrac{\Dentr}{\cp}=\dfrac{F\cz \gravcz d^4}{\rhocz\tmpcz\cp\nu\kappa^2}$ &  $10^7$ & $10^7$ & $\sn{7.0}{24}$ & $\sn{9.5}{24}$ \\
$\roc$ & $\sqrt{\gravcz(\Dentr/\cp)}/\twoOmzero$ & 0.2 & 0.2 & $\sn{4.3}{-4}$ & $\sn{9.2}{-5}$ \\
$\ek$ & $\nu/4d\Omzero^2$& $\sn{4.47}{-5}$ & $\sn{4.47}{-5}$ & $\sn{1.1}{-16}$ & $\sn{1.2}{-17}$ \\
$\sigma$ & $(N\wl/\twoOmzero)\sqrt{\nu/\kappa}$ & - & 2.93 & 10 & 3.2 \\
$\bu$ & $(N\wl/\twoOmzero)^2$ & - & 17.2 & 230 & 68 \\
\hline
$n_\phi$ & Number of longitudinal grid points & 1536 & 2304 & - & - \\
$n_\theta$ &Number of latitudinal grid points & 768 & 1152 & - & - \\
$n_r$ & Number radial grid points & 96 & (96, 32) & - & - \\
\hline
\end{tabular}
\end{table*}

\subsection{Equations of Motion}\label{sec:equations}
The anelastic fluid equations are solved in a rotating frame with constant angular velocity $\Omzerovec=\Omzero\ez$. The Coriolis force is kept but the oblateness and centrifugal force are ignored. {\rayleigh} computes the evolution with respect to time $t\dimm$ of the three components of vector velocity $\vecu\dimm=u_\phi\dimm\ephi + u_\theta\dimm\etheta + u_r\dimm\erad$ ($ = u_\lambda\dimm\elambda + u_\phi\dimm\ephi + u_z\dimm\ez$) and the thermal perturbations $\prshat\dimm$ and $\entrhat\dimm$, where the hats denote deviations of the thermal variables from the corresponding background-state profiles.

The dimensional anelastic fluid equations are then (e.g., \citealt{Clune1999})
\begin{align}
	\nabla\dimm\cdot(\rhotilde\vecu)\dimm &\equiv  0,
	\label{eq:contdim}
\end{align}
\begin{subequations}
	\begin{align}
		\left(\rhotilde\frac{D\vecu}{Dt}\right)\dimm &= -2\rhotilde\dimm\Omzerovec\times\vecu\dimm -\left[\rhotilde\nabla \left(\frac{\prshat}{\rhotilde}\right)\right]\dimm \nonumber\\
        &\ \ \ +\frac{(\rhotilde\,\gravtilde \entrhat)\dimm}{\cp}\erad + \nabla\dimm\cdot (D_{ij}\dimm\e_i\e_j),\label{eq:momdim}\\
		\where D_{ij}\dimm &\define \left\{ 2\rhotilde\,\nutilde \left[e_{ij} - \frac{1}{3}(\nabla\cdot\vecu) \delta_{ij} \right]\right\}\dimm\label{eq:vstress}\\
		\andd e_{ij}\dimm &\define\frac{1}{2}\left(\pderiv{u_i}{x_j} + \pderiv{u_j}{x_i} \right)\dimm,\label{eq:ratestrain}
	\end{align}
\end{subequations}
and
\begin{align}
	\left(\rhotilde\tmptilde\frac{D\entrhat}{Dt}\right)\dimm =\ & -\cp\left[\rhotilde\tmptilde\frac{\brunttildesq}{\gravtilde} u_r\right]\dimm + \heattilde\dimm \nonumber\\
    &+ 
    \nabla\dimm\cdot\left(\kappatilde\,\rhotilde\tmptilde\nabla \entrhat\right)\dimm
    + D_{ij}\dimm e_{ij}\dimm.
	\label{eq:endim}
\end{align}

The indices $i$ and $j$ run over the three directions in a Cartesian coordinate system, $\delta_{ij}$ is the Kronecker delta, and repeated indices are summed. Note that in Equation \eqref{eq:momdim}, the diffusion of momentum is ``Laplacian," in that the highest-order term in $\nabla\cdot(D_{ij}\e_i\e_j)$ is $\nabla^2\vecu$. This further differentiates the 3D global simulations from 2D models, wherein ``hyperdiffusion" (terms like $-\nabla^4\vecu$ or even higher-order) is often employed (e.g., \citealt{Maltrud1991,Lian2008}). In Equation \eqref{eq:endim}, the diffusion of heat, which is also Laplacian, is assumed to respond to entropy gradients instead of temperature gradients, in accordance with mixing-length theory (e.g., \citealt{Gilman1972}).

\subsection{Nondimensionalization}\label{sec:nondim}
We nondimensionalize the coordinates and fields using the following system of units:
\begin{subequations}\label{eq:nond}
\begin{align}
[\nabla\dimm] = d^{-1}\\
[\partial/\partial t\dimm] = \twoOmzero\\
[\vecu\dimm] = \twoOmzero d\\
[\prshat\dimm] = \rhocz(\twoOmzero d)^2\\
[\entrhat\dimm] =\Dentr\est \define \frac{F\cz d}{\rhocz\tmpcz\kappa}.
\end{align}
\end{subequations}
We use the subscripts ``cz" or ``wl" to denote the volume-average of a quantity over the CZ or WL, respectively. For the background-state quantities, we omit the tildes (for example, $\rhocz$ denotes the volume-average of $\rhotilde\dimm$ over the CZ). Note that $\Dentr\est$ is the \textit{estimated} entropy difference across the CZ due to the imposed heat flux. The actual entropy difference $\Dentr$ after the convection has been established is not known a priori and the Rayleigh number derived below is thus ``flux-based."

We nondimensionalize the background state by scaling $\rhotilde\dimm$, $\prstilde\dimm$, $\tmptilde\dimm$, and $\gravtilde\dimm$ by their volume-averages over the CZ, $\heattilde\dimm$ by $F\cz/d$, and $\brunttilde\dimm$ by $N\wl$.

With these scalings, the nondimensional equations of motion are
\begin{align}
	\Div(\rhotilde\vecu) &\equiv 0\label{eq:cont},
\end{align}
\begin{align}\label{eq:mom}
    \rhotilde\left(\matderiv{\vecu}\right) =\ &-\rhotilde\ez\times\vecu-\rhotilde\nabla\left(\frac{\prshat}{\rhotilde} \right) +\rocsq \rhotilde\, \gravtilde \entrhat\erad \nonumber\\
    &+\sqrt{\frac{\pr}{\raf}}\roc \Div (D_{ij}\e_i\e_j),
\end{align}
and
\begin{align}\label{eq:heat}
	\rhotilde\tmptilde \matderiv{\entrhat} =\ &- \frac{\sigma^2}{\pr\rocsq} \rhotilde\tmptilde \frac{\brunttildesq}{\gravtilde} u_r+ \frac{\roc}{\sqrt{\pr\raf}} \heattilde\nonumber\\
    +\frac{\roc}{\sqrt{\pr\raf}}&\Div(\rhotilde \tmptilde \kappatilde \nabla \entrhat) + \sqrt{\frac{\pr}{\raf}}\frac{\di}{\roc} D_{ij}e_{ij}.
\end{align}


The nondimensional control parameters appearing in Equations \eqref{eq:cont}--\eqref{eq:heat} are
\begin{subequations}\label{eq:control}
\begin{align}
     \pr &\define \frac{\nu}{\kappa} \five \text{(thermal Prandtl number)},\\
     \raf &\define \frac{\gravcz d^3}{\nu\kappa}\left(\frac{\Dentr\est}{\cp}\right)= \frac{F\cz \gravcz d^4}{\rhocz\tmpcz\cp\nu\kappa^2}\nonumber\\
     &\five \text{(flux-based Rayleigh number)},\\
      \roc &\define \frac{\sqrt{(\gravcz/d)(\Dentr\est/\cp)}}{\twoOmzero} =\ek\sqrt{\frac{\ra}{\pr}}\nonumber\\
      &\five \text{(convective Rossby number)}\label{eq:controlroc},\\
      &\nonumber\\
      &\text{and}\nonumber\\
       \sigma &\define \left(\frac{\bruntrz}{\twoOmzero}\right)\sqrt{\frac{\nu}{\kappa}}=\sqrt{\bu\pr}\nonumber\\
       &\five\text{($\sigma$-parameter),}\label{eq:controlsigma}
\end{align}
\end{subequations}
where in Equations \eqref{eq:controlroc} and \eqref{eq:controlsigma} we have implicitly defined the Ekman number
\begin{align}\label{eq:ek}
    \ek&\define \frac{\nu}{\twoOmzero d^2}
\end{align}
and the buoyancy number
\begin{align}\label{eq:fr}
    \bu\define\frac{\brunt\wl^2}{4\Omzero^2}.
\end{align}

The five independent parameters in Equations \eqref{eq:control} fully characterize the system and we thus treat $\ek$ and $\bu$ as dependent parameters. The dissipation number
\begin{equation}\label{def:di}
\di\define\frac{\gravcz d}{\cp\tmpcz}\equiv 0.599
\end{equation} 
is not an independent control parameter, but is a property of the background state (see Equation \ref{eq:di}).


\newtext{These nondimensional control parameters are also presented in Table \ref{tab:input_nond}. The value $\roc=0.2$ was chosen to match CZ-only simulation of \citet{Wulff2022}. The value of $\bu=17.2$ should be $\gg1$ for the giant planets and is about as high as we can currently achieve before the gravity-wave time scale becomes prohibitively short. The values of $\raf$ and $\pr$ were chosen to match the CZ-only simulation of \citealt{Wulff2022} and it should be noted that $\raf$ differs by many orders of magnitude from what is expected in a real giant planet. Taken together, our simulations are thus most similar to Saturn in some ways (say in the value of $\alphacz$), but are wildly different from any giant planet in many other ways (say in the values of $\raf$, $\roc$, or $\alphawl$). We thus emphasize that our simulations should not be thought of as representing a real gas giant like Jupiter or Saturn with substantial fidelity. This is, unfortunately, a general aspect of most state-of-the-art simulations of real astrophysical systems, and arises fundamentally from computational constraints. For instance, increasing $\raf$ to its astrophysical value would result in extreme turbulence and scales of motion that are far too small to resolve by even the most powerful supercomputer.  Future work could partially address this issue (short of computers becoming many times more powerful) by implementing systematic parameter studies to describe the asymptotic behavior of the dynamics (e.g., \citealt{Julien1998,Aurnou2020,vanKan2025}).}


We report all results in nondimensional form, but we can always consider an equivalent dimensional system by a process outlined in Appendix \ref{ap:redim}. To convert between nondimensional and dimensional (CGS) units, use Table \ref{tab:redim}. 

\subsection{Initial and boundary conditions}
Each simulation is initialized from random noise in $\entrhat$ of amplitude roughly $10^{-3}$, with $\vecu$ and $\prshat$ both $\equiv 0$. The convection then develops and we wait for the system to reach a statistically steady state. 

At both boundaries, each case is impenetrable, stress-free, and has zero conductive flux: 
\begin{align}\label{eq:bc}
\urad=\pderiv{}{r}\left(\dfrac{\utheta}{r}\right)=\pderiv{}{r}\left(\dfrac{\uphi}{r}\right)=\pderiv{\entrhat}{r}\equiv0.
\end{align}

\subsection{Notation for averages}
We denote instantaneous longitudinal (or zonal) averages of quantities with angular brackets, e.g., $\av{\uphi}$. Note that for a given field $\psi$, we often utilize the Reynolds decomposition  $\psi=\av{\psi}+\psi^\prime$ and we call $\av{\psi}$ the mean $\psi$ and $\psi^\prime$ the fluctuating $\psi$. We append the subscript ``sph" to the brackets to denote a horizontal average over a spherical surface (e.g., $\avsph{\hat{s}}$). We append the subscript ``v" to indicate an instantaneous volume-average over a domain that will be specified in the text (e.g., $\avvol{(\vecuprime)^2}$, where the averaging might occur over the volume of the CZ). We denote an instantaneous combined longitudinal and axial average (i.e., taking an average of the longitudinally averaged quantity along $z$ at a fixed $\lambda$) by adding a ``$z$" subscript (e.g., $\avz{\uphi}$). Inside the tangent cylinder, we perform this $z$-average separately for each hemisphere. Similarly, we denote a combined longitudinal and radial average in the CZ--WL case (i.e., taking an average of a longitudinally averaged quantity along $r$ at a fixed $\theta$, where the radius range is always taken to be $r=\rc$ to $r=\rout$, i.e., over the WL) by appending an ``$r$" subscript (e.g., $\avr{\uphi}$). Note that the radial average is weighted by $r^2$ to take into account the variation of the local volume element along a ray at fixed $\theta$. For a combined spatial and temporal average, we append the subscript ``$t$" to the averaged quantity (e.g., $\avt{\uphi}$ or $\avrt{\uphi}$). The specific interval for the temporal average will be noted in the text. 

\section{Initial Flow Structures}\label{sec:jets_structure}
The mean zonal flow in our simulations takes a very long time to grow and finally reach equilibration. We say that the zonal flow has reached equilibrium if the following requirements are met: (a) the kinetic energy in the mean zonal flow evolves with time in a statistically stationary manner, (b) the latitudinal structure of the zonal flow is not obviously changing (i.e., no migration/merging of the jets), and (c) the right-hand sides of each of the equations of motion are near zero. Figure \ref{fig:etrace} shows the evolution of the shell-averaged mean zonal kinetic energy $\rhotilde\av{\uphi}^2/2$ during the full run of each simulation. There is initially very rapid growth in the kinetic energy (for example, over the interval $t=(0,25000)$ highlighted in Figure \ref{fig:etrace}a), followed by very slow secular growth until the kinetic energy curve finally flattens out, roughly at the time $\tequil$. We estimate $\tequil$ by eye from the growth curves and separately confirm that requirements (b) and (c) above are met. We find that multiple jets only appear at early times. To make contact with prior global simulations, which were also mostly analyzed at early times before equilibration, we examine each driving mechanism by averaging over the early time interval $t=5000\pm250$ ($= (4750,5250)$) highlighted in Figure \ref{fig:etrace}(b). In Sections \ref{sec:basicflow}--\ref{sec:superrot}, we analyze this early time interval exclusively. 

\begin{figure*}
	\centering
	\includegraphics[width=7.25in]{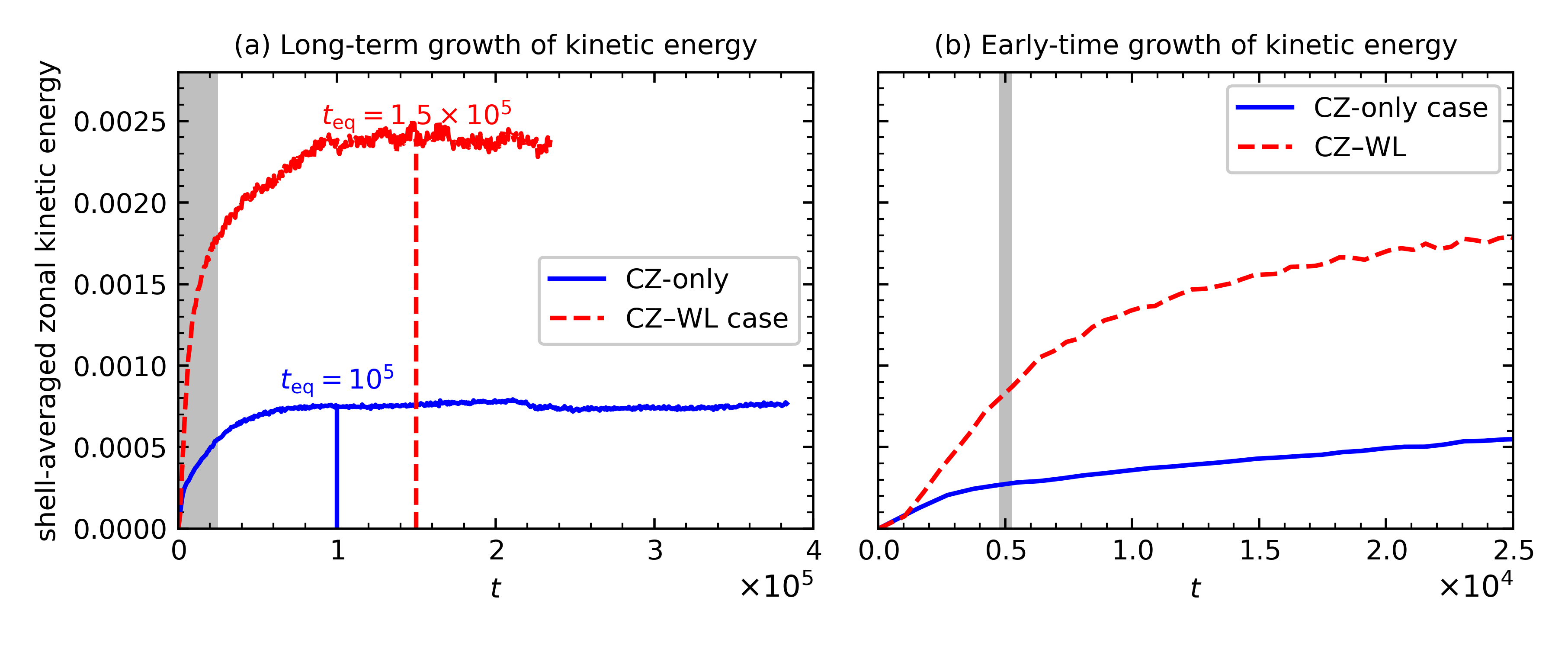}
	\caption{Trace of the mean zonal kinetic energy density, $\rhotilde\av{\uphi}^2/2$, instantaneously volume-averaged over the full shell for each case. (a) Full evolution of the kinetic energy from $t=0$ to $t=\trun$, with the interval $t=(0,25000)$ highlighted, and the estimated time at which the simulations finally equilibrate ($\tequil$) labeled by vertical lines. The total run-time of each simulation, $\trun$, is $\sn{3.84}{5}$ for the CZ-only case and $\sn{2.35}{5}$ for the CZ--WL case. (b) Early-time evolution of the kinetic energy during the interval $t=(0,25000)$ highlighted in panel (a). The early-time averaging interval $t=5000\pm250$ analyzed in Sections \ref{sec:basicflow}--\ref{sec:superrot} is highlighted. Recall that our dimensional unit of time is $(2\Omega_0)^{-1}$ (rotation period divided by $4\pi$) and that in these units, the diffusion timescales are $\ek^{-1}=22400$ (viscous) and $\pr/\ek=11200$ (thermal).}
	\label{fig:etrace}
\end{figure*}

\subsection{Basic flow properties}\label{sec:basicflow}
Figure \ref{fig:cutout3d} shows the fluctuating axial vorticity $\omzprime$ at $t=5000$. There are many thin, cylindrically aligned columns of vorticity, which basically extend throughout the full CZ in each case. On a given spherical surface near the equator, the cross sections of these columns display the canonical outward spiral associated with Busse columns (e.g., \citealt{Busse2002} and Figure \ref{fig:eqslice_initial}). The flows in the CZ have largely comparable structure in both cases, but in the WL of the CZ--WL case, the flows are clearly larger-scale overall and the symmetry with respect to the rotation axis is broken. Instead, because of the strong stable stratification, the flows in the WL consist of pancake vortices of predominantly horizontal flow with small vertical extent (compare to Figure \ref{fig:schematic}).

\begin{table}
    \caption{Nondimensional output parameters for the CZ-only case, the CZ of the CZ--WL case, and the WL of the CZ--WL case. Here the temporal averaging interval is $t=5000\pm250$. See Equations \eqref{eq:rof}--\eqref{eq:rem}.}
\label{tab:output_nond}
\centering
\begin{tabular}{*{4}{l}}
\hline
Parameter & CZ-only & CZ--WL (CZ) & CZ--WL (WL) \\
\hline
$\ro_{\rm f}$ & 0.19 & 0.24 & 0.16\\
$\ro_{\rm m}$ & 0.087 & 0.14 & 0.083\\
$\re_{\rm f}$ & 180 & 270 & 140\\
$\re_{\rm m}$ & 590 & 1000 & 1500\\
$u_{\rm f}$ & 8.1e-03 & 0.012 & 6.1e-03\\
$u_{\rm m}$ & 0.026 & 0.045 & 0.069\\
\hline
\end{tabular}
\end{table}

\begin{figure*}
	\centering
	\includegraphics[width=7.25in]{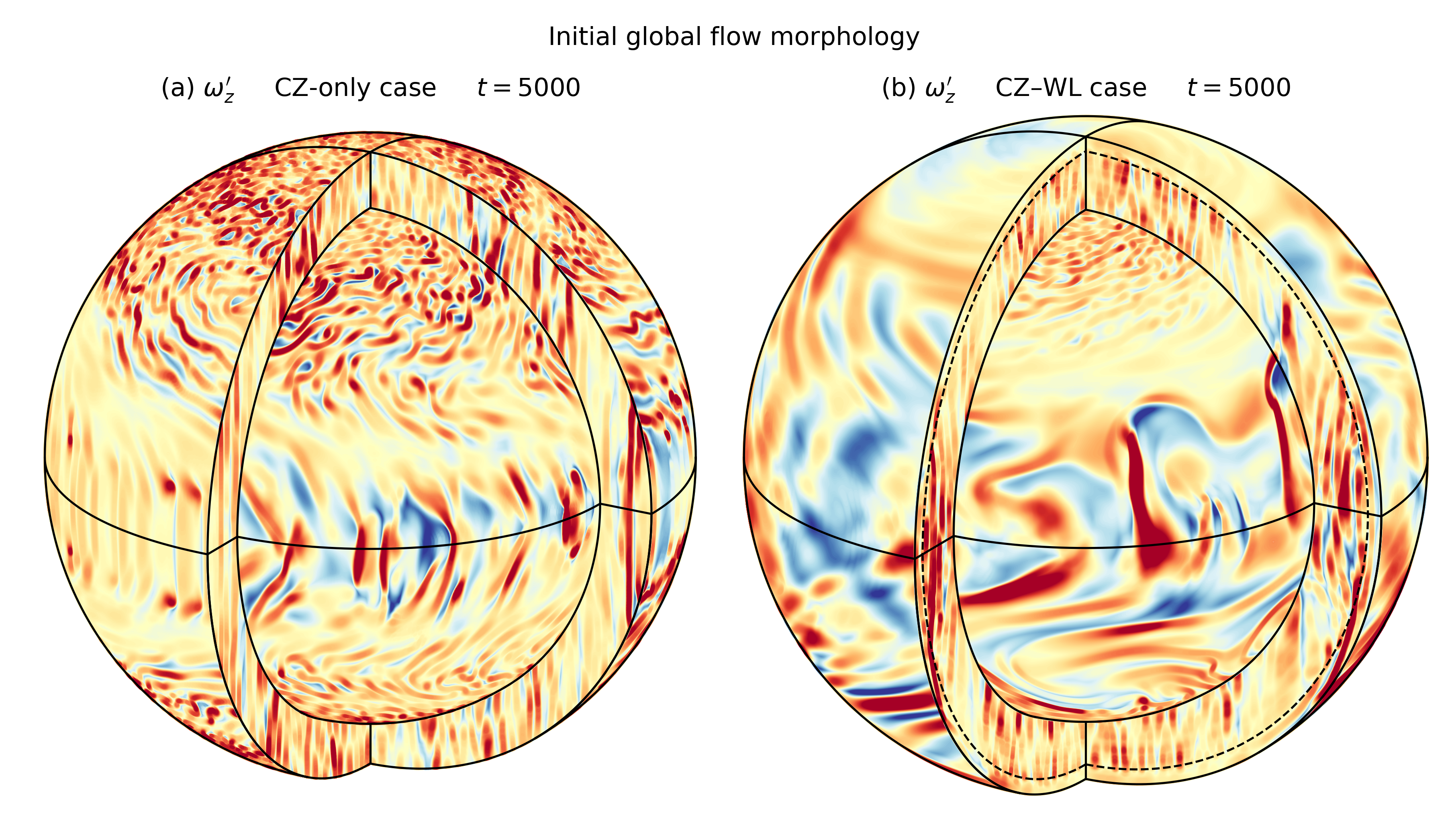}
	\caption{Spherical and meridional cutouts of our 3D shellular simulation domain, showing snapshots of the fluctuating axial vorticity $\omzprime$ in (a) the CZ-only case and (b) the CZ--WL case. Each snapshot is taken near the beginning of the simulation when there are multiple jets ($t=5000$). The inner and outer spherical surfaces bounding the cutout are taken just above and below the inner and outer domain boundaries, respectively. The meridional surfaces bounding the cutouts are located $90^\circ$ apart, at $\phi=-30^\circ$ and $\phi=60^\circ$, with $\phi=0^\circ$ being the central viewing longitude (each cutout is viewed from the latitude $\theta=20^\circ$). Each radial level is normalized separately by the rms of $\omzprime$ at that level. In panel (b), the interface $r=\rc$ is marked by dashed black curves. The relative sizes of the plots reflect the larger domain size for the CZ--WL case. An animated version of this figure is available in the online journal. In the 30-second animation, the evolution of $\omzprime$ for the two cases is shown side by side for an interval of length $\approx$ 720 (in simulation time units) centered about $t=5000$. In each animation, time is measured from beginning of the animation. Animated versions of the separate panels are also available for download at \url{https://doi.org/10.5281/zenodo.20332552} and for viewing at \url{https://youtube.com/shorts/YSKIkOU8nYI} (CZ-only case), \url{https://youtube.com/shorts/hl5Ipgm9unU} (CZ--WL case), and \url{https://youtu.be/9rYgFe9U_ag} (both cases side-by-side). }
	\label{fig:cutout3d}
\end{figure*}

Some nondimensional diagnostics characterizing the simulated flows are the Rossby number of the fluctuating flows, which here we define as
\begin{align}\label{eq:rof}
	\ro_{\rm f} &\define \frac{\avvol{(\vecom^{*\prime})^2}^{1/2}}{\twoOmzero} = \avvol{(\vecom^{\prime})^2}^{1/2},
\end{align}
the Rossby number of the mean flows,
\begin{align}\label{eq:rom}
	\ro_{\rm m} &\define \frac{\avvol{\av{\vecom\dimm}^2}^{1/2}}{\twoOmzero} = \avvol{\abs{\av{\vecom}}^2}^{1/2},
\end{align}
the Reynolds number of the fluctuating flows,
\begin{align}\label{eq:ref}
	\re_{\rm f} &\define \frac{\avvol{(\vecu^{*\prime})^2}^{1/2}H}{\nu} = \frac{\avvol{\abs{\vecu^{\prime}}^2}^{1/2}}{\ek},
\end{align}
the Reynolds number of the mean flows,
\begin{align}\label{eq:rem}
	\re_{\rm m} &\define \frac{\avvol{\av{\vecu}^2}^{1/2}H}{\nu} = \frac{\avvol{\av{\vecu}^2}^{1/2}}{\ek},
\end{align}
and the nondimensional amplitudes of the fluctuating and mean flows, which we denote by $u_{\rm f}\define \avvol{(\vecuprime)^2}^{1/2}$ and $u_{\rm m}\define \avvol{\av{\vecu}^2}^{1/2}$, respectively. Here, the volume-average is over either the CZ or WL.

Table \ref{tab:output_nond} shows these output parameters separately for each case and region for the interval $t=5000\pm250$. Separate volume-averages are taken for the full shell of the CZ-only case, the CZ of the CZ--WL case, and the WL of the CZ--WL case. All the flows are rotationally constrained, with the various Rossby numbers never exceeding 0.24. The simulations are still reasonably turbulent (even in the WL of the CZ--WL case), with the various Reynolds numbers all exceeding 140. The flow amplitudes themselves are quite small, with dimensional values typically only a few percent of $\twoOmzero H$. 

\begin{figure}
	\centering
	\includegraphics[width=3.4375in]{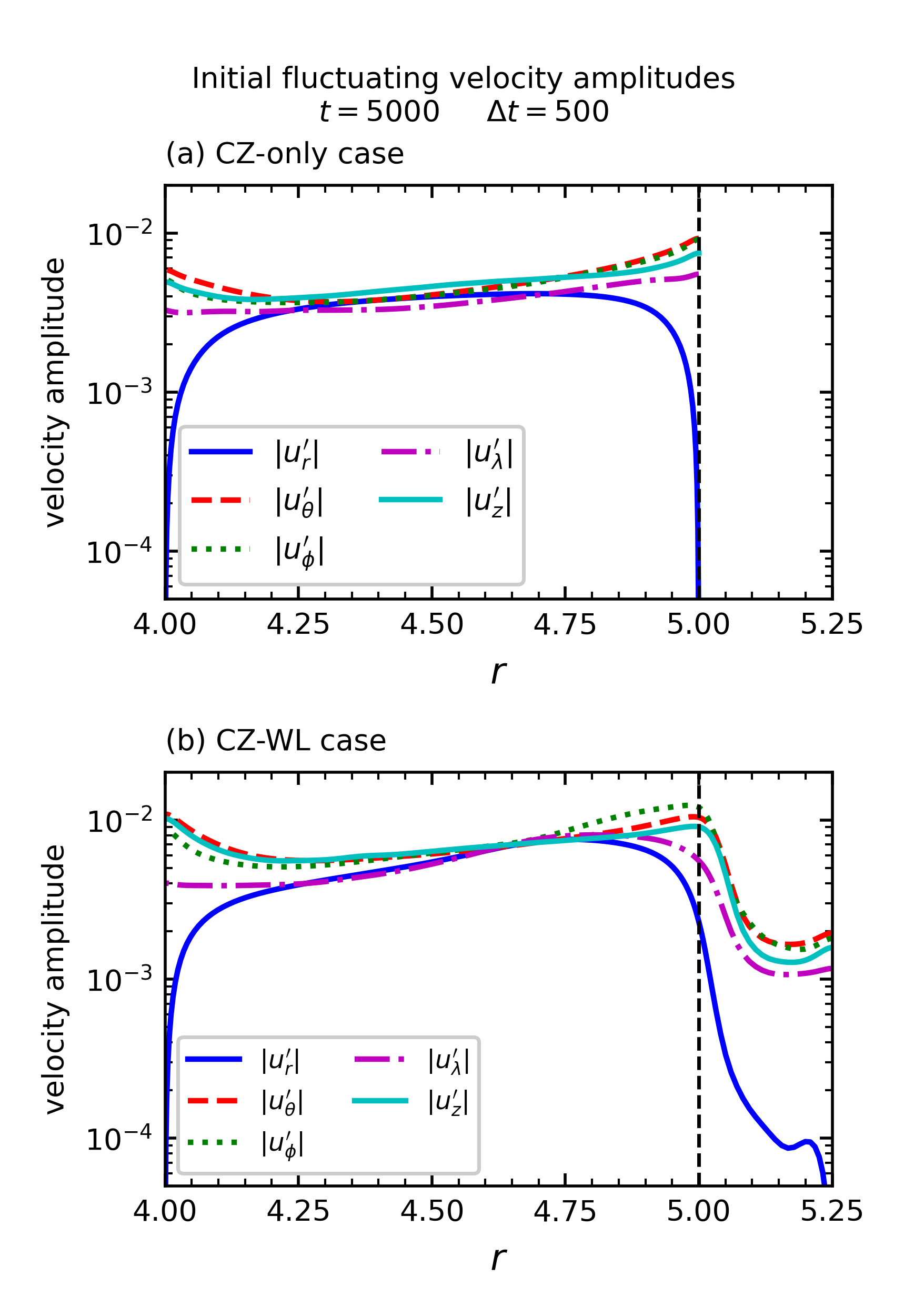}
	\caption{Fluctuating flow amplitudes for the various velocity components as functions of radius for (a) the CZ-only case and (b) the CZ--WL case, averaged in time over $t=5000\pm250$. Here, the absolute values denote the rms of the enclosed quantity, with the mean taken over time and horizontal (i.e., spherical) surfaces. For example, $\lvert u_r^\prime\rvert\define\avspht{(u_r^\prime)^2}^{1/2}$. }
	\label{fig:uamp_initial}
\end{figure}

Note that despite the prominent organization of the convective flow structures into axially aligned columns, the motion in each simulation's CZ is fully 3D. We emphasize this point in Figure \ref{fig:uamp_initial}, which shows the fluctuating flow amplitudes of each velocity vector component (both cylindrical and spherical) separately for both the CZ-only and CZ--WL cases. Except near the boundaries, all flow components have comparable amplitudes in the CZ. In particular, $|u_z^\prime|$ is significant, which we would not expect for a thin stationary column of axial vorticity. This can be understood because the columns are \textit{not} stationary. Instead, they are advected by neighboring columns and move in the $\lambda$ direction, undergoing significant stretching and compression because of the tilted spherical boundaries, and thus generating significant $\uzprime$. In the WL of the CZ--WL case, the radial velocity amplitude drops off significantly and becomes about 20 times weaker than the horizontal velocity components. This indicates that unlike the motion in the CZ, the motion in the WL \textit{is} primarily 2D, being composed of pancake structures of vertical vorticity $\omradprime$.

\subsection{Initial jets}\label{sec:initial_jets}
Both types of eddies---Busse columns and pancake vortices---initially drive multiple jets, the structures of which are shown in Figure \ref{fig:jets_initial}. The two cases are largely comparable: in the CZ-only case there are seven jets in the northern hemisphere and five jets in the southern hemisphere, whereas in the CZ--WL case, there are five jets in each hemisphere (the equatorial jet gets double-counted). In each case, there is one large superrotating jet just outside the tangent cylinder.

Despite this similarity in the overall structure of the jets, the jets in the CZ--WL case are of significantly higher amplitude and have less cylindrically aligned isocontours of zonal flow in the bulk of the CZ. Indeed, in the WL, particularly at low to mid latitudes, the isocontours are closer to radial alignment than cylindrical alignment. We now verify that this is due to a modification of the thermal wind balance by the presence of our \newtext{simplified} WL, which results in a large cell of meridional circulation in the WL which transports heat.

\begin{figure*}
	\centering
	\includegraphics[width=5in]{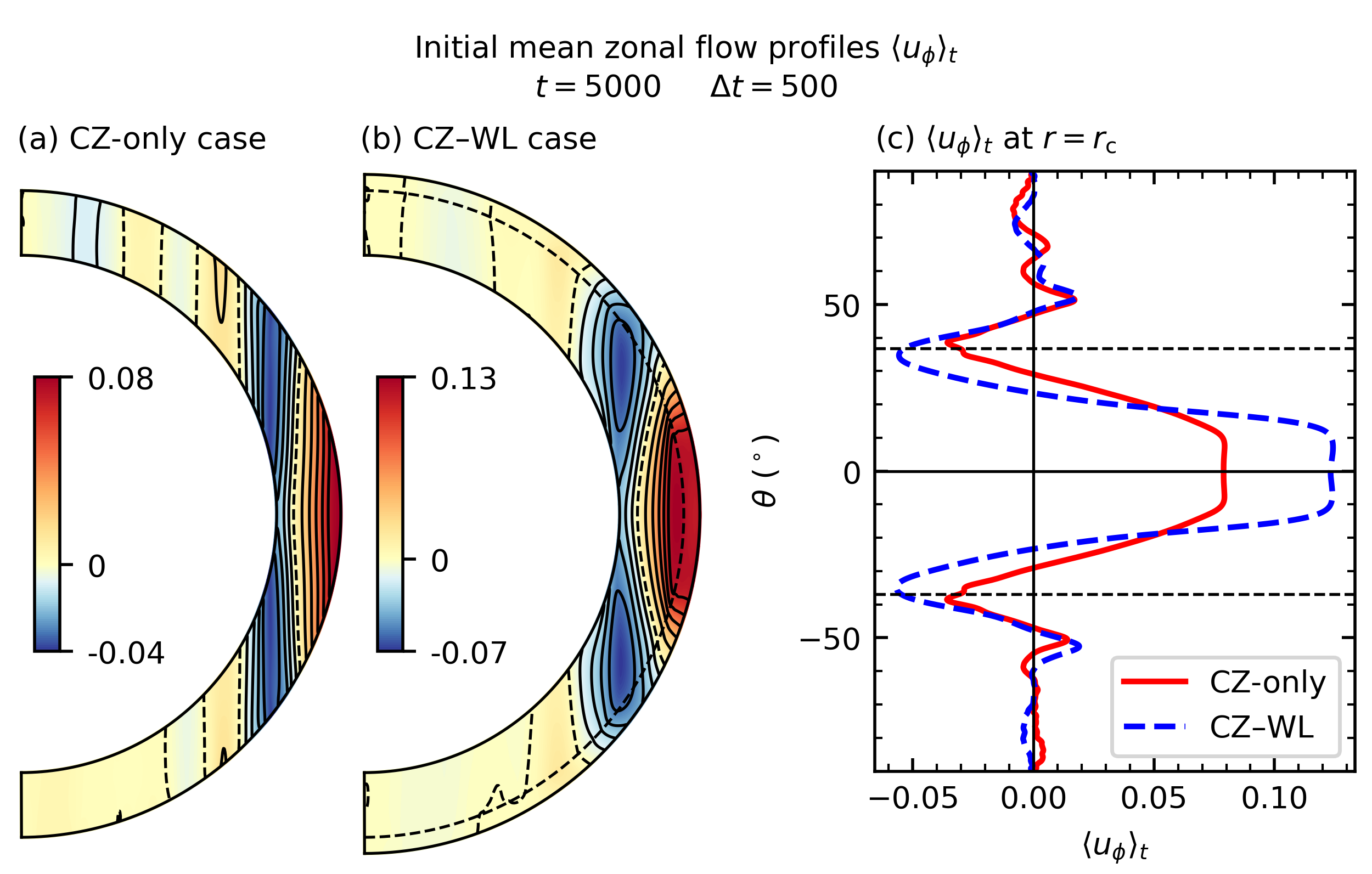}
	\caption{Temporally averaged mean zonal flow $\avt{\uphi}$ in the meridional plane, showing the initial multiple jet structure in (a) the CZ-only case and (b) the CZ--WL case, averaged in time over the same interval $t=5000\pm250$ as in Figure \ref{fig:uamp_initial}. Positive values are normalized separately from negative values and the nonzero contours (solid black curves) are equally spaced within each region of positive and negative values (the $\avt{\uphi}\equiv0$ contours are dashed). In panel (b), the CZ--WL boundary ($r=\rc$) is marked by the dashed semicircle. (c) Comparison of the latitudinal profiles of zonal flow at the top of the CZ ($r=\rc$) in the two cases. The horizontal dashed black lines denote the location of the tangent cylinder at $r=\rc$ ($\theta=36.9^\circ$).  }
	\label{fig:jets_initial}
\end{figure*}

\begin{figure}
	\centering
	\includegraphics[width=3.4375in]{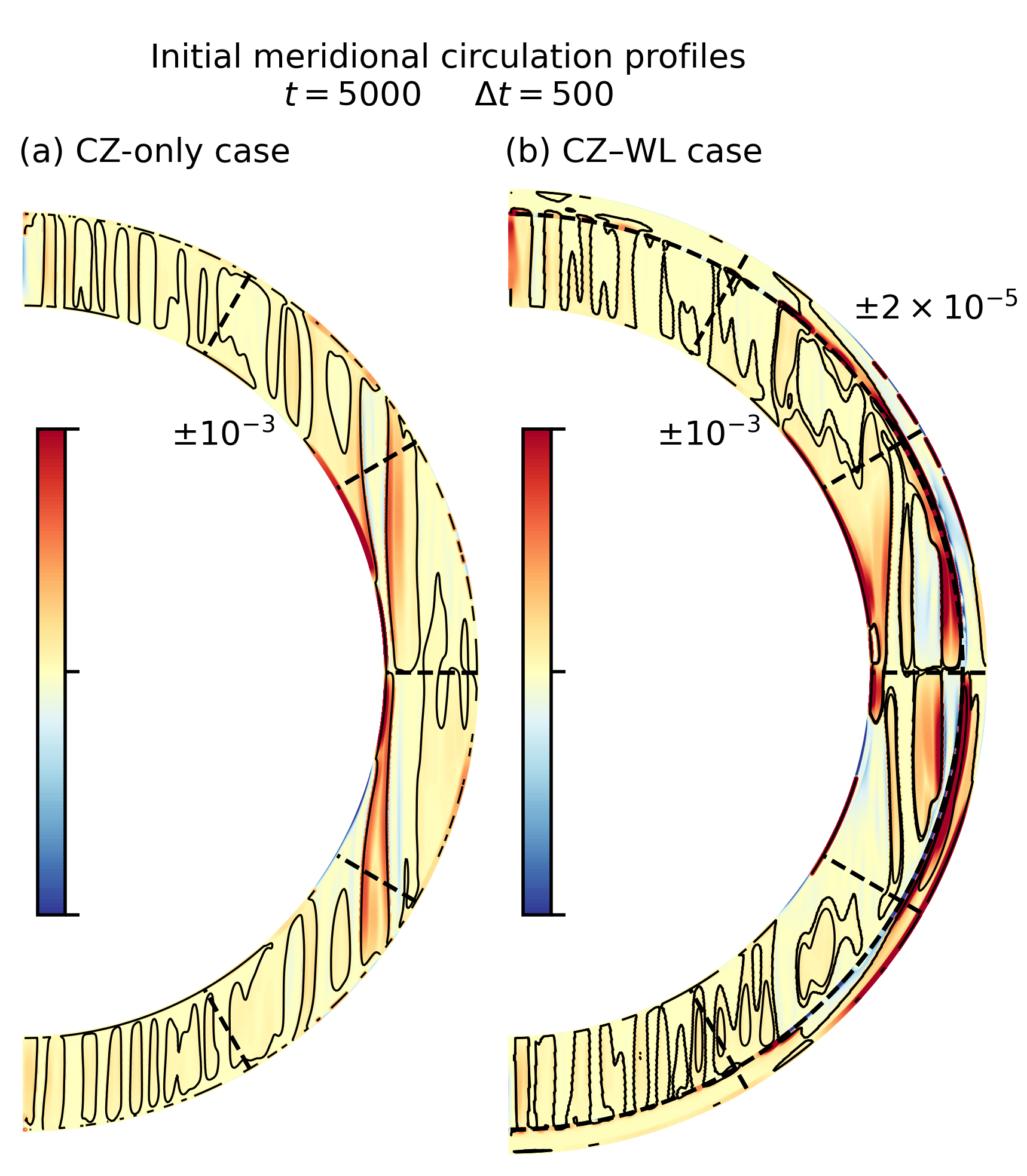}
	\caption{Meridional circulation profiles $\lvert\rhoumer_t\rvert\sgn{\avt{\Psi}}$ (see Equations \ref{eq:umer} and \ref{eq:psi}) for (a) the CZ-only case and (b) the CZ--WL case, plotted in the meridional plane and averaged over $t=5000\pm250$. Red tones indicate clockwise circulation and blue tones indicate counterclockwise circulation. Contours of $\avt{\Psi}\equiv0$ (the circulation cell boundaries) are also shown.  In panel (b), the color table is saturated separately for the CZ (extreme values $\pm 10^{-3}$, also used for the CZ-only case) and the WL (extreme values $\pm\sn{2}{-5}$). This allows the rendering of the small-amplitude flow in the WL. The dashed semicircle denotes $r=\rc$. }
	\label{fig:mercirc}
\end{figure}

\begin{figure*}
	\centering
	\includegraphics[width=7.25in]{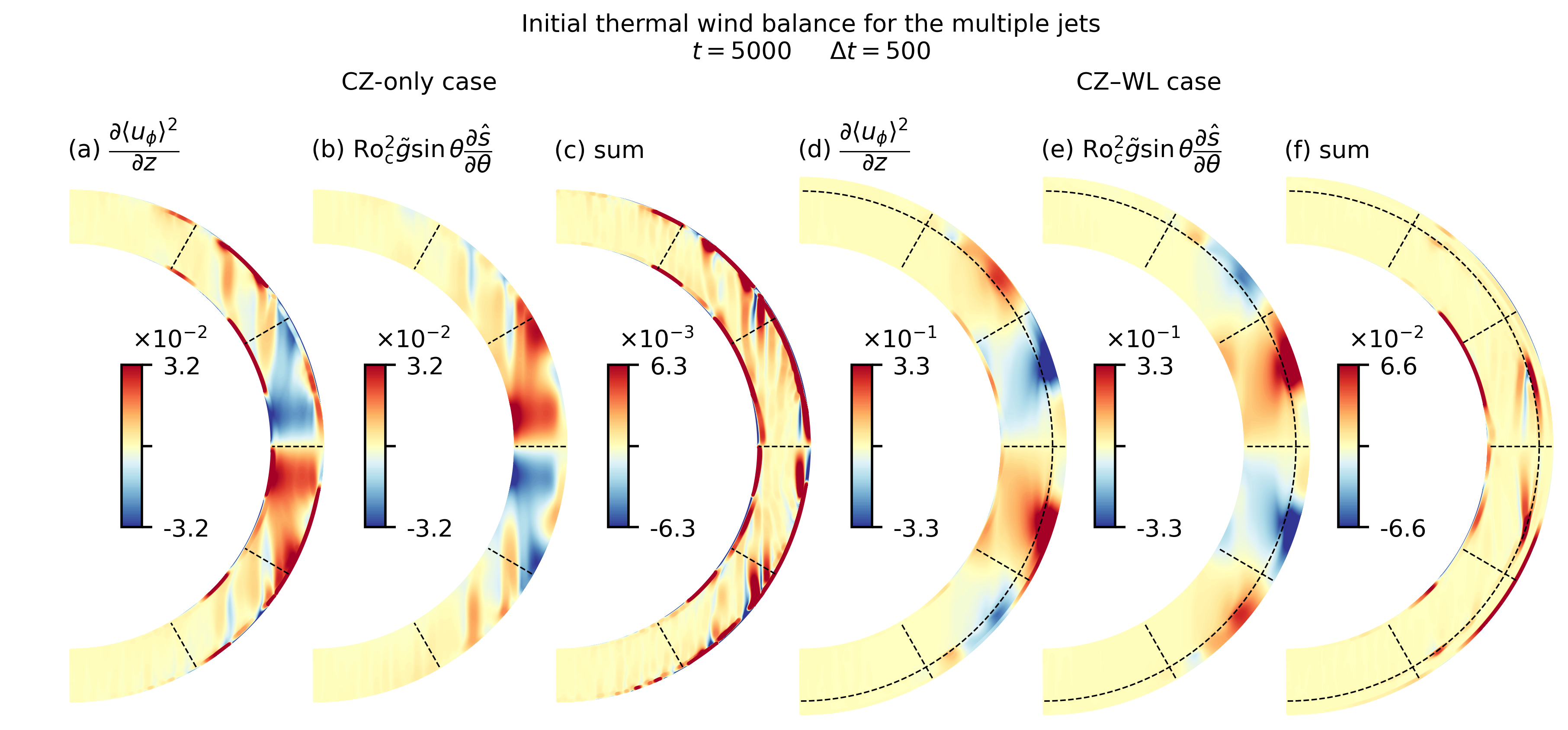}
	\caption{Thermal wind balance (Equation \ref{eq:tw}), averaged in time over $t=5000\pm250$. The centrifugal term, baroclinic term, and their sum are plotted in the meridional plane for (a--c) the CZ-only case and (d--f) the CZ--WL case. As in other figures, the dashed semicircle in panel (b) denotes the CZ--WL interface. }
	\label{fig:twbalance}
\end{figure*}

\subsection{Initial meridional circulation}\label{sec:mercirc}
In addition to the mean zonal flow, there is also a mean meridional circulation, which is generally described by the vector mean mass flux $\rhoumer$, where
\begin{equation}\label{eq:umer}
\umer\define\urad\erad + \utheta\etheta = \ulambda\elambda + \uz\ez
\end{equation}
represents the flow components in the meridional plane. 

Because $\Div \rhoumer \equiv0$, we can define the meridional-flow streamfunction $\Psi(r,\theta,t)$, which satisfies $\rhoumer=\nabla\times[\Psi(\ephi/\lambda)]$, or
\begin{equation}\label{eq:psi}
\rhourad=\frac{1}{r^2\sin\theta}\pderiv{\Psi}{\theta}\ \ \text{and}\ \ \rhoutheta=- \frac{1}{r\sin\theta}\pderiv{\Psi}{r}.
\end{equation}
The contours of $\Psi$ then correspond to the streamlines of $\av{\umer}$ and the density of the contours corresponds to the magnitude of the mass flux $\lvert\rhoumer\rvert$. We choose the integration constant such that $\Psi$ vanishes on the boundaries of the meridional plane, which must also be boundaries of circulation cells. Thus, the contours $\Psi\equiv0$ correspond to the boundaries of the circulation cells and the sign of $\Psi$ indicates the sense of circulation within a cell. Looking along positive $\phi$ (i.e., into the page), $\Psi>0$ indicates clockwise circulation and $\Psi<0$ indicates counterclockwise circulation. 

Figure \ref{fig:mercirc} shows the initial meridional circulation profiles ($\lvert\rhoumer_t\rvert\sgn{\avt{\Psi}}$, where $\sgn$ is the sign function) for both cases. In the CZ, except near the boundaries, the circulation cells are largely aligned with the rotation axis and there are many alternating clockwise/counterclockwise cells, commensurate with the multiple jets. The two circulation profiles are largely similar in the CZ.

In the WL of the CZ--WL case, the meridional flow is quite different. It is aligned primarily along spherical rather than cylindrical surfaces. Although there is some imprinting of the CZ's multiple circulation cells upward, the bulk of the WL is dominated by a single counterclockwise cell (occupying low to mid-latitudes) in the northern hemisphere and a clockwise cell in the southern hemisphere. These two cells are also the locations of the significant contour tilts of mean zonal flow in Figure \ref{fig:jets_initial}(b). For each hemisphere, this monocellular flow allows significant transport of heat in latitude and resultant large-scale baroclinicity. We remark that the absence of substantial multicellular meridional circulation in the \newtext{simplified} WL is interesting in light of Juno observations, which indicate the presence of many circulation cells (Ferrel-like cells) in the Jovian WL \citep{Duer2021}. 

\subsection{Thermal Wind Balance of the Initial Jets}\label{sec:twbalance}
To elucidate why the contour tilts of mean zonal flow are so different in each case, we consider the maintenance of the meridional flow. For high rotational constraint, this often takes the form of a special type of thermal wind balance appropriate for deep and shallow atmospheres alike (e.g., \citealt{Matilsky2023}).\footnote{The traditional form of thermal wind balance (e.g., \citealt{Vallis2017}, p. 87) is only applicable to shallow and/or strongly stratified fluid layers, such as the Earth's atmosphere and oceans, for which the vertical extent of the motion is much less than the horizontal extent.} Taking the mean zonal component of the curl of the momentum equation \eqref{eq:mom} and dropping the viscous and Reynolds-stress terms, we find
\begin{align}\label{eq:tw}
    \pderiv{\av{\uphi}^2}{z}+\rocsq \gravtilde \sin\theta \pderiv{\entrhat}{\theta}\approx 0.
\end{align}
which is the (nondimensional) canonical form thermal wind balance for deep atmospheres. The left-most term is called the centrifugal term because of its resemblance to a centrifugal force and the right-most term is the baroclinic term arising from latitudinal entropy gradients. 

Figure \ref{fig:twbalance} shows the thermal wind balance during $t=5000\pm250$ for each case, i.e., the two terms in Equation \ref{eq:tw} and their sum. Overall, the balance is fairly good in each case, with the relative residual between the two terms only about 20\% in the bulk of the layer (close to the boundaries, thermal wind balance is broken because Ekman layers result in a large contribution from the viscous term). In large portions of the CZ--WL case's bulk, the relative residual is even smaller, on the order of only a few percent. Analyses at subsequent times (not shown) show that the balance improves as the steady state is approached, with the residual approaching just a few percent everywhere in the bulk for both cases in the equilibrated state.

In the WL of the CZ--WL case, the large tilts in the zonal flow's isocontours show up as an increased amplitude of the centrifugal term at low to mid-latitudes, which is in turn balanced by enhanced baroclinicity. Analysis of the heat transport equation (not shown) reveals an advective--diffusive balance in the mean entropy equation \eqref{eq:heat}, showing that the large baroclinicity is caused by advective transport from the two cells of meridional circulation at low to mid latitudes in the WL (Figure \ref{fig:mercirc}b). In summary, even our \newtext{simplified} WL can significantly alter the dynamics of the zonal jets by leading to enhanced heat transport by meridional circulation and a sizable modification of the thermal wind balance.

\newtext{As noted in Section \ref{sec:intro}, inversions of the gravimetric data typically suggest that the jets on Jupiter and Saturn extend through the CZ along cylinders rather than radial lines (e.g., \citealt{Kaspi2023,Galanti2025}). However, as these inversions are forward models, they do not indicate the precise tilt angles of the jets, thus allowing for finite deviation from cylindrical symmetry. To assess whether the jets in our simulations are consistent with observations, we define the average tilt angle of the jets in a given volume $\rm v$ as follows (e.g., \citealt{Matilsky2020b}):}
\begin{align}\label{eq:tilt}
    \text{tilt} \define \tan^{-1}\left[\dfrac{\avvol{|\pderivline{\avt{\uphi}}{z}|}} {\avvol{|\pderivline{\avt{\uphi}}{\lambda}|}}\right].
\end{align}
\newtext{We do not distinguish between positive and negative tilt angles here. Table \ref{tab:tilts} shows the tilt angles for the whole domain of the CZ-only case and separately for the CZ and WL of the CZ--WL case, at different times throughout the simulation. There is some evolution toward higher (lower) tilt angles in the CZ (WL), but in general the following is true at all times: the jets in both CZs remain quite close to cylindrical alignment, deviating by at most 2--4$^\circ$ in the CZ-only case and up to $\approx$10$^\circ$ in the CZ--WL case. The simplified WL, however, consistently achieves large tilt angles close to 30$^\circ$, perhaps closer to radial alignment than cylindrical alignment. \newnewtext{Such tilts do not seem to be obviously excluded by the gravimetric observations. For instance,} a tilt between $3^\circ$ and $10^\circ$ in the CZ does not seem sufficient to significantly perturb the mass distribution; meanwhile, the overall mass of each planet's WL is so small that the inversions give essentially no constraint on the contour tilts in the WL. Of course, full consistency with observations could only be confirmed by explicitly computing a forward model inspired by our results.}
\begin{table}
    \caption{\newtext{Bulk tilt angles (as defined in Equation \ref{eq:tilt}) for the following regions: the CZ-only case (whole domain), the CZ--WL case (just the CZ), and the CZ--WL case (just the WL). The rows correspond to the different time intervals examined in Section \ref{sec:migration} (for instance, see Figure \ref{fig:migration}, panels b--g).}}
\label{tab:tilts}
\centering
\begin{tabular}{*{4}{l}}
\hline
time & CZ-only & CZ--WL (CZ) & CZ--WL (WL) \\
\hline
5,000 & 2.0$^\circ$ & 7.8$^\circ$ & 33.1$^\circ$ \\
10,000 & 2.4$^\circ$ & 9.0$^\circ$ & 29.9$^\circ$ \\
20,000 & 2.9$^\circ$ & 9.7$^\circ$ & 27.3$^\circ$ \\
30,000 & 3.2$^\circ$ & 10.0$^\circ$ & 26.5$^\circ$ \\
40,000 & 3.3$^\circ$ & 10.0$^\circ$ & 25.6$^\circ$ \\
eq. state & 3.6$^\circ$ & 10.2$^\circ$ & 26.3$^\circ$ \\
\hline
\end{tabular}
\end{table}

\section{Formation of Jets in the CZ}\label{sec:jets_cz}

Figure \ref{fig:cutout3d} shows that in both cases, the convection largely takes the form of columnar rolls aligned with the rotation axis. We find that these columnar structures share many properties explored by (e.g.) \citet{Busse2002}, which is why we refer to them as Busse columns. As discussed in Section \ref{sec:driving_mechanisms}, Busse columns can drive cylindrically aligned jets in a quasi-2D framework through the topographic $\beta$-effect. In this section, we thus consider the homogenization of $\avt{Q_z}$ (over the early time interval $t=5000\pm250$), the formation of PV staircases, and the width of the jets relative to the topographic Rhines scale. In terms of the nondimensional variables, Equation \eqref{eq:pvz_dim} gives
\begin{align}
Q_z &= \frac{\omz +1}{H} = \frac{\omz}{H} + \ftopo\label{eq:pvz}>0,\\
\ftopo &\define \frac{1}{H},\\
\andd \betatopo &\define -H\frac{d\ftopo}{d\lambda} = \frac{1}{H}\frac{dH}{d\lambda},
\end{align}
where, from Figure \ref{fig:schematic}, we explicitly calculate
\begin{equation}\label{eq:hlambda_dim}
H(\lambda) \define 
\begin{cases}
 2\sqrt{r_{\rm c}^2-\lambda^2}   & \lambda\geq\rin \\
 2(\sqrt{r_{\rm c}^2-\lambda^2}-\sqrt{r_{\rm in}^2-\lambda^2}),    & \lambda <\rin
\end{cases}
\end{equation}
To avoid a discontinuity in $H$, we take $H$ inside the tangent cylinder to be double the axial distance between the shells. 

\begin{figure*}
	\centering
	\includegraphics[width=5in]{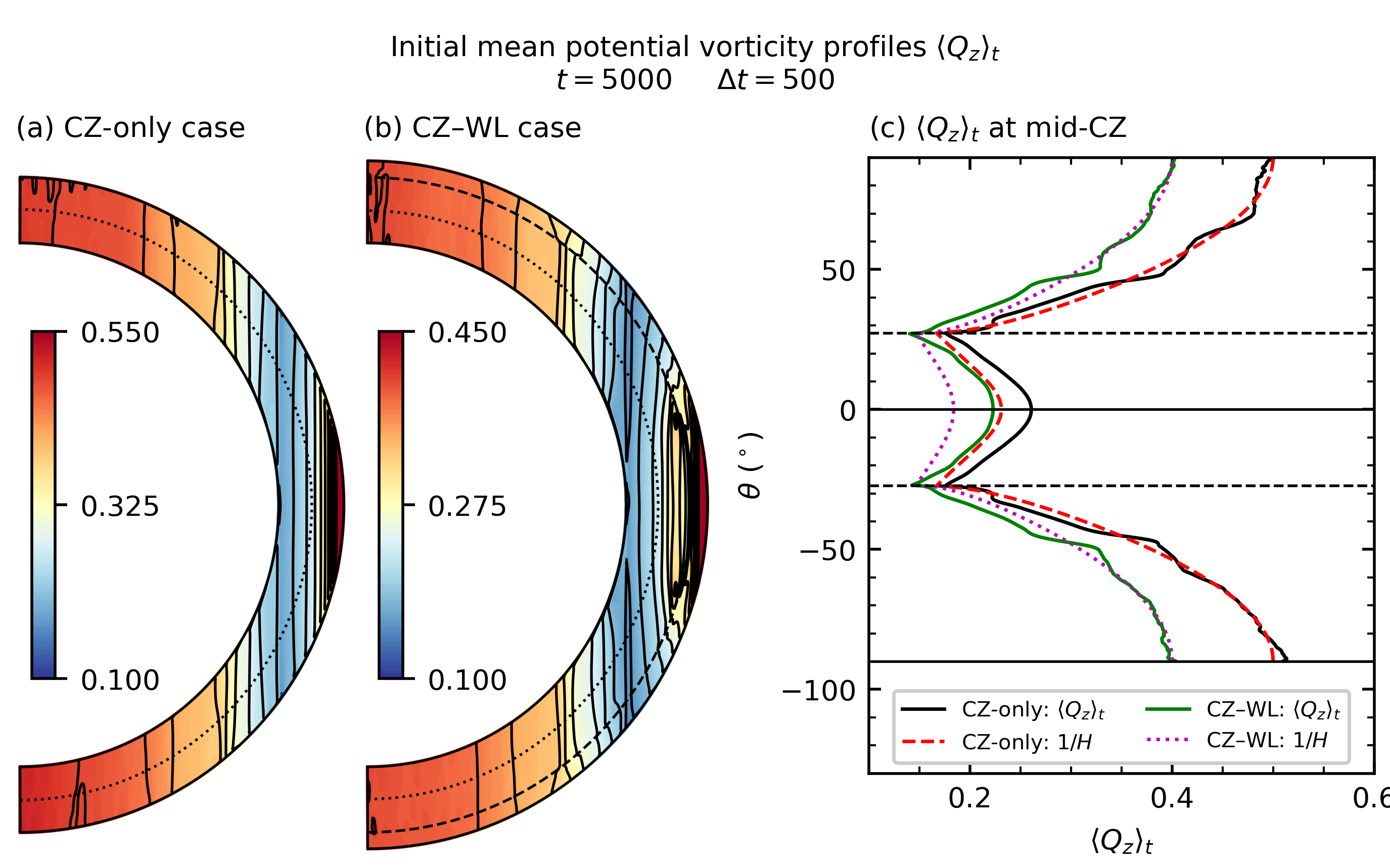}
	\caption{Temporally averaged mean PV $\avt{Q_z}$ in (a) the CZ-only case and (b) the CZ--WL case, averaged over the same interval $t=5000\pm250$ as in Figures \ref{fig:uamp_initial} and \ref{fig:jets_initial}. All contours (solid black curves) are equally spaced. The radii $r=4.5$ and $r=5$ are marked by the dotted and dashed semicircles, respectively. (c) Latitudinal profiles of $\avt{Q_z}$ and the planetary PV $\ftopo=1/H$ for both cases at mid-CZ ($r=4.5$). The horizontal dashed black lines denote the location of the tangent cylinder at $r=4.5$ ($\theta=27.3^\circ$).}
	\label{fig:pv_initial}
\end{figure*}

Figure \ref{fig:pv_initial} shows the mean PV $\avt{Q_z}$ associated with the initial jets.\footnote{Note that ideally $H(\lambda)$ would always represent the axial distance between the bounding spheres of the CZ, $r\equiv\rin$ and $r\equiv\rc$ (as in Figure \ref{fig:schematic}) However, this would lead to undefined $H$ and $\betatopo$ in the WL of the CZ--WL case. To simplify the calculations, in the CZ--WL case, we take $H$ to be the axial distance between the bounding spheres of the full domain (including the WL).} The PV contours are are almost entirely cylindrically aligned in the CZ of each case, but tilt significantly in the WL of the CZ--WL case at low to mid latitudes, just like the contours of $\avt{\uphi}$. This again suggests that $Q_z$ is the right form for PV, but only in the CZ.  In Figure \ref{fig:pv_initial}(c), it is clear that there are some suggestions of staircase-like structures in PV outside the tangent cylinder, with regions of roughly homogeneous $\avt{Q_z}$ in between the high-latitude jets.

To increase the signal of the initial staircase structure in PV (and to respect the axial symmetry of the columns in the CZ), we plot the axially averaged flow profiles for mean zonal flow velocity (Figure \ref{fig:jets_avz_initial}) and PV (Figure \ref{fig:staircases_initial}). To capture the hemispheric asymmetry of the flow inside the tangent cylinder, we average in $z$ over each hemisphere separately and plot the results as a function of the coordinate
\begin{align}\label{eq:lambda}
    \Lambda(\lambda,\theta)\define (\rout - \lambda)\sgn{\theta}.
\end{align}
The coordinate $\Lambda$ falls in the range $(-\rout,\rout)$ and is positive in the northern hemisphere and negative in the southern hemisphere, with $\Lambda=0$ marking the equator at the outer surface. Outside the tangent cylinder (the region $\lvert\Lambda\rvert\leq\rout-\rin$), we average over the full range of $z$, so all $z$-averaged functions of $\Lambda$ are symmetric outside the tangent cylinder but asymmetric inside it. 

In Figure \ref{fig:jets_avz_initial}, we locate the maxima and minima of the zonal flow $\avzt{\uphi}$ and mark these extrema with thin dotted lines in Figures \ref{fig:jets_avz_initial}--\ref{fig:rhines_initial}. As is apparent from Figure \ref{fig:staircases_initial}, inside the tangent cylinder, PV is partially homogenized between each pair of extrema, which is what gives rise to the staircase structure in $\avzt{Q_z}$ (Figure \ref{fig:pv_initial}c). Because of the high level of turbulence and fully 3D character of the flow, the homogenization of PV between the jets is imperfect, but surprisingly good after the axial average. 

Outside the tangent cylinder in the region of superrotation, the PV remains very close to the planetary PV $\ftopo=1/H$. We believe that this is because the Busse columns are strongly constrained by the convex boundaries to transport angular momentum outward outside the tangent cylinder. If this outward angular momentum transport is the dominant effect, it interferes with the full homogenization of PV. Note, however, that there is still partial homogenization of PV, with both profiles of $\avzt{Q_z}$ slightly higher than $\ftopo$. This is most obvious for the CZ--WL case, as evidenced by the raised ``shoulders" in $\avzt{Q_z}$ around $\Lambda=\pm0.5$. We confirm these ideas explicitly in Section \ref{sec:superrot}.

\begin{figure}
	\centering
	\includegraphics[width=3.4375in]{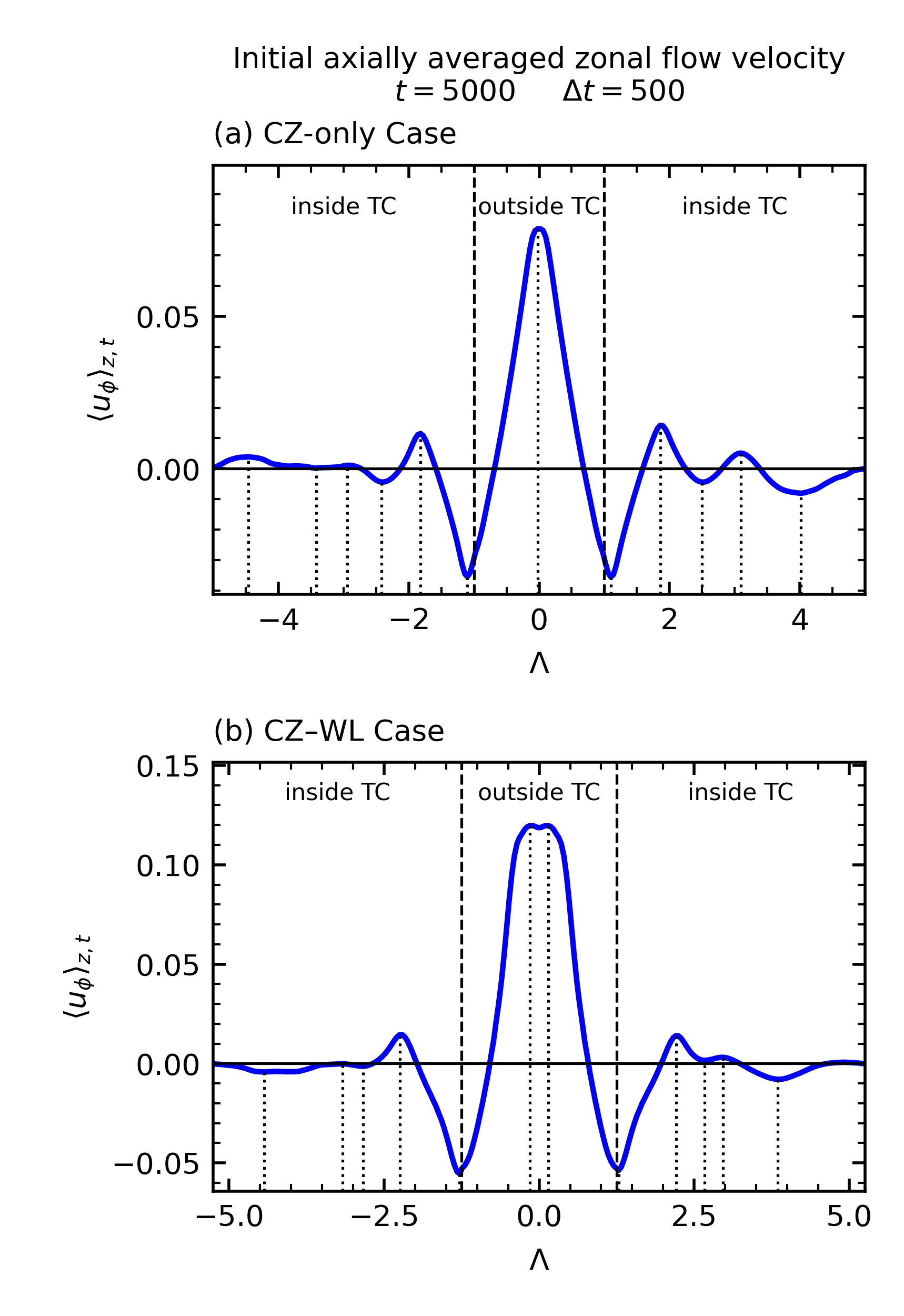}
	\caption{Temporally  and axially averaged mean zonal flow velocity $\avzt{\uphi}$ in (a) the CZ-only case and (b) the CZ--WL case, averaged over the same interval $t=5000\pm250$ as in Figures \ref{fig:uamp_initial}--\ref{fig:pv_initial}. The vertical dotted lines show the locations of the zonal flow's maxima and minima and the dashed lines show the location of the tangent cylinder at $\Lambda=\pm(\rout - \rin)$. Note that there are two minima at the tangent cylinder, obscuring the distinction between the lines there. All profiles are plotted with respect to the cylindrical coordinate $\Lambda$ (see Equation \ref{eq:lambda}). For clarity, the inside and outside of the tangent cylinder (TC) are labeled in this figure. } 
	\label{fig:jets_avz_initial}
\end{figure}

\begin{figure}
	\centering
	\includegraphics[width=3.4375in]{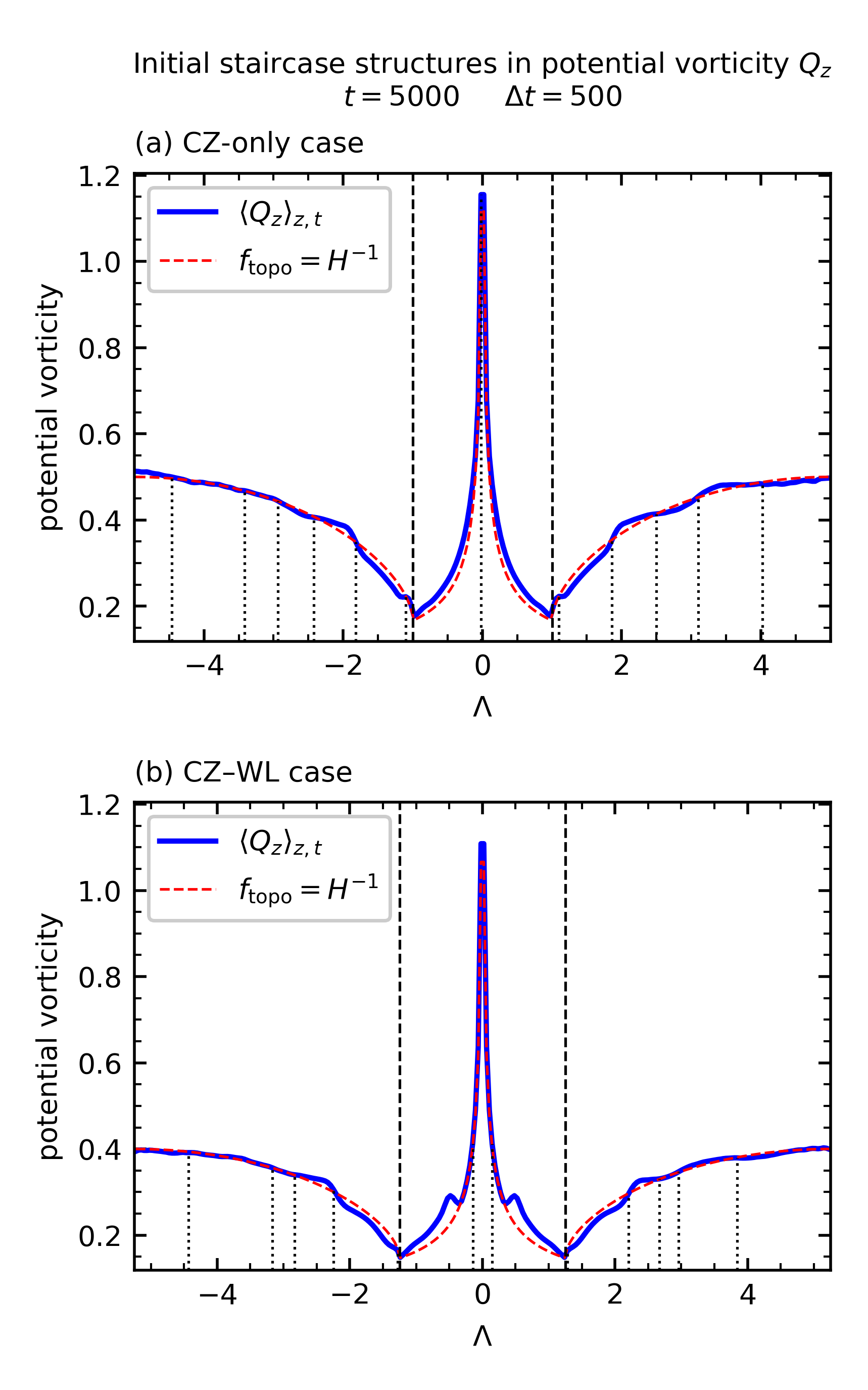}
	\caption{Temporally  and axially averaged mean PV $\avzt{Q_z}$ and the planetary PV $1/H$ in (a) the CZ-only case and (b) the CZ--WL case, averaged over $t=5000\pm250$. The vertical dotted lines show the location the maxima and minima in the zonal flow profile $\avzt{\uphi}$ and the dashed lines show the location of the tangent cylinder.}
	\label{fig:staircases_initial}
\end{figure}

To estimate the topographic Rhines scale $\lrtopo$, we note that if $Q_z$ is conserved ($D\omz/Dt\equiv0$), then
\begin{equation}\label{eq:cons_pvz}
    \frac{D\omz}{Dt}-\betatopo\ulambda\approx0,
\end{equation}
where we have assumed low Rossby number ($|\omega_z|\ll 1$) to drop one of the nonlinear terms and we explicitly compute
\begin{equation}\label{eq:betatopo}
\betatopo(\lambda) \define
\begin{cases}
 -\dfrac{\lambda}{r_{\rm c}^2-\lambda^2}   & \lambda\geq\rin \\
  \dfrac{\lambda}{\sqrt{r_{\rm c}^2-\lambda^2}\sqrt{r_{\rm in}^2-\lambda^2}}    & \lambda <\rin. 
\end{cases}
\end{equation}

Note that there is a sign change and discontinuity in $\betatopo$ at the tangent cylinder. There is also a singularity in $\betatopo$ as the tangent cylinder is approached from within, and another singularity at the equator. Physically, these singularities arise from locations where one of the bounding spheres becomes infinitely steep. 

For eddies of a typical (nondimensional) length scale $L$ and velocity amplitude $U$, we estimate $D\omz/Dt\sim U^2/L^2$ and $\beta\utheta\sim |\beta|U$. As expected, the nonlinear advection term dominates for small $L$. The Rhines scale (the crossover scale larger than which effects from the background PV gradient becomes important) is thus given by equating the magnitudes of the two terms in Equation \eqref{eq:cons_pvz}, i.e., by setting
\begin{equation}\label{eq:lrtopo}
    L=\lrtopo(\Lambda) \define \sqrt{\frac{U(\Lambda)}{|\betatopo|}},
\end{equation}
where $U(\Lambda)$ is the typical speed (at a given $\Lambda$) of the eddies under consideration. Eddies at different length scales will have different turnover speeds $U$ and hence different topographic Rhines scales. In our simulations, however, $U$ is typically dominated by the jet speed $\av{\uphi}$. Thus, to compute the Rhines scale of the jets themselves, we explicitly define
\begin{align}\label{eq:lrtopo2}
    \lrtopo(\Lambda)\define \sqrt{ \frac{\avzt{|\vecu|^2}^{1/2}}{\abs{\betatopo}}},
\end{align}
where we have replaced $U(\lambda)$ in Equation \eqref{eq:lrtopo} with the typical speed of the total flow, including both turbulent and mean components. We now assess whether the Rhines scale defined in this way matches the actual local cylindrical width of the jets $L_{{\rm jet},\lambda}(\lambda)$ in each case's CZ. Considering the presence of the second spatial derivative of the mean flow in the term $D\av{\omz}/Dt$, we define the latter cylindrical width as
\begin{align}\label{eq:jetwidth_lambda}
L_{{\rm jet},\lambda}(\Lambda)\define \sqrt{\frac{\avz{\abs{\avt{\uphi}}}}{\avz{\abs{\pderivline{^2\avt{\uphi}}{\lambda^2}}}}}.
\end{align}

\begin{figure}
	\centering
    	\includegraphics[width=3.4375in]{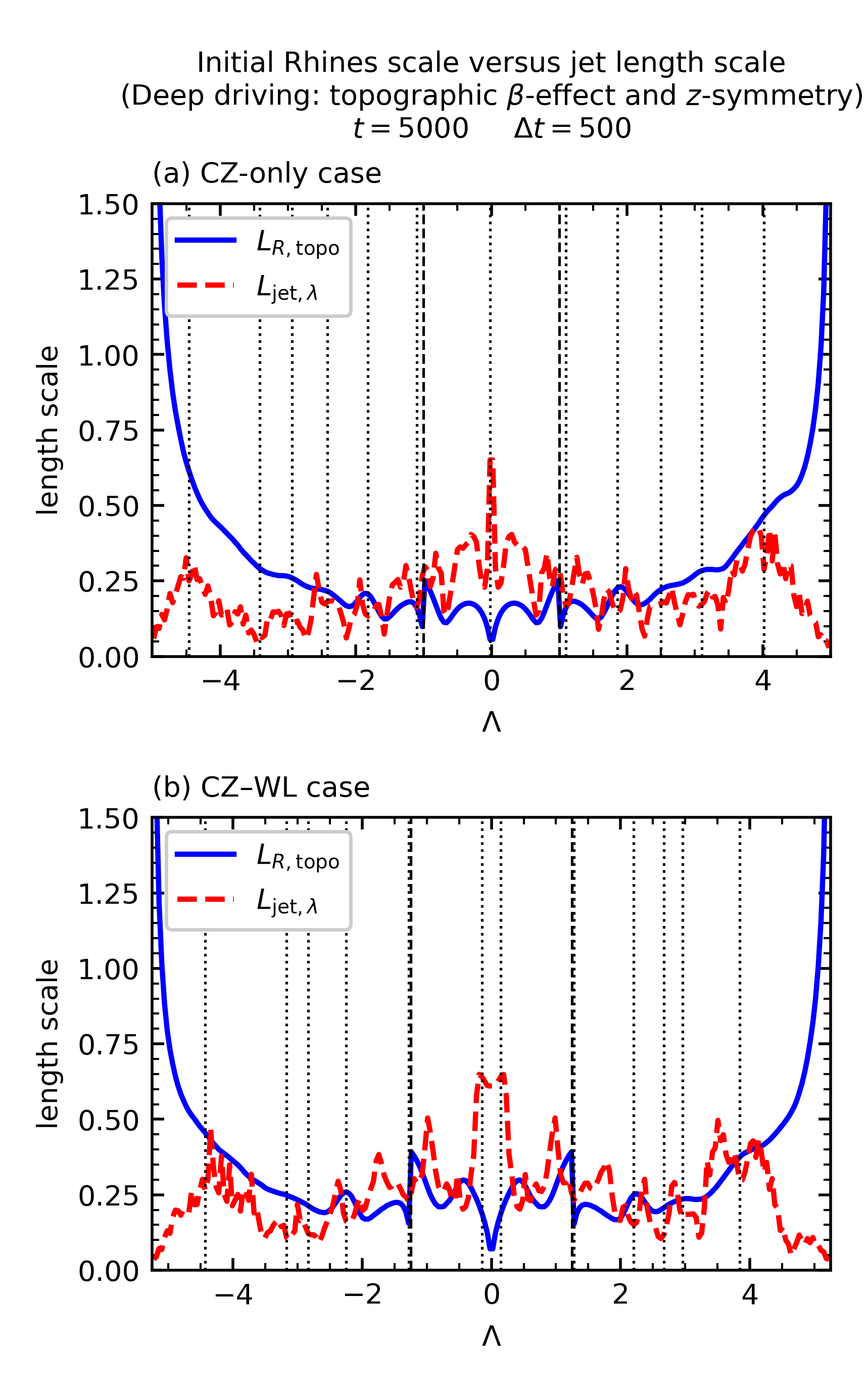}
	\caption{ Rhines scale versus jet length scale (Equations \ref{eq:lrtopo2} and \ref{eq:jetwidth_lambda}), based on the temporally  and axially averaged flows, in (a) the CZ-only case and (b) the CZ--WL case, averaged over $t=5000\pm250$ and plotted as functions of $\Lambda$. The dotted lines show the locations of the maxima and minima in the zonal flow profile $\avzt{\uphi}$ and the dashed lines show the location of the tangent cylinder.}
	\label{fig:rhines_initial}
\end{figure}

Figure \ref{fig:rhines_initial} shows the comparison between the local length scales $\lrtopo(\Lambda)$ and $L_{{\rm jet}}(\Lambda)$. They match the best for the CZ--WL case, which also has slightly more convincing staircase-like structures in PV. Especially inside the tangent cylinder at high latitudes, the local jet width oscillates about the local Rhines scale as $\Lambda$ varies in each case.

Overall, these results suggest that in the CZ region, quasi-homogenization of $Q_z$ leads to multiple jets with a width comparable to an appropriately defined Rhines scale, as arguments from 2D fluid flow would suggest. However, there are significant departures from this simple picture owing to the fully 3D nature of the flow, with the biggest discrepancies occurring in the region of superrotation outside the tangent cylinder. 

\section{Formation of Jets in the WL}\label{sec:jets_wl}
In the WL, we have already seen that the flows consist of flattened pancake structures of radial vorticity, which suggest from Section \ref{sec:driving_mechanisms} that the appropriate form of PV is $Q_r$. By analogy with Section \ref{sec:jets_cz}, the nondimensional form of Equation \eqref{eq:pvr_dim} leads to
\begin{align}
	Q_r &\define \omrad + \sin\theta = \omrad + f\label{eq:pvr}\\
    f &\define \sin\theta,\label{eq:f}\\
    \frac{D\omrad}{Dt}+\beta\utheta &\approx0 \label{eq:cons_pvr},\\
    \beta&\define \frac{df}{d\theta}=\frac{\cos\theta}{\rout},\label{eq:beta}\\
    \andd L_R(\theta)&\define \sqrt{\frac{U(\theta)}{|\beta|}},
\end{align}
where $U(\theta)$ is the typical speed (at a given $\theta$) of the eddies under consideration and $L_R(\theta)$ is the locally varying (geometric) Rhines scale. In analogy with Section \ref{sec:jets_cz}, we explicitly define the jet Rhines scale through
\begin{align}\label{eq:lr}
	L_R(\theta)\define \sqrt{\frac{\avrt{|\vecu|^2}^{1/2}}{\beta}}
\end{align}
and this should be comparable to the actual local latitudinal width of the jets in the WL,
\begin{align}\label{eq:jetwidth_theta}
L_{{\rm jet},\theta}(\theta)\define \sqrt{\frac{\avr{\abs{\avt{\uphi}}}\rout^2}{\avz{\abs{\pderivline{^2\avt{\uphi}}{\theta^2}}}}},
\end{align}
where we recall that $\avr{\cdots}$ denotes a radial volume-weighted average over the WL.

Figure \ref{fig:staircaes_r_initial} shows the initial latitudinal structure of the WL's mean zonal flow $\avrt{\uphi}$, along with the initial staircases in $\avrt{Q_r}$. It is clear from Figure \ref{fig:staircaes_r_initial}(a) that the superrotating jet has a local minimum \newtext{or ``dimple"} at the equator, \newtext{which is qualitatively similar to the dimple in the superrotating jet observed on Jupiter.} Furthermore, the staircase structure in $\avrt{Q_r}$ is slightly more convincing than that of the CZ, with flatter steps in the staircase. However, very close to the equator ($\theta|\leq10^\circ$ or so), very little homogenization of PV occurs, and $\avrt{Q_r}$ basically matches the planetary PV $f=\sin\theta$. 

This suggests that the local length scales $L_R(\theta)$ and $L_{{\rm jet},\theta}(\theta)$ should match slightly better than the CZ length scales $\lrtopo(\Lambda)$ and $L_{{\rm jet},\lambda}(\Lambda)$ shown in Figure \ref{fig:rhines_initial}, at least away from the equator. This conclusion is borne out in Figure \ref{fig:rhines_wl_initial}, which shows the comparison of $L_R(\theta)$ and $L_{{\rm jet},\theta}(\theta)$ at $t=5000\pm250$. The matching is very good at all latitudes except the very-near-equator region $|\theta|\leq10^\circ$, i.e., the region of strongest superrotation. There, the jet width is a factor of several larger than the local Rhines scale would suggest. Figure \ref{fig:rhines_wl_initial}, along with the fact that very little homogenization of PV seems to occur across the equator (Figure \ref{fig:staircaes_r_initial}b), once again suggests the superrotation in our simulations cannot be fully explained by quasi-2D arguments, even in the WL, where the motion basically \textit{is} 2D (see Figure \ref{fig:uamp_initial}b). 

\begin{figure}
	\centering
	\includegraphics[width=3.4375in]{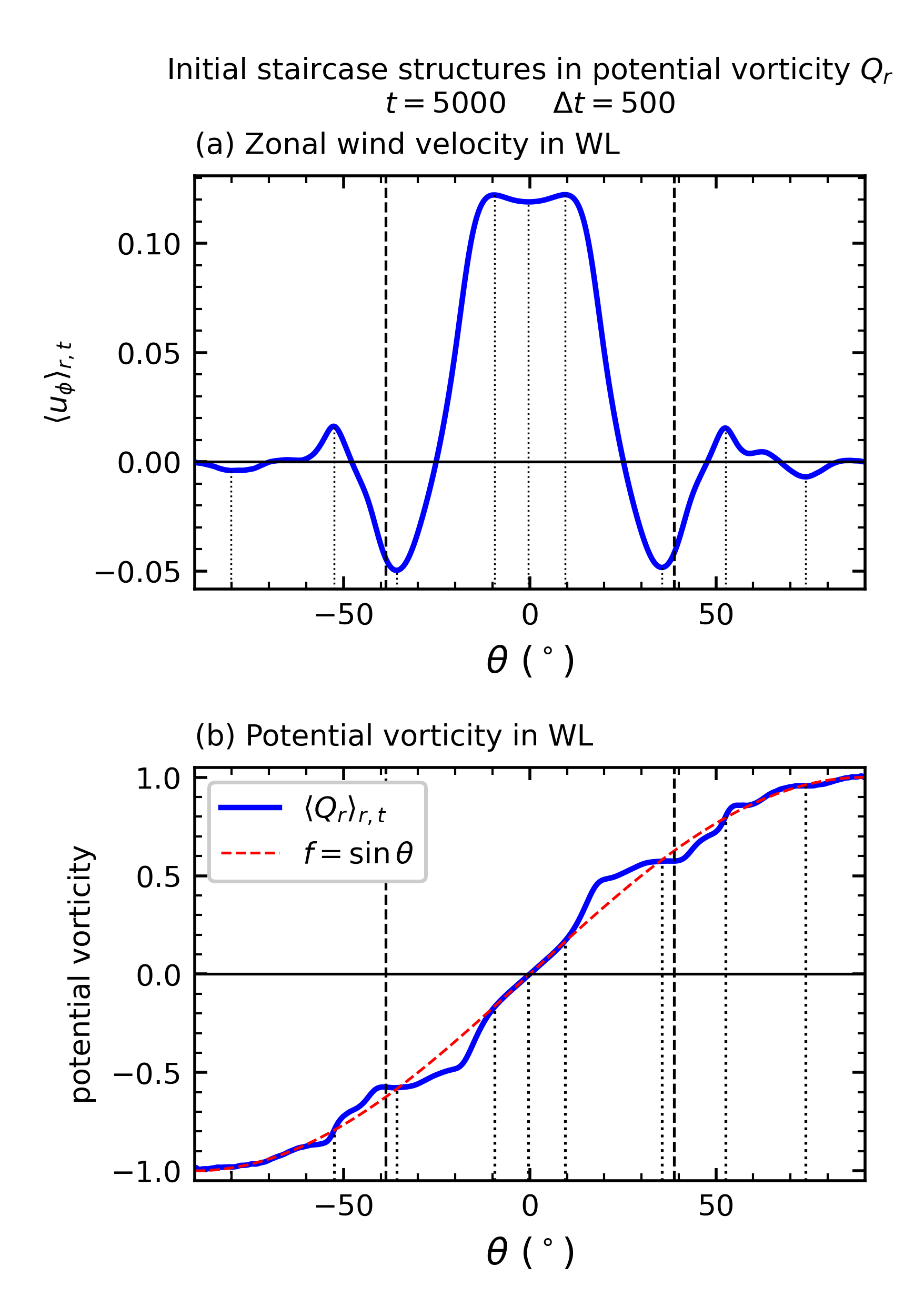}
	\caption{The CZ--WL case's (a) mean zonal flow $\avrt{\uphi}$ and (b) mean PV $\avrt{Q_r}$ (along with the planetary PV $\sin\theta$), both temporally averaged over the interval $t=5000\pm250$. Recall that the radial average is always taken over the WL only. The thin dotted lines show the location the maxima and minima in the zonal flow profile $\avrt{\uphi}$ and the dashed lines show the location of tangent cylinder (here interpreted to be $|\theta|=\arccos{(\rin/r_{\rm mwl})}=38.7^\circ$, where $r_{\rm mwl}\define(\rc+\rout)/2$). }
	\label{fig:staircaes_r_initial}
\end{figure}

\begin{figure}
	\centering
    	\includegraphics[width=3.4375in]{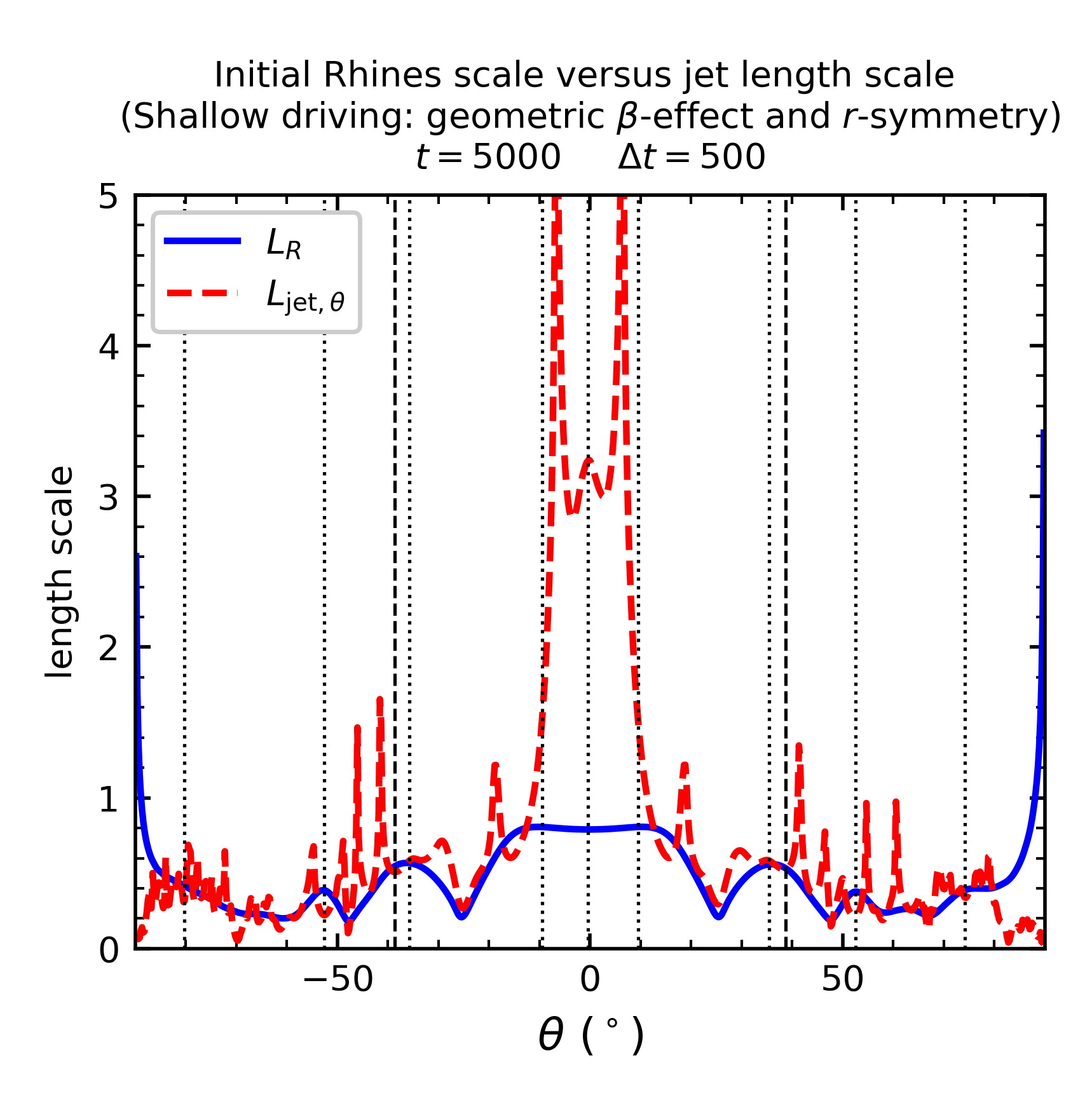}
	\caption{Rhines scale versus jet length scale (Equations \ref{eq:lr} and \ref{eq:jetwidth_theta}), based on the temporally  and radially averaged flows in the CZ--WL case's WL, averaged over $t=5000\pm250$. The thin dotted lines show the location the maxima and minima in the zonal flow profile $\avrt{\uphi}$ and the dashed lines show the location of tangent cylinder $|\theta|=38.7^\circ$. }
	\label{fig:rhines_wl_initial}
\end{figure}

We note that \citet{Heimpel2022} also performed an analysis of PV homogenization (for both $Q_r$ and $Q_z$) for several models with a CZ underlying a stable layer similar to our CZ--WL case. In that work, $Q_r$ and $Q_z$ were plotted as functions of latitude for one radius in the CZ and one in the stable layer (see their Figs. 13 and 14). The staircase structures discussed in this work (see Figures \ref{fig:staircases_initial} and \ref{fig:staircaes_r_initial}) are qualitatively similar to those of \citet{Heimpel2022}, however, the influence of the simplified WL seems to be much stronger in our case due to its effects on thermal wind balance. This tilts the isocontours of $\avt{\uphi}$ significantly more radially at low to mid latitudes in our WL.

\section{Superrotation}\label{sec:superrot}
The presence of negative $\betatopo$ outside the tangent cylinder leads to an elegant argument for the generation of superrotation. Namely, for a jet of constant PV, $\av{Q_z} = (\av{\omz}+1)/H = C > 0$ (where $C$ is some positive constant) and hence $\partial\av{\omz}/\partial\lambda = C dH/d\lambda$ or $\pderivline{^2\av{\uphi}}{\lambda^2}\approx CH\betatopo <0$. Thus, the single large jet outside the tangent cylinder is expected to have a local maximum in $\av{\uphi}$, i.e., be superrotating, if it has constant PV. However, the results of Sections \ref{sec:jets_cz} and \ref{sec:jets_wl} show that PV (either $Q_z$ or $Q_r$) is rather far from constant in the region of superrotation. It thus appears that quasi-2D theory (namely the topographic $\beta$-effect) is insufficient to explain the form of superrotation achieved in our simulations.

Instead, superrotation may be studied by examining the transport of angular momentum, which is conserved in our models \newtext{to about 1 part in $10^4$ (it is not conserved strictly because the equations are solved on a discrete grid)}. Busse columns tend to transport angular momentum cylindrically outward and thus to low latitudes outside the tangent cylinder, causing equatorial superrotation (e.g., \citealt{Brun2002,Hotta2015,Matilsky2026c}). This is because the columns, in the presence of convex spherical-shell boundaries, typically have cross sections that are flared outward, leading to a $\lambda$-varying tilt in the prograde direction---a result long expected from quasi-nonlinear stability analysis (e.g., \citealt{Busse2002}). Figure \ref{fig:eqslice_initial} shows an example of this flaring pattern for our two simulations. The pattern is obvious, despite the strong nonlinearity of our simulations. Because of the prograde cross-sectional tilts, the fluctuating cylindrical velocity $\ulambda^\prime$ is correlated with the fluctuating zonal velocity $\uphiprime$ such that $\av{\ulambdaprime\uphiprime}>0$. Because of the flare, this means there is a cylindrically outward transport of angular momentum that is strongest close to the boundaries, resulting in a positive Reynolds-stress torque near the equator (thus driving superrotation) and a negative one near the tangent cylinder. 

Note that both mechanisms for generating superrotation (the homogenization of PV $Q_z$ and the flaring of Busse columns discussed further below) arise from the tilt of the bounding spherical surfaces. However, in the mechanism of PV homogenization, only the sign of the tilt ($dH/d\lambda<0$) matters for superrotation, whereas for the flaring of Busse columns, the curvature of the boundaries ($d^2H/d\lambda^2$) is the reason for superrotation. The two mechanisms are thus related but may be distinct. 

To quantify how Busse columns transport angular momentum in our simulations, as well as determine how the jets imprint into the WL, we analyze the full torque equation explicitly. The specific angular momentum density of the fluid is 
\begin{equation}\label{eq:amom}
	\amom\define \lambda\left(\frac{\lambda}{2} + \av{\uphi}\right)=\lambda^2\Omega,
\end{equation}
where 
\begin{equation}\label{eq:omega}
\Omega\define \frac{1}{2}+\frac{\av{\uphi}}{\lambda}
\end{equation}
is the mean rotation rate of the fluid.

Multiplying the $\phi$-component of the zonal mean of Equation \eqref{eq:mom} by $\lambda$ yields the evolution equation for $\amom$ (or equivalently, for $\av{\uphi}$ or $\Omega$):
\begin{subequations}\label{eq:torque}
	\begin{align}
		\rhotilde\pderiv{\amom}{t} &=\rhotilde\lambda\pderiv{\av{\uphi}}{t} = \rhotilde\lambda^2\pderiv{\Omega}{t}\nonumber\\
		&= -\Div[\rhotilde (\lambda \av{u^\prime_\phi \umerprime} 
		+\amom\av{\umer} -\ek\nutilde\lambda^2\nabla\Omega)]\\
		&= \taurs + \taumc + \tauv = \tau_{\rm tot},
	\end{align}
\end{subequations}
where we have defined the appropriate time-dependent torque densities:
\begin{subequations}\label{eq:torques}
	\begin{align}
		\taurs \define &\ -\Div (\rhotilde (\lambda \av{u^\prime_\phi \umerprime})\nonumber\\
		&\ \text{(Reynolds-stress torque)}, \label{eq:taurs}\\
		\taumc \define  &\ -\rhoumer\cdot\nabla\amom \nonumber\\
		&\ \text{(meridional-circulation torque)},\label{eq:taumc}\\
		\tauv \define  &\ \ek \Div(\rhotilde\nutilde\lambda^2\nabla\Omega)\nonumber\\ 
		&\ \text{(viscous torque)},\label{eq:tauv}\\
		\andd \tau_{\rm tot} \define  &\ \taurs+\taumc+\tauv\nonumber\\ 
		&\ \text{(total torque).}\label{eq:tautot}
	\end{align}
\end{subequations}

\begin{figure*}
	\centering
    	\includegraphics[width=7.25in]{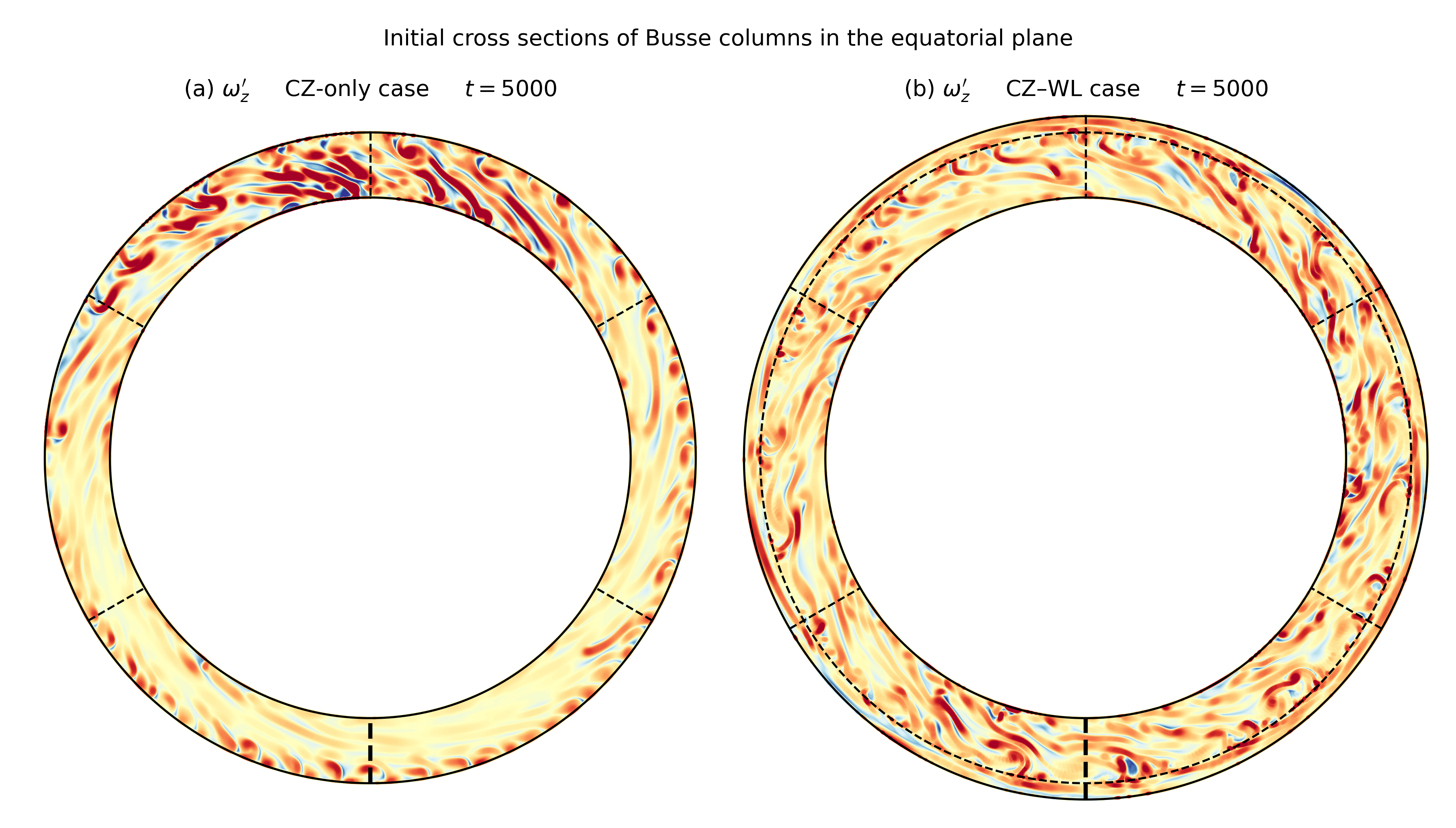}
	\caption{Cross sections of the Busse columns (columns of $\omzprime$) in the equatorial plane for (a) the CZ-only case and (b) the CZ--WL case at $t=5000$ (the view is from the north pole). Red tones indicate cyclonic vorticity and blue tones indicate anticyclonic vorticity. The outward spirals of the columns in the CZ is obvious, while in the WL of the CZ--WL case, the columns have much larger horizontal extent and have no clear spiraling tendency. Just as in Figure \ref{fig:cutout3d}, the fields are normalized separately at each radial level by the rms of $\omzprime$. }
	\label{fig:eqslice_initial}
\end{figure*}

Figure \ref{fig:torque_initial} shows $\avt{\rhotilde\lambda\uphi}$ in the meridional plane, along with the three torques and their sum, for both cases over $t=5000\pm250$. In the CZs of both cases, outside the tangent cylinder, the Reynolds-stress torque $\taurs$ is clearly responsible for driving the superrotation, being negative in a columnar pattern close to the edge of the tangent cylinder and positive in the equatorial regions further away. The other two torques (from viscosity and meridional circulation) tend to oppose the production of superrotation. At higher latitudes inside the tangent cylinder, there is a complicated three-way balance between the torques that leads to a banded pattern for the total torque $\tau_{\rm tot}$, which is offset slightly cylindrically inward from the actual jet structure exhibited by $\avt{\rhotilde\lambda\uphi}$. It is this offset, whose origins are not obvious, which is responsible for driving the high-latitude jets poleward with time, causing them to eventually disappear, as discussed in the following section. \newtext{It is also possible that the banded pattern of total torque is related in some way to the staircase-like structures in PV discussed in Section \ref{sec:jets_cz}, in which case the offset in the total torque could be explained using quasi-2D theory. }

\newtext{We note that \citet{Duer2023} also acknowledged the importance of eddy momentum flux convergence (essentially our $\taurs$ term) in driving multiple high-latitude jets and superrotation. They also found a strong relationship between the meridional circulation torque (there interpreted in the context of Ferrell-like cells on giant planets) and the spatial distribution of the jets. We suggest that Figure \ref{fig:torque_initial} reaffirms the conclusions of \citet{Duer2023}, namely that the Reynolds stress and meridional circulation torques are both essential for establishing the jets and superrotation, while viscosity plays a much lesser \newnewtext{role}. However, examining the relative positions of the positive and negative bands of $\taurs$, $\taumc$, and $\tauv$, it is not obvious that a simple interpretation in terms of Ferrell cells (e.g., \citealt{Fletcher2020}) is possible in our case, there being no clear balance between $\taumc$ and $\tauv$ (as in the terrestrial context) or between $\taumc$ and $\taurs$ (as in the gas-giant context explored by \citealt{Duer2023}).   }

In the WL of the CZ--WL case at low latitudes, the Reynolds-stress torque is extremely weak due to the small radial velocity. Instead, superrotation is driven here by both the viscous torque $\tauv$ (at very low latitudes) and the meridional-circulation torque $\taumc$ (at mid-latitudes). This type of dual imprinting by the viscosity and circulation (especially for $\sigma$ near unity) is expected from prior analyses of coupled convecting layer/stable layer systems in the solar context, for which the CZ's meridional circulation and viscous stresses tend to ``burrow" into the stable layer, despite being inhibited by the background stable stratification, carrying angular momentum with them \citep{Clark1973,Spiegel1992,Garaud2025,Matilsky2025,Matilsky2026,Matilsky2026d}. It shows that the WL's superrotation is not driven primarily by nonlinear effects (either homogenization of PV or Reynolds stresses from Busse columns) but instead is the result of linear imprinting of the CZ's superrotation upward. At higher latitudes, by contrast, the effects of the Reynolds stresses are stronger and multiple jets form mostly according to the quasi-2D ideas outlined in Section \ref{sec:jets_wl}, namely the homogenization of $Q_r$ on the scale $L_R$ and subsequent formation of staircase structures.
\begin{figure*}
	\centering
    	\includegraphics[width=7.25in]{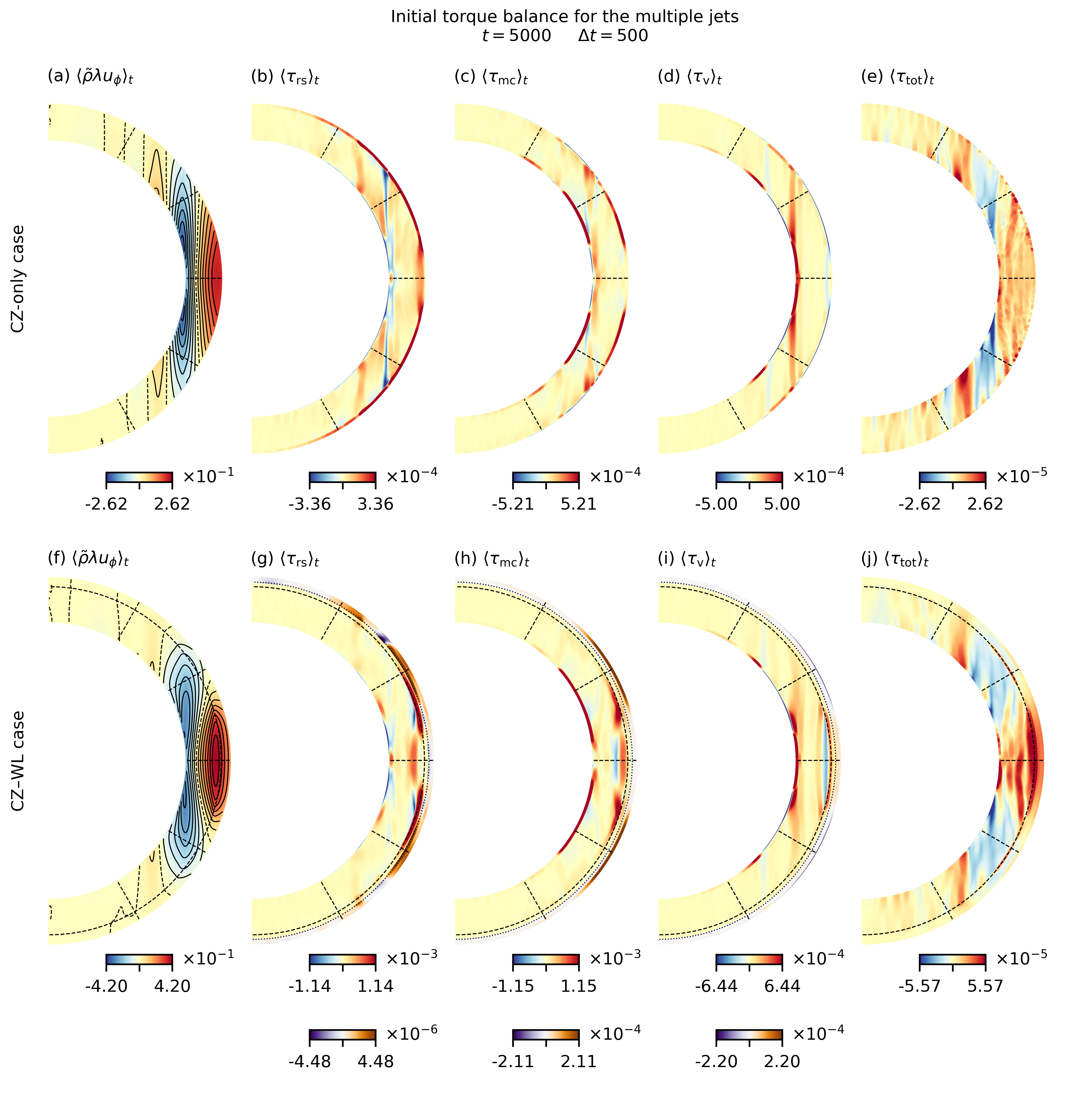}
	\caption{All terms in the torque equation (see Equations \ref{eq:torque} and \ref{eq:torques}) in the meridional plane, zonally and temporally averaged (over $t=5000\pm250$) for (a--e) the CZ-only case and (f--j) the CZ--WL case. Red tones denote positive values for the torque (i.e., tending to produce prograde zonal flow in the rotating frame) and blue tones denote negative values (tending to produce retrograde zonal flow). In the plots of $\rhotilde\lambda\avt{\uphi}$ (left-most column), the contours are equally spaced and the zero contour is dashed. In the bottom row, the dashed semicircles denote $r=\rc$. In the plots for $\avt{\taurs}$, $\avt{\taumc}$, and $\avt{\tauv}$ (middle three columns) a separate color table (purple/orange) is used for the upper WL ($r_{\rm mwl} < r < \rout$) and the dotted semicircles mark $r=r_{\rm mwl}$, where $r_{\rm mwl}\define(\rc+\rout)/2$. }
	\label{fig:torque_initial}
\end{figure*}

\section{Jet Merger and Migration}\label{sec:migration}
Up to this point, we have only analyzed the simulations for the relatively early time interval $t=5000\pm250$ (see Figure \ref{fig:etrace}). However, examining the full simulations, we find that the high-latitude jets migrate poleward and/or merge on a very long timescale (but the superrotation outside the tangent cylinder remains basically steady). The end state for both simulations is a banded ``fast-slow-fast" structure, consisting of three jets per hemisphere: the fast superrotating jet outside the tangent cylinder, and one pair of alternating slow and fast jets per hemisphere inside the tangent cylinder. This same fast-slow-fast structure of the zonal flow is often encountered for low-Rossby-number simulations in the solar context as well, where it is referred to as ``Jupiter-like" (e.g., \citealt{Brun2017}).

\begin{figure*}
	\centering
	\includegraphics[width=7.25in]{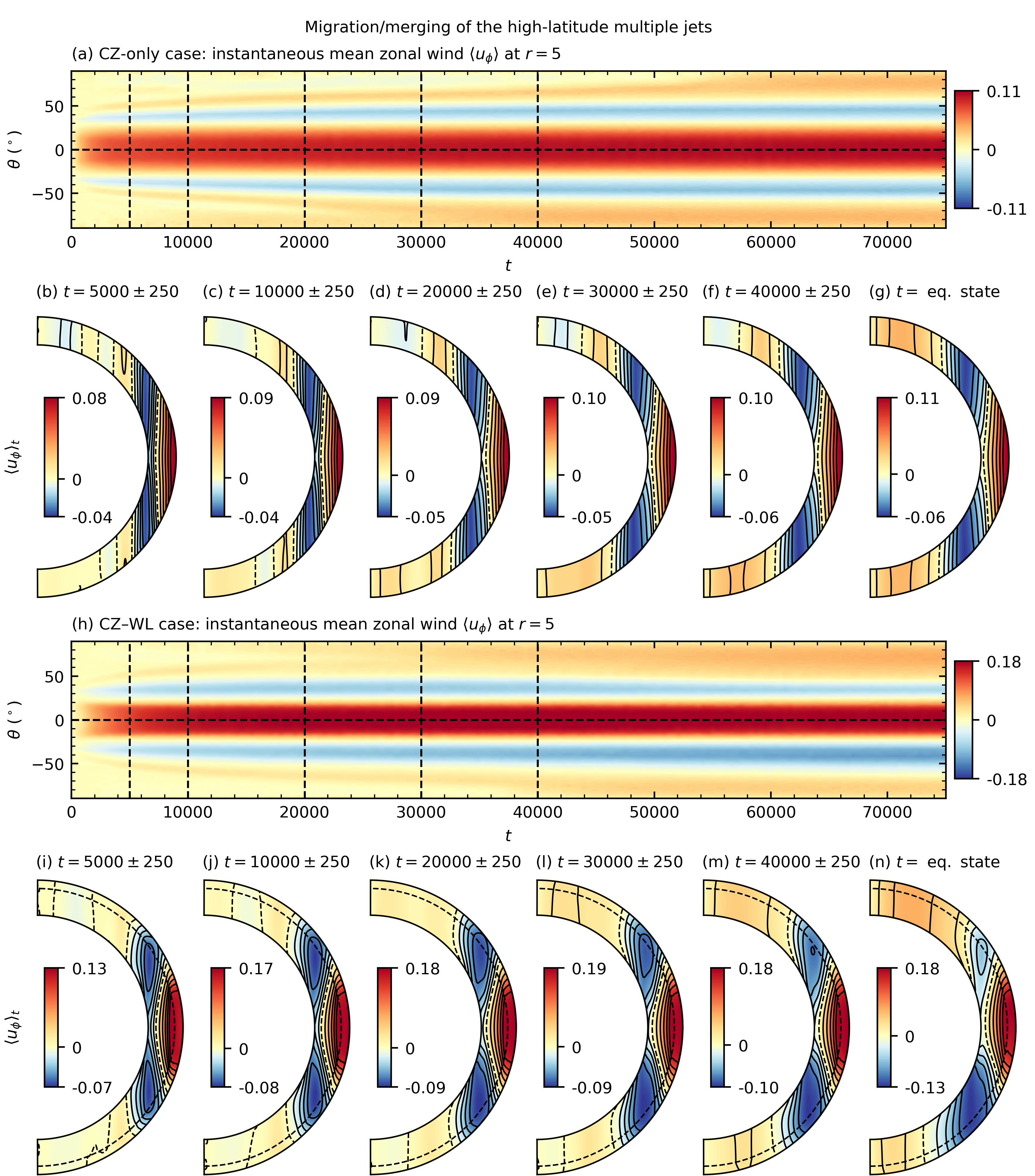}
	\caption{(a) Mean zonal flow velocity $\av{\uphi}$ at $r=\rc$ for the CZ-only case, plotted as a function of time and latitude. (b--f) Sequence of mean zonal flow profiles $\avt{\uphi}$, averaged over the indicated time-intervals. In (b--e), the center of the indicated time intervals correspond to the vertical dashed lines in (a). In (g), the interval corresponds to the equilibrated state, which is roughly $t=(100000,384000)$ for the CZ-only case. The contours are equally spaced separately for the positive and negative regions of the (binormalized) color table and the zero contour is dashed. (h--n) The same as (a--g), but for the CZ--WL case. In (i--n) the interface $r=\rc$ is shown as the dashed semicircle. In (n), the equilibrated state for the CZ--WL case is roughly $t=(150000,235000)$. Note that in our chosen units, the diffusion timescales are $\ek^{-1}=22400$ (viscous) and $\pr/\ek=11200$ (thermal). }
	\label{fig:migration}
\end{figure*}

Figure \ref{fig:migration} shows the slow evolution of the jet structure in a time--latitude plot for the instantaneous mean zonal flow $\av{\uphi}$ at $r=\rc$ and as a sequence of profiles of the temporally averaged zonal flow $\avt{\uphi}$ in the meridional plane. Interestingly, the superrotating jet outside the tangent cylinder does not significantly evolve and equilibrates relatively rapidly, after $O(1)$ diffusion time. The initially multiple high-latitude jets, however, slowly widen, migrate poleward, and/or merge, in a hemispherically asymmetric fashion, over $O(10)$ diffusion timescales \newtext{(also about 10,000 planetary rotation periods or 20,000 convective turnover times)}, ultimately leaving only two high-latitude jets per hemisphere in the equilibrated state. \newtext{Given that Saturn has far fewer high-latitude jets than Jupiter (e.g., \citealt{Kaspi2020}), we might speculate our simulations are better representative of Saturn based on the migration result. However, given the extreme gap in parameter space between our simulations and any real giant planet (see Section \ref{sec:nondim}), we must treat such speculations with caution.  } 

Overall, these results on jet migration in our simulations are similar to the work of \citet{Takehiro2024}, who found that the mid- and high-latitude jets in long-time integrations of a rapidly rotating Boussinesq fluid eventually merged/migrated and disappeared, leaving a still-evolving final state similar to the fast-slow-fast banded structure we find as well. Those models were significantly more rotationally constrained than our cases and had a thinner shell (in that work, $\ek=\sn{1.5}{-6}$, $\roc=0.11$, and $\beta=0.9$ compared to our $\ek=\sn{5}{-5}$, $\roc=0.2$, and $\beta=0.8$) and also utilized various hyperdiffusive parameterizations (whereas we have exclusively considered explicit Laplacian diffusivities). Here, we have thus affirmed the core result suggested by \citet{Takehiro2024}, that the final asymptotic steady state of global, 3D Jovian jet simulations may contain very few high-latitude jets. We do so for the actual asymptotic state and operate in the slightly different context of anelastic, more slowly rotating simulations, one of which includes a WL. We also suggest that the superrotating jet outside the tangent cylinder may be \textit{part} of the asymptotic state and does not undergo significant evolution.

The true cause of jet migration/merging in the simulations is not well understood. Comparing Figures \ref{fig:torque_initial}(a,e) (or equivalently, Figures \ref{fig:torque_initial}(f,j)) we see that the total torque inside the tangent cylinder at early times has the same sign structure as the angular momentum associated with the multiple jets, but is shifted slightly inward from the jets. This may at least suggest jet migration (as opposed to jet widening/merging), although similar analyses of the torque balance at later times (not shown) complicates this picture. In any event, it is far from obvious how (or why) the complicated three-way balance of torques at high latitudes serendipitously gives rise to a banded sign structure clearly correlated with, but displaced from, the jets. In a simpler model, \citet{Morin2006} found that steady multiple jets arose only in the presence of an Ekman drag, which halted the inverse cascade of energy to large scales. The lack of any such drag in our simulations thus may be affecting the stability of the jets.

Because of our current lack of clarity regarding jet migration, we postpone a detailed analysis to future work. We only note that the same quasi-2D arguments for high-latitude jet production, involving staircase-like structures of PV and associated Rhines scales, which were explored for $t=5000\pm250$, appear to remain valid at subsequent times and in the equilibrated state. Figure \ref{fig:rhines_migration} shows the evolution of the global jet width and Rhines scale for high latitudes as the jets migrate and/or merge, as well as successive latitudinal profiles of the local jet widths and Rhines scales at different times (each like a transposed Figure \ref{fig:rhines_initial}). 

Because the local Rhines scale, in both the CZ and WL, substantially underestimates the actual jet width in the region of superrotation, we exclude this region in the definition of the global length scales. Namely, we define the high-latitude globally averaged (time-dependent) topographic Rhines scale by
\begin{align}\label{eq:L_R_t}
    \lrtopo(t) = \avvol{\sqrt{\frac{\av{|\vecu|^2}^{1/2}}{|\betatopo|}}},
\end{align}
where the volume-average here is over the high-latitude CZ ($|\theta|>45^\circ$, $\rin<r<\rc$). Similarly, we define the globally averaged high-latitude CZ jet width by
\begin{align}\label{eq:L_jet_t}
    L_{{\rm jet},\lambda}(t) = \avvol{\sqrt{\frac{|\av{\uphi}|}{|\pderivline{^2\av{\uphi}}{\lambda^2}}}}.
\end{align}

\begin{figure*}
	\centering
	\includegraphics[width=7.25in]{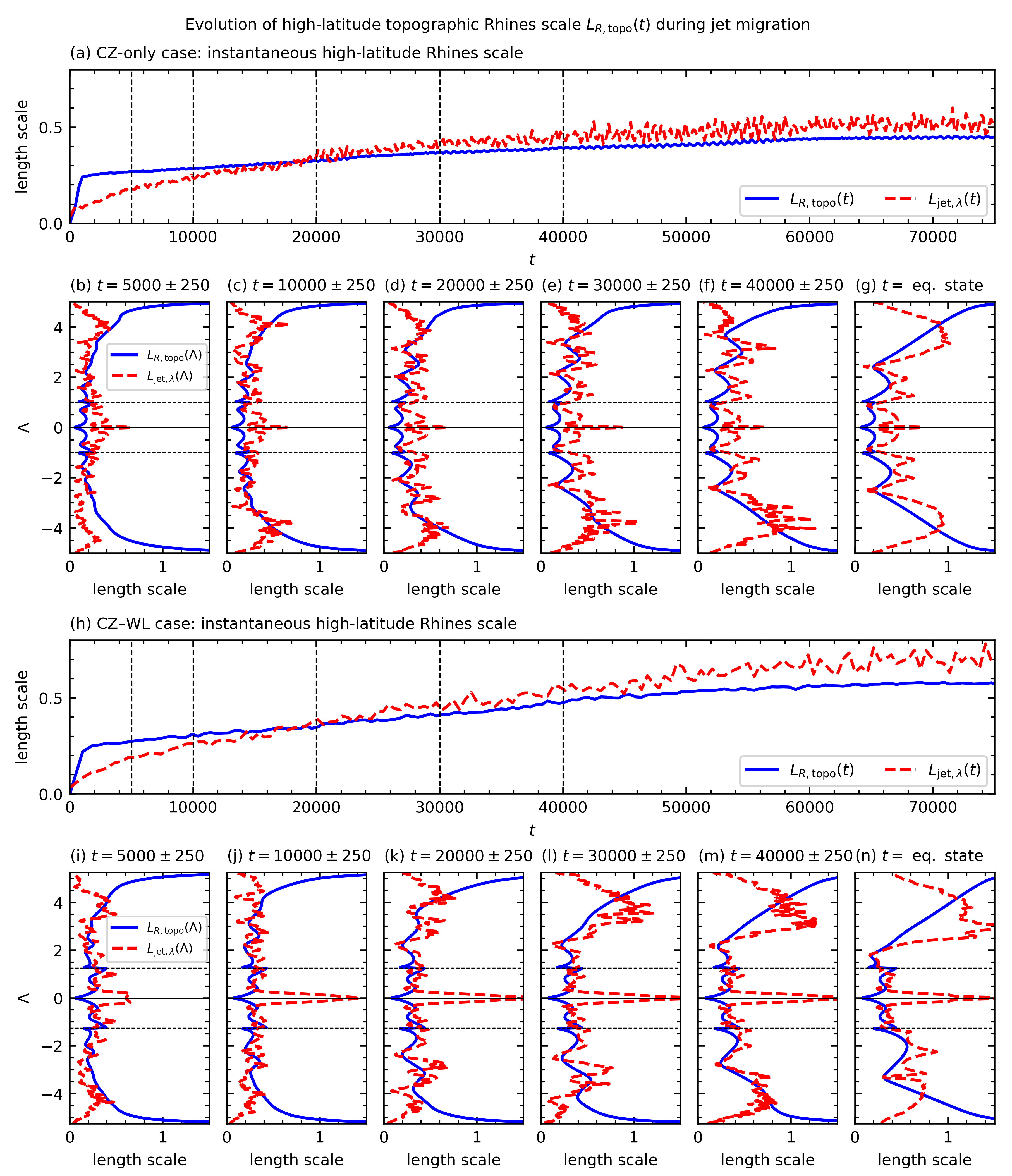}
	\caption{(a) Comparison of the time-dependent high-latitude length scales $L_{R,\lambda}(t)$ and $L_{{\rm jet},\lambda}(t)$ for the CZ-only case (see Equations \ref{eq:L_R_t} and \ref{eq:L_jet_t}). (b--f) Sequence of comparisons between the local CZ length scales $L_{R,\lambda}(\Lambda)$ and $L_{{\rm jet},\lambda}(\Lambda)$, averaged over the indicated time-intervals (see Equations \ref{eq:lrtopo2} and \ref{eq:jetwidth_lambda}). The horizontal solid line denotes $\Lambda=0$ and the dashed lines denote the edge of the tangent cylinder at $\Lambda=\pm(\rout - \rin)$. In (b--e), the center of the indicated time intervals correspond to the vertical dashed lines in (a). In (g), the interval corresponds to the equilibrated state (h--n). The same as (a--g), but for the CZ--WL case. }
	\label{fig:rhines_migration}
\end{figure*}

From Figure \ref{fig:rhines_migration}, it is clear that the high-latitude length scales match fairly well at all times and both exhibit slow monotonic growth with time. (Note that if the low latitudes---the superrotation---are included in the definitions \ref{eq:L_R_t} and \ref{eq:L_jet_t}, the jet length scale $L_{{\rm jet},\lambda}(t)$ is systematically about twice as large as the Rhines scale $L_{R,\lambda}(t)$ at all times.) Furthermore, at any given time, the local length scales match reasonably well, except in the region of superrotation outside the tangent cylinder, and grow locally in concert with one another. In fact, the comparison between $\lrtopo(\Lambda)$ and $L_{{\rm jet},\lambda}(\Lambda)$ generally improves as time progresses. 

Overall, we interpret Figure \ref{fig:rhines_migration} as follows. The initial crossover scale $\lrtopo(t)$ for the high-latitude jets is achieved early on (say around $t=1000$) in both simulations. This funnels energy from the turbulent convection into the large-scale zonal flow in the manner of \citet{Vallis1993}. The jets then slowly build up enough energy (after about $t=20000$, or $\sim2$ thermal diffusion times) for their widths to catch up to the Rhines scale. Both length scales then continue to grow with time as the jets receive more and more energy from the turbulent cascade. The process only stops once the final equilibrium (set by a complicated three-way balance of torques, as well as thermal wind balance and equilibration of the heat equation) is achieved. Apparently for our simulations, and those of \citealt{Takehiro2024} at least, there is no process in the equilibrated state to dissipate the energy of the jets at sufficiently low energy levels for the eddy speeds (and hence the Rhines scales) to remain small. The jet width thus cannot remain small enough to sustain multiple high-latitude jets in the domain.

The absence of a dissipation mechanism for the inverse cascade at large scales greater than the crossover scale is well known. For example, \citet{Vallis1993} found that the $\beta$-effect only redirects the inverse energy cascade into the zonal flows, but it does not actually stop the cascade at the crossover scale. A scale-independent linear drag to rigid rotation, say of the form $-r_d\uphi$ on the right-hand side of the specific zonal momentum equation (where $r_d$ has the dimensions of inverse time) has been implemented in 2D models in order to sustain steady high-latitude jets (e.g., \citealt{Williams1978,Vallis1993,Williams2002,Lian2008,Zeng2026}). The absence of any such drag in 3D models like ours and those of \citet{Takehiro2024} thus may be responsible for the ultimate disappearance of multiple high-latitude jets. 

\section{Discussion}\label{sec:disc}
We have analyzed two global, 3D anelastic simulations of the fluid equations in rapidly rotating spherical shells in order to assess the influence of a \newtext{simplified} weather layer (in this case, simply a stably stratified layer) atop a convective zone, as well as to verify whether quasi-2D arguments involving PV homogenization/Reynolds stresses from Rossby waves could explain the formation of multiple jets and superrotation. To summarize our results, we address \newtext{some aspects of the deep versus shallow driving debate} as follows: 
\begin{enumerate}
    \item The final steady state of zonal flow in our simulations consists of a ``fast-slow-fast" pattern with very few high-latitude jets, similar to \citet{Takehiro2024}. This is due to a monotonically increasing Rhines scale and subsequent migration/merging of the jets.  
    \item There appear to be two distinct forms of relevant PV: $Q_z$ for the CZ and $Q_r$ for the WL, arising physically from the distinct topologies of the two types of eddies pictured in Figure \ref{fig:schematic}. Inside the tangent cylinder (high latitudes), the appropriate PV is largely homogenized to form staircase-like structures and associated multiple jets. Correspondingly, the local jet width matches appropriate Rhines scale ($\lrtopo$ or $L_R$) quite well at high latitudes. 

    \item Outside the tangent cylinder, the superrotating jet appears to be driven primarily by \newtext{a 3D effect, namely,} the Reynolds stresses associated with flared Busse columns in the CZ (see Figures \ref{fig:eqslice_initial} and \ref{fig:torque_initial}), where the flare is a result of the convex spherical boundaries.
    
    \item A \newtext{simplified} WL (a stable layer atop the CZ) has a strong effect on the dynamics, significantly altering the transport of heat and subsequent thermal wind balance, thus forcing the jets away from cylindrical alignment. This effect is most pronounced in the WL itself, but is also relevant in the bulk of the CZ. 
\end{enumerate}

\newtext{Because we exclude realistic driving of the WL (we have no differential solar heating, no moist convection, and no radiative cooling), one may rightfully question how much these results should be interpreted at face value for future studies of giant planets' zonal flows, especially where the WL is concerned. We emphasize that quantitative comparisons should indeed be made cautiously at this stage. However, even for our simplified WL, there are several reasons we might expect at least qualitative relevance of our core dynamics to those of the real systems. First, although there is no imposed differential radiative heating (which is thought to be one of the leading drivers of 2D turbulence in the WL), there is an upward-diffusing baroclinicity from the thermal-wind-balance jets in the CZ, which may have a similar effect, as in \citet{Schneider2009}. Second, the other main driver of WL turbulence is the overshooting convection, which is reasonably well captured by our models. Finally, the result that the mere presence of a stable layer significantly tilts the isocontours of zonal flow is not new (e.g., \citealt{Wulff2022,Matilsky2022,Matilsky2024}) and suggests that this tilting phenomenon may be robust. In future work, we plan to introduce more of the missing physics, for example by adding differential heating at the top of the WL, a representation of drag effects, and the inclusion of two molecular species to better capture cloud dynamics.}

These conclusions have broad consequences for how the giant planets' zonal flows might be studied in the future. Overall, our results suggest that the dynamics of high-latitude jets can usefully be analyzed using quasi-2D theories separately for the CZ (staircase structures in $Q_z$) and WL (staircase structures in $Q_r$). In that sense, there may be no real resolution to the ``deep versus shallow driving" debate, because the two layers (each of which is driven by roughly equal amounts of energy flux) can \newnewtext{each support jet-forming dynamics}. However, the coupling between the CZ and WL is important, especially for the cylindrical alignment (or lack thereof) of the zonal flows. This is true even without the enhanced baroclinicity provided by equatorially concentrated heating by the Sun (which was excluded in our work). Making progress on the fine structure of the high-latitude jets will thus likely require modeling both layers simultaneously.

At low latitudes, in the region of superrotation outside the tangent cylinder, there seems to be more physics than can be included in a simple quasi-2D framework. In the CZ, there is significant outward transport of angular momentum by the Reynolds stresses from flared Busse columns. PV ($Q_z$ in this case) is only weakly homogenized across the superrotating jet, and thus the flare of the Busse columns seems to be essential for the generation of superrotation. What remains unclear is whether this same outward flare would be reproduced in quasi-2D simulations on the disc incorporating the topographic $\beta$-effect. The first such 2D investigations \citep{Zeng2026} did consistently yield superrotation, but the possible connection between negative $\betatopo$ and flared Busse columns was not explored. 

\newtext{Finally, we believe that our results on jet migration have laid bare a potential problem in current state-of-the-art 3D simulations of Jupiter and Saturn's jets, namely that the high-latitude jets in those models were very likely still evolving,  possibly toward an asymptotic state with very few high-latitude jets. To be clear, this ``migration problem" is only an issue for the \textit{simulations}; there is no observational indication of secular jet migration on either of the giant planets. Thus, until the detailed physical origins of jet migration in simulations are understood, these models would appear to face a significant hurdle in describing the interior dynamics of Jupiter and Saturn self-consistently. For now, we must be content to admit that some essential physics at work in the interiors of Jupiter and Saturn, which is responsible for keeping the observed zonal jets in a quasi-steady state, is absent from our global simulations. }   

\begin{acknowledgments}
We thank P. Wulff, A. van Kan, N. Lewis, P. Marcus, E. Knobloch, H. Quan, W. Kang, G. Flierl, S. Tobias, Y. Kaspi, and J. Pedlosky for helpful discussions. This work was primarily supported by L. Matilsky's NSF Astronomy \& Astrophysics Postdoctoral Fellowship award AST-2202253 and the COFFIES DRIVE Science Center (NASA grant 80NSSC22M0162), with additional support from NASA grants 80NSSC18K1127 and 80NSSC24K0125. M. Browning was supported by the UK Science and Technology Facilities Council under grant agreement ST/Y0021561/1. Part of this work was performed during the Woods Hole Oceanographic Institution's Geophysical Fluid Dynamics Programs of 2024 and 2025, which were funded by the National Science Foundation grant OCE-1829864. Computational resources were provided by the NASA High-End Computing (HEC) Program through the NASA Advanced Supercomputing (NAS) Division at Ames Research Center. {\rayleigh} is supported by the Computational Infrastructure for Geodynamics (CIG) through NSF awards NSF-0949446 and NSF-1550901. For reproducibility, the {\rayleigh} input files and final checkpoints for each simulation are publicly accessible at \url{https://doi.org/10.5281/zenodo.20332552}
\end{acknowledgments}

\appendix
\twocolumngrid
\restartappendixnumbering

\section{Background State}\label{ap:ref}
Here we describe how the background state of the CZ--WL case is constructed in the radius range $r=(\rin,\rout)$. (The CZ-only case is the same, but considers only the radius range $r=(\rin,\rc)$.) The implementation is similar to the solar simulations of \citet{Matilsky2021,Matilsky2022,Matilsky2024,Matilsky2025,Matilsky2026}, except now the stable layer (the WL) is above the CZ instead of below it. 

In terms of the dimensional background state, the perfect gas law is
\begin{align}\label{eq:idgasdim}
	\prstilde\dimm = \left[\frac{(\gamma-1)\cp}{\gamma}\right]\rhotilde\dimm\tmptilde\dimm,
\end{align}
hydrostatic balance is
\begin{align}\label{eq:hydrdim}
	\frac{d\prstilde\dimm}{dr\dimm}=-\rhotilde\dimm\gravtilde\dimm,
\end{align}
and the first law of thermodynamics is
\begin{align}\label{eq:firstlawdim}
	\frac{1}{\cp}\left(\frac{d\entrtilde}{dr}\right)\dimm=\frac{d\ln\tmptilde\dimm}{dr\dimm}-\left(\frac{\gamma-1}{\gamma}\right)\frac{d\ln\rhotilde\dimm}{dr\dimm}.
\end{align}
After the nondimensionalization described in Section \ref{sec:equations}, Equations \eqref{eq:idgasdim}--\eqref{eq:firstlawdim} take the form
\begin{align}\label{eq:idgas}
	\prstilde&= \rhotilde\tmptilde,
\end{align}
\begin{align}\label{eq:hydr}
	\frac{d\prstilde}{dr}=-\di\left(\frac{\gamma}{\gamma-1}\right)\rhotilde\,\gravtilde,
\end{align}
and 
\begin{align}\label{eq:firstlaw}
	\dsdrtilde=\frac{1}{\gamma}\frac{d\ln\tmptilde}{dr}-\left(\frac{\gamma-1}{\gamma}\right)\frac{d\ln\rhotilde}{dr},
\end{align}
where $\entrtilde\define\entrtilde\dimm/\cp$ and $\prstilde\define\prstilde^*/p\cz$. Recall that $\di\define\grav\cz H/\cp T\cz$ (Equation \ref{def:di}). We combine Equations \eqref{eq:idgas}--\eqref{eq:firstlaw} to find
\begin{align}
	\frac{d\tmptilde}{dr}-\left(\dsdrtilde\right)\tmptilde &= -\di\,\gravtilde,
\end{align}
which has the exact solution 
\begin{align}\label{eq:tmptilde}
	\tmptilde &= e^{\entrtilde}\left[\tmptilde(\rin) - \di\int_{\rin}^r \gravtilde(x)  e^{-\entrtilde(x)}dx\right].
\end{align}
Note that the actual background entropy $\entrtilde$ is only relevant up to an integration constant. In Equation \eqref{eq:tmptilde}, we have therefore chosen, without loss of generality, 
\begin{equation}\label{eq:entrtilde}
\entrtilde(r)\define \int_{\rin}^r\frac{d\entrtilde}{dr^\prime}dr^\prime
\end{equation}
such that $\entrtilde\equiv0$ in the CZ.

We then eliminate $\prstilde$ from Equations \eqref{eq:idgas}  and \eqref{eq:hydr} to yield
\begin{align}\label{eq:rhotilde}
	\rhotilde &= \rhotilde(\rin) \exp{\left[-\left(\frac{\gamma}{\gamma-1}\right)\entrtilde\right]}\tmptilde^{1/(\gamma-1)}. 
\end{align}

For $\gravtilde\ofr\propto1/r^2$ and the normalization condition $(4\pi/V\cz)\int_{\rc}^{\rout}\gravtilde(r)r^2dr=1$ (where $V\cz\define (4\pi/3)(\rout^3-\rc^3)=(4\pi/3)[(1-\alphacz^3)/(1-\alphacz)^3]$ is the nondimensional volume of the CZ), we require
\begin{align}
	\gravtilde(r) = \left[ \frac{1-\alphacz^3}{3(1-\alphacz)^3}    \right]\frac{1}{r^2}.
\end{align}
To model the transition from convective instability to stability at $r=\rc$, we choose $\dsdrtildeline$ to be $\equiv0$ in the CZ, $\equiv1$ in the WL, and continuously matched in between over the distance $\delta=0.05$. We accomplish this via quartic matching: 
\begin{equation}\label{eq:dsdrquart}
	\dsdrtilde = \begin{cases}
		0 & r\leq \rc \\
		1 - \left[1 - \left(\dfrac{r-\rc}{\delta}\right)^2\right]^2 & \rc  < r < \rc + \delta\\
		1 & r\geq \rc + \delta.
	\end{cases}
\end{equation} 
Note that the amplitude of $\dsdrtildeline$ in the WL is technically arbitrary and choosing $\dsdrtildeline\equiv1$ is really another way to specify $\Nrhowl$, given Equations \eqref{eq:tmptilde}--\eqref{eq:rhotilde}.

There are three equations relating $\rhotilde(\rin)$, $\tmptilde(\rin)$, $\di$, $\gamma$, $\alphacz$, and $\Nrhocz$: two from our  choice of nondimensionalization---$(4\pi/V\cz)\int_{\rc}^{\rout}\rhotilde(r)r^2dr=1$ and $(4\pi/V\cz)\int_{\rc}^{\rout}\tmptilde(r)r^2dr=1$---and one from the definition $\Nrhocz\define\ln{[\rhotilde(\rin)/\rhotilde(\rc)]}$. Thus, $\rhotilde(\rin)$, $\tmptilde(\rin)$, and $\di$ may all be regarded as functions of $\gamma$, $\alphacz$, and $\Nrhocz$. 

Because $\dsdrtildeline\equiv0$ in the CZ, explicit formulae can be derived for $\di$ and $\tmptilde(\rc)$:
\begin{align}
	& \di \define \frac{\gravcz H}{\cp\tmpcz} = \nonumber\\
    &\frac{3 \alphacz (1 - \alphacz)^2 (1 - e^{-\Nrhocz/n})} 
	{ \dfrac{3\alphacz}{2} (1 - \alpha\cz^2) (1 - e^{-\Nrhocz/n}) - (1-\alpha\cz^3)(\alphacz-e^{-\Nrhocz/n})}\label{eq:di}    
\end{align}
and
\begin{align}
	&\tmptilde(\rc)=\nonumber\\
   & \frac{(1-\alpha\cz^3)(1-\alphacz)}{ \dfrac{3\alphacz}{2} (1 - \alpha\cz^2) (1 - e^{-\Nrhocz/n}) - (1-\alpha\cz^3)(\alphacz-e^{-\Nrhocz/n})},\label{eq:tmp0}
\end{align}
where 
\begin{equation}\label{eq:polyn}
n\define \frac{1}{\gamma-1}=2
\end{equation}
is the ``polytropic index" (but note that only the convective zone has the polytropic stratification $\rhotilde=\tmptilde^2$; e.g., \citealt{Featherstone2016a,Matilsky2018,Matilsky2019,Matilsky2020a,Matilsky2020c,Bice2020}). The equation for $\rhotilde(\rc)$ is transcendental and requires a numerical solution. For the cases considered here, note that $\di=0.599$, $\tmptilde(\rin)=1.35$, and $\rhotilde(\rin)=1.77$. 

The background $\rhotilde(r)$ and $\tmptilde(r)$ are now completely determined via Equations \eqref{eq:tmptilde}--\eqref{eq:polyn} and the control parameters $\gamma$, $\alphacz$, $\Nrhocz$, and (for the CZ--WL case) $\alphawl$ and $\Nrhowl$. 


The background heating $\heattilde$ is chosen to input a heating layer at the base of the CZ and a cooling layer at the top of the CZ, each of width $\dheat=0.1$. This is accomplished again via quartics. 
\begin{equation}\label{eq:heating}
	\heattilde =
    \begin{cases}
    A\,B_1\left[1 - \left(\dfrac{r-\rin}{\dheat}\right)^2\right]^2     
    \five &r\leq\rin+\dheat \\
    -A\,B_2\left[1 - \left(\dfrac{r-\rc}{\dheat}\right)^2\right]^2  & \rc-\dheat \leq r \leq \rc\\
    0 &\text{otherwise}
    \end{cases}
\end{equation}
For the CZ--WL case, we also multiply the expression in Equation \eqref{eq:heating} by $[1+\tanh((r-\rc)/0.005)]/2$ to ensure that the transition to $\heattilde\equiv0$ in the WL is not completely abrupt. The coefficients $B_1$ and $B_2$ are chosen such that the energy injected by the heating layer is exactly removed by the cooling layer, i.e., 
\begin{align}\label{eq:heatingnormalization}
4\pi\int_{\rin}^{\rc}\heattilde(\rprime)\rprime^2d\rprime=0.
\end{align}
\\

Since $\heattilde\dimm=(F\cz/H)\heattilde$, $\fluxscalartilde\dimm=F\cz\fluxscalartilde$, and $\fluxscalartilde\dimm\define(H/r^2)\int_r^{\rout}\heattilde\dimm\ofrprime\rprime^2d\rprime$ (recall Equation \eqref{eq:fnrad}),  normalization of the heat flux (i.e., $(4\pi/V\cz)\int_{\rc}^{\rout}\fluxscalartilde(r) r^2dr=1$) requires that
\begin{equation}\label{eq:heatnorm}
  \frac{4\pi}{V\cz}\int_{\rin}^{\rc}\int_r^{\rc}\heattilde\ofrprime \rprime^2d\rprime dr=1,
\end{equation}
which sets the proportionality constant $A$ in Equation \eqref{eq:heating}.

\section{Comparison of compressional and topographic $\beta$-effects}\label{ap:beta}
\newnewtext{Throughout this work, have assumed that the relevant PV in the high-latitude CZ of each case takes the form $Q_z=(\omega_z+1)/H$, which is true as long as topographic effects are larger than compressional effects. In the WL, compressional effects are necessarily very weak because of the small radial velocity and the anelastic constraint \eqref{eq:cont}, which implies that $\Div\vecu = u_r/H_\rho = \beta_{\rm comp}u_r$. To justify the neglect of $\beta_{\rm comp}$ in the CZ, here we estimate the two effects directly for our models. To do this, we compare the appropriate terms in the $z$-vorticity equation directly, following the approach of \citet{Gastine2012}. The nondimensional $z$-vorticity equation comes from the $z$-component of curl of Equation \eqref{eq:mom}, neglecting baroclinicity and viscosity and assuming small Rossby number ($|\vecom|\ll1$): }
\begin{align}\label{eq:zvort}
    \frac{D\omz}{Dt} \approx \underbrace{-\nabla\cdot\vecu}_{\text{compression}} + \underbrace{\pderiv{u_z}{z}}_{\text{topography}}. 
\end{align}
\newnewtext{The first term $-\nabla\cdot\vecu= -\beta_{\rm comp}u_r$ is the compressional $\beta$-effect and the second term is the topographic $\beta$-effect (as we demonstrate explicitly below). As long as $|\Div\vecu|\ll|\pderivline{u_z}{z}|$, the compressional $\beta$-effect can be neglected. Figure \ref{fig:beta} shows the ratio of compressional to topographic $\beta$-effects as a function of latitude for the equilibrated state of each case in the CZ. Inside the tangent cylinder at high latitudes, this ratio is $\sim$0.25--0.35. Outside the tangent cylinder, however, the ratio is larger (up to $\sim$0.55 for the CZ-only case). This once again suggests that analyzing homogenization of $Q_z$ and calculating topographic Rhines scales is appropriate at high latitudes (where we have now shown that we can largely neglect the compressional $\beta$-effect), but not at low latitudes in the region of superrotation.}

\newnewtext{To explicitly see that $\pderivline{\uz}{z}$ corresponds to the topographic $\beta$-effect term $\betatopo\ulambda$, we average in $z$ and assume axial symmetry ($z$-independence) in $\ulambda,\uphi,$ and $\omz$, finding:}
\begin{align}
    \avz{\pderiv{\uz}{z}}&=\frac{1}{\ztop - \zbot}(\uz|_\ztop - \uz|_\zbot)\nonumber\\
    &=\frac{\ulambda}{\ztop - \zbot}\frac{d}{d\lambda}(\ztop - \zbot)\nonumber\\
    &= \frac{\ulambda}{H}\frac{dH}{d\lambda} = \betatopo\ulambda,
\end{align}
\newnewtext{where we have applied the impenetrable boundary condition to calculate $\uz$ at the spherical-shell boundaries, defined $\ztop$ and $\zbot$ as the $z$-locations of the top and bottom, respectively, of the columns in Figure \ref{fig:schematic}, and noted that $\ztop-\zbot=H$ (outside the tangent cylinder) or $H/2$ (inside the tangent cylinder). Thus, neglecting the compressional $\beta$-effect term in Equation \eqref{eq:zvort}, averaging in $z$, and assuming axial symmetry in $\ulambda$ and $\uphi$ recovers Equation \eqref{eq:cons_pvz} directly and justifies assuming that $Q_z$ is the appropriate form of PV in the CZ. }

\begin{figure}
	\centering
    	\includegraphics[width=3.4375in]{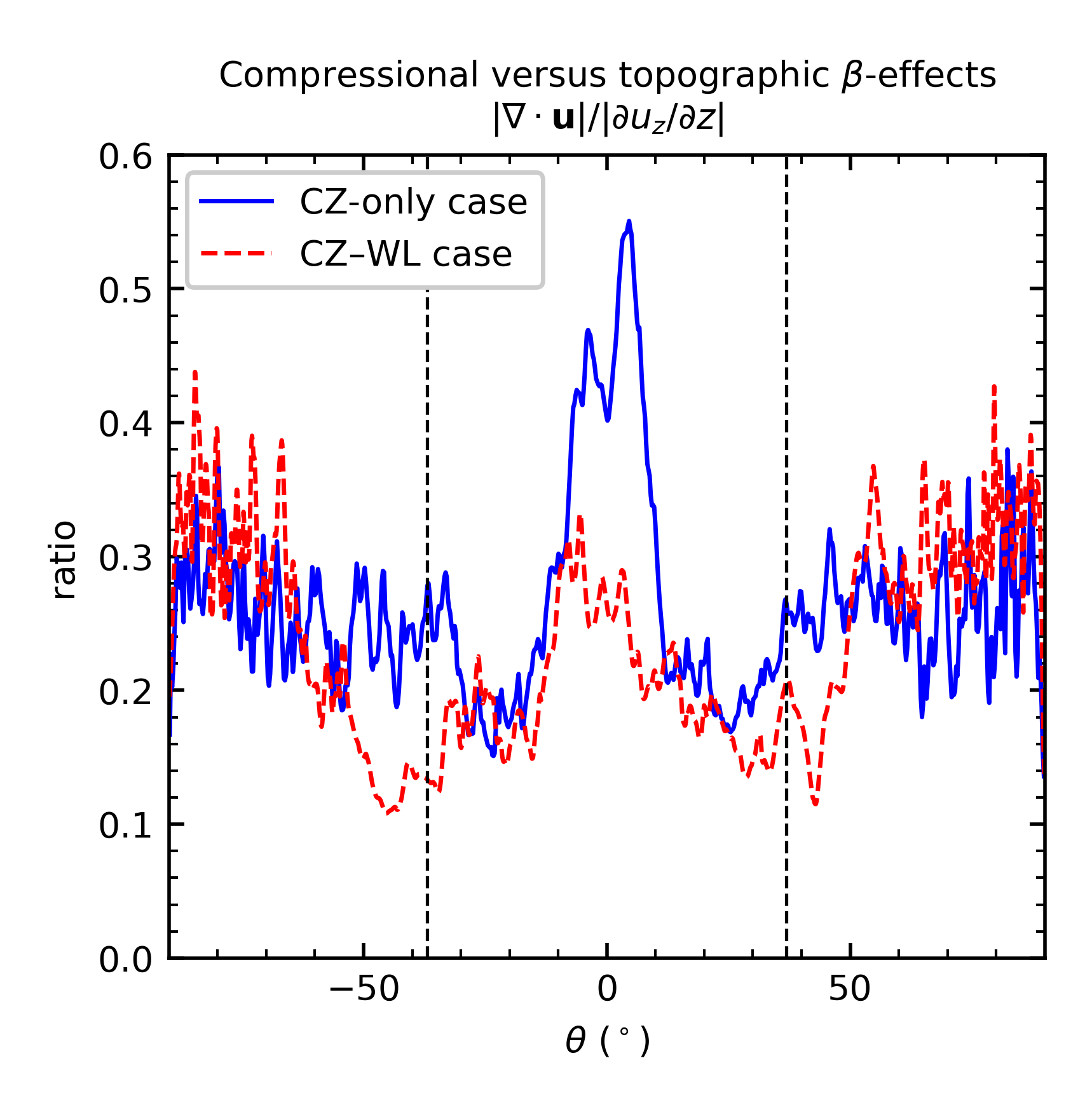}
	\caption{\newnewtext{Ratio of compressional $\beta$-effect to topographic $\beta$-effect for each case as a function of latitude. Here, each term in Equation \eqref{eq:zvort} has been evaluated at an instant of time toward the end of each simulation's equilibrated state. More precisely, we plot $\avr{|\Div\vecu|}/\avr{|\pderivline{u_z}{z}|}$, where the average in this case is over longitude and radius over the CZ (from $\rin$ to $\rc$).  } }
	\label{fig:beta}
\end{figure}


\section{Redimensionalization of the simulations}\label{ap:redim}
\begin{table*}
    \caption{Redimensionalization parameters for our simulations, compared with real values from the gas giants. For the real values, we take $a$ and $\Omega_0$ from \citet{Guillot2005}, $\rhocz$ from \citet{French2012} and \citet{Preising2023} (representative of, respectively, Jupiter and Saturn's $\rhotilde\dimm$ at mid-CZ, or $R=0.975a$ and $R=0.925a$), $d$ from \citet{Kaspi2020}, $d\wl$ from Equation \ref{eq:dwl}, $N\wl$ from Equation \eqref{eq:Nwl}, $F\cz$ from \citet{Guillot2005}, and the diffusivities $\nucz$ and $\kappacz$ (at mid-CZ) from ab-initio structure simulations of Jupiter \citep{French2012} and Saturn \citep{Preising2023}. For the simulations, the values are computed according to the redimensionalization scheme outlined in Equations \eqref{eq:redim}. The time scales are computed according to Equations \eqref{eq:timescalesdim} and the units (for the simulations only) according to Section \ref{sec:nondim}. Note that the dimensions of the torque densities (see, Figure \ref{fig:torque_initial}) are the same as the unit of kinetic energy density $[\rhotilde\vecu^2/2]$ (see Figure \ref{fig:etrace}). }
\label{tab:redim}
\centering
\begin{tabular}{*{5}{l}}
\hline
Parameter & Jupiter value (real) & Jupiter Value (sim) & Saturn value (real) & Saturn Value (sim)\\
\hline
$a$ (km) & $70$,$000$ & $70$,$000$ & $58$,$000$ & $58$,$000$ \\
$\Omega_0$ ($\rm{rad\ s^{-1}}$)& $\sn{1.8}{-4}$ & $\sn{1.8}{-4}$ & $\sn{1.6}{-4}$ & $\sn{1.6}{-4}$ \\
$\rhocz$ ($\gram\ \cm^{-3}$) & 0.08 & 0.08 & 0.14 & 0.14  \\
$d$ (km) & $3$,$000$ & $13$,$000$  & $9$,$000$ & $11$,$000$ \\
$d\wl$ (km) & 110 & $3$,$300$  & 190 & $2$,$800$ \\
$N\wl$ ($\rm{rad\ s^{-1}}$) & $\sn{5.5}{-3}$ & $\sn{1.5}{-3}$ & $\sn{2.6}{-3}$ & $\sn{1.3}{-3}$ \\
$F\cz$ (W m$^{-2}$) & $5.4$ & $\sn{5.3}{7}$ & $2.0$ & $\sn{3.7}{7}$  \\
$\nucz$ ($\stoke$) & $\sn{3.5}{-3}$ & $\sn{2.9}{10}$ & $\sn{3.0}{-3}$ & $\sn{1.7}{10}$  \\
$\kappacz$ ($\stoke$) & $\sn{8.0}{-3}$ & $\sn{5.7}{10}$ & $\sn{2.0}{-2}$ & $\sn{3.5}{10}$ \\
\hline
$\tomegadim$ (hr) & 0.77 & 0.77 & 0.87 & 0.87 \\
$\tffdim$ (hr) & $\sn{1.8}{3}$ & 3.9 & $\sn{9.4}{3}$ & 4.3 \\
$\tbruntdim$ (hr) & 0.05 & 0.19 & 0.11 & 0.21 \\
$\tnudim$ (yr) & $\sn{2.0}{13}$ & 2.0 & $\sn{2.1}{14}$ & 2.2  \\
$\tkappadim$ (yr)  & $\sn{8.6}{12}$ & 1.0 & $\sn{3.1}{13}$ &  1.1 \\
\hline
$[\vecu]$ (${\rm m\ s^{-1}}$) & - & $\sn{4.8}{3}$ & - & $\sn{3.5}{3}$\\
$[\rhotilde\vecu^2/2]$ ($\erg\ \cm^{-3}$) & & $\sn{1.8}{10}$ & - & $\sn{1.7}{10}$ \\
$[Q_z]$ ($\second^{-1}\ \cm^{-1}$) & - & $\sn{2.7}{-13}$ & - & $\sn{2.9}{-13}$  \\
$[\rhotilde\vecu]$ ($\gram^{-1}\ \cm^{-2}\ \second^{-1}$) & - & $\sn{3.8}{2}$ & - & $\sn{4.9}{2}$  \\
$[\pderivline{\av{\uphi}^2}{z}]$ ($\cm\ \second^{-2}$) & - & $\sn{1.7}{-2}$ & - & $\sn{1.1}{-2}$  \\
$[\rhotilde\lambda\av{\uphi}]$ ($\gram\ \cm^{-1}\ \second^{-1}$) & - & $\sn{5.1}{11}$ & - &  $\sn{5.5}{11}$ \\
\hline
\end{tabular}
\end{table*}

\newtext{Although we report all results in nondimensional units in this work, we can always consider dimensional analogs of the simulations to more easily compare with observations and prior work, a process called redimensionalization. Because there are many more dimensional parameters than nondimensional ones, each nondimensional simulation corresponds to an infinite number of dimensional analogs (for more details, see \citealt{Matilsky2026}, their Appendix E). To choose two particular suggestive analogs, we arbitrarily set the parameters $\Omega_0$, $a$, and $\rhocz$ to either Jovian-like or Saturn-like values. We give the real dimensional parameters of Jupiter and Saturn, alongside their simulation counterparts, in Table \ref{tab:redim}. }

\newtext{For the real parameters, we take some quantities directly from the sources referenced in Table \ref{tab:redim}'s caption. Values for the thickness of the WL ($d\wl$) and $N\wl$ are difficult to come by, however. To compute $d\wl$ (simply to verify that the real WLs are very thin), we define each planets WL as the region $0.1\ {\rm bar} < \prstilde\dimm < 10\ {\rm bar}$ and assume an isothermal, hydrostatically balanced atmosphere, which yields}
\begin{align}\label{eq:dwl}
d\wl \define 4.6\frac{\mathcal{R}\overline{T}}{g},
\end{align}
\newtext{where $\ln(100)\approx4.6$, $\mathcal{R}=\sn{3.5}{7}\ \unitentr$ is appropriate for a molecular gas of hydrogen and helium at solar abundances, $g=GM/a^2$ ($G=\sn{6.67}{-8}\ \cm^3\ \gram^{-1}\ \second^{-2}$ is the gravitational constant and the planetary mass $M$ is taken from \citealt{Guillot2005}), and $\overline{T}$ is the temperature (at $\prstilde\dimm=1\ \rm bar$) taken from \citet{Seiff1996} (Jupiter) and \citet{Fletcher2007} (Saturn).  }

\newtext{For $N\wl$, we define}
\begin{align}\label{eq:Nwl}
    N\wl \define \sqrt{\frac{g}{\overline{T}}\left(\pderiv{T}{r} + \frac{g}{\cp}\right)}, 
\end{align}
\newtext{where $\cp=\gamma\mathcal{R}/(\gamma-1)=\sn{1.2}{8}\ \unitentr$ ($\gamma=7/5$ for diatomic hydrogen) and the temperature gradients $\pderivline{T}{r}$ (at $\prstilde\dimm=1\ \rm bar$) are obtained from \citet{Magalhaes2002} (Jupiter) and \citet{Fletcher2007} (Saturn). }

\newtext{To redimensionalize the simulations, we set}
\begin{align}
    d \define \frac{a}{5.25} \five\andd d\wl\define 0.25d,
\end{align}
\newtext{defining $d\wl$ as the dimensional depth of the WL (for the simulations) and recalling that $a$ is defined as the radius of the top of the WL.}

\newtext{Following the procedure of \citet{Matilsky2026}, we then must choose}
\begin{subequations}\label{eq:redim}
\begin{align}
    N\wl &\define \bu^{1/2}(\twoOmzero),\\
    F\cz &\define \frac{\raf\ek^3}{\pr^2\di}\rhocz(2\Omega_0d)^3,\\
    \nucz &\define \ek (\twoOmzero d^2),\\
    \andd \kappacz &\define \frac{\ek}{\pr} (\twoOmzero d^2).   
\end{align}
\end{subequations}

\newtext{Both real and simulated (dimensional-analog) dimensional parameter values are given in the first block of Table \ref{tab:redim} for Jupiter and Saturn.}

\newtext{We also compute various dimensional time scales (second block of Table \ref{tab:redim}) from the dimensional parameters:}

\begin{subequations}\label{eq:timescalesdim}
\begin{align}
    \tomegadim &\define (\twoOmzero)^{-1}\five\text{(rotational time scale)},\\
    \tbruntdim &\define N\wl^{-1}\five\text{(gravity-wave time scale)},\\
\tffdim&\define {\rm Ro_c^{-1}}(\twoOmzero)^{-1} \five\text{(convective free-fall time scale)},\\
    \tnudim &\define  \frac{d^2}{\nucz}\five\text{(viscous diffusion time scale)},\\
    \tkappadim &\define  \frac{d^2}{\kappacz}\five\text{(thermal diffusion time scale)}.
\end{align}
\end{subequations}

\newtext{Note that to estimate $\tffdim$ for Jupiter and Saturn, we first define}
\begin{align}
    u_{\rm ff}\dimm \define \left(\frac{F\cz}{\rhocz}\right)^{1/3}
\end{align}
\newtext{as the convective free-fall velocity according to mixing-length theory and then set $\tffdim\define d/u_{\rm ff}\dimm$.}

\newtext{Finally, the dimensional units for various quantities according to our adopted choices in Section \ref{sec:nondim} are given in the third block of Table \ref{tab:redim} (for the simulations only). To use this part of the table, simply multiply the reported nondimensional values in the simulation by the corresponding nondimensional CGS unit for the appropriate planet. For instance, if we are interested in how the simulations compare dimensionally to Jupiter, we use $[\vecu]=\sn{4.8}{3}\  {\rm m\ s^{-1}}$ to find that the superrotation velocities shown in Figure \ref{fig:jets_initial} correspond to dimensional amplitudes of about $0.08[\vecu]= 380\ {\rm m\ s^{-1}}$ (for the CZ-only case) and $0.13[\vecu]=620\ {\rm m\ s^{-1}}$ (for the CZ--WL case). }

\newtext{We note that the redimensionalization process forces us to choose very unrepresentative values for the energy flux $F\cz$ driving the convection, which ends up being much higher in the simulations than in either Jupiter or Saturn. We are also forced to choose much higher values for the dimensional diffusivities $\nucz$ and $\kappacz$ than would be expected for any giant planet. This mismatch between the simulation and the real system in terms of the dimensional parameters (which also appears as a mismatch in the nondimensional parameters; see Table \ref{tab:input_nond} and Section \ref{sec:nondim}) is, unfortunately, an unavoidable consequence of limited computing power: in order to capture the extreme parameters of a real giant planet, an extreme range of dynamical length scales and time scales would be required, corresponding to a required grid resolution and time-step orders of magnitude more expensive than even the most powerful supercomputers can currently achieve.}



\end{document}